\documentclass[ reprint,  amsmath,amssymb,  aps]{revtex4-1}%
\usepackage{graphicx}
\usepackage{dcolumn}
\usepackage{bm}
\usepackage{amsmath}
\usepackage{amsfonts}
\usepackage{amssymb}
\usepackage{color}
\usepackage{lineno}
\newcommand*\patchAmsMathEnvironmentForLineno[1]{
\expandafter\let\csname old#1\expandafter\endcsname\csname #1\endcsname
\expandafter\let\csname oldend#1\expandafter\endcsname\csname end#1\endcsname
\renewenvironment{#1}
{\linenomath\csname old#1\endcsname}
{\csname oldend#1\endcsname\endlinenomath}}
\newcommand*\patchBothAmsMathEnvironmentsForLineno[1]{
\patchAmsMathEnvironmentForLineno{#1}
\patchAmsMathEnvironmentForLineno{#1*}}
\AtBeginDocument{
\patchBothAmsMathEnvironmentsForLineno{equation}
\patchBothAmsMathEnvironmentsForLineno{align}
\patchBothAmsMathEnvironmentsForLineno{flalign}
\patchBothAmsMathEnvironmentsForLineno{alignat}
\patchBothAmsMathEnvironmentsForLineno{gather}
\patchBothAmsMathEnvironmentsForLineno{multline}
}

\providecommand{\U}[1]{\protect\rule{.1in}{.1in}}

\begin{document}

\title{
Field-induced electric polarization and elastic softening caused by parity-mixed $d$-$p$ hybridized states with electric multipoles in Ba$_2$CuGe$_2$O$_7$
}

\author{
R. Kurihara$^1$, 
Y. Sato$^2$,
A. Miyake$^2$,
M. Akaki$^3$,
K. Mitsumoto$^4$,
M. Hagiwara$^5$,
H. Kuwahara$^6$,
and
M. Tokunaga$^2$
}

\affiliation{
$^1$Department of Physics and Astronomy, Faculty of Science and Technology, Tokyo University of Science, Noda, Chiba 278-8510, Japan	
}
\affiliation{
$^2$The Institute for Solid State Physics, The University of Tokyo, Kashiwa, Chiba 277-8581, Japan
}
\affiliation{
$^3$Institute for Materials Research, Tohoku University, Sendai, Miyagi 980-8577, Japan
}
\affiliation{
$^4$Liberal Arts and Sciences, Toyama Prefectural University, Imizu, Toyama 939-0398, Japan
}
\affiliation{
$^5$Center for Advanced High Magnetic Field Science, Graduate School of Science, Osaka University, Toyonaka, Osaka 560-0043, Japan
}
\affiliation{
$^6$Department of Physics, Sophia University, Chiyoda, Tokyo 102-855, Japan
}

\begin{abstract}
We performed high-magnetic-field magnetization, polarization, and ultrasonic measurements in Ba$_2$CuGe$_2$O$_7$ to investigate field-induced multiferroic properties arising from a cross-correlation between electric dipoles and electric quadrupoles in addition to cross-correlation between magnetic dipoles and electric dipoles.
Magnetization $M$ shows saturation behavior above 20 T for several magnetic field directions, however, electric polarization $P_c$ exhibits an increase, and elastic constants show a softening above 20 T.
Based on quantum states with a crystalline electric field for the $D_{2d}$ point group and $d$-$p$ hybridization between Cu-$3d$ and O-$2p$ electrons, we confirmed that the matrix of an electric dipole $P_z$ was proportional to that of an electric quadrupole $O_{xy}$.
Furthermore, considering the spin-orbit coupling of $3d$ electrons and the Zeeman effect, we showed that $P_z$ and $O_{xy}$ simultaneously exhibited field-induced responses.
These findings indicate that the orbital degrees of freedom, in addition to the spin degrees of freedom, contribute to the high-field multiferroicity in Ba$_2$CuGe$_2$O$_7$.
\end{abstract}

\maketitle


\section{
\label{sect_intro}
Introduction
}
A cross-correlation between magnetic dipoles and electric dipoles has been focused on solid-state physics
\cite{Fiebig}.
Because this phenomenon describes how magnetic (electric) fields generate electric (magnetic) fields in solids, scientists and engineers aim for application to some electronic devices.
Furthermore, a piezoelectric effect due to a cross-correlation between electric dipoles and elastic quadrupoles, which describes how electric fields (elastic strains) induce elastic strains (electric fields)
\cite{Cady_Text},
has also been applied to a variety of items like motors, injectors, speakers, stages, etc.
These phenomena arising from the interplay of different types of multipoles are known as multiferroicity.
Understanding the multiferroicity mechanism is an important scientific subject for further applications.

Several mechanisms have been proposed to describe multiferroic properties between magnetic dipoles and electric dipoles.
One is an exchange striction model described as
$\left| \boldsymbol{P}_{ij} \right| \propto \boldsymbol{S}_i \cdot \boldsymbol{S}_j$
\cite{Arima_PRL96, Tokunaga_PRL101, Ishiwata_PRB81}.
Here, $\boldsymbol{P}_{ij}$ is the polarization vector, and $\boldsymbol{S}_i$ ($\boldsymbol{S}_j$) is the spin at site $i$ ($j$).
Another is an inverse Dzyaloshinskii-Moriya (DM) model written as
$\boldsymbol{P}_{ij} \propto \boldsymbol{e}_{ij} \times (\boldsymbol{S}_i \times \boldsymbol{S}_j)$,
where $\boldsymbol{e}_{ij}$ denotes the unit vector connecting the neighboring site $i$ and $j$
\cite{Katsura_PRL95}.
These two mechanisms are based on inter-site interactions.
Therefore, the magnetic structure plays a key role in such polarizations.
In contrast, a spin-dependent $d$-$p$ hybridization model
\cite{Arima_JPSJ76, Murakawa_PRL105},
described as $\boldsymbol{P}_{l} \propto (\boldsymbol{S} \cdot \boldsymbol{r}_{l} )^2 \boldsymbol{r}_{l}$, is based on a cluster consisting of a magnetic ion and non-magnetic ions.
Here, $\boldsymbol{r}_l$ is the bonding vector between the transition metal and the $l$-th ligand oxygen ion.
This model indicates that the rotation of the magnetic moments in real space carries a polarization change.
As long as magnetization is increased by magnetic fields, a field-induced electric polarization (FIEP) can be expected in such a model.

\begin{figure}[t]
\begin{center}
\includegraphics[clip, width=0.5\textwidth, bb=0 0 580 280]{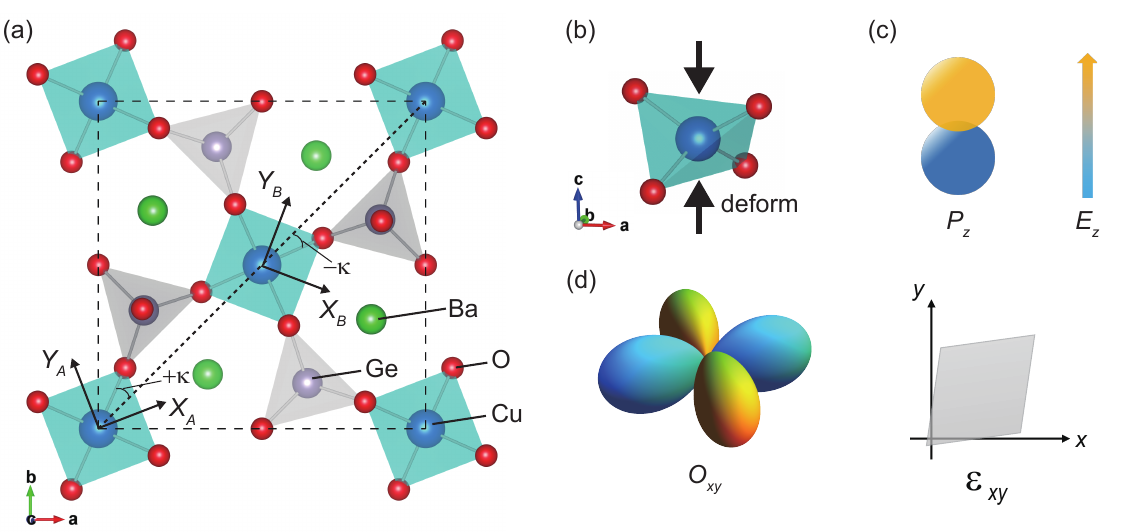}
\end{center}
\caption{
(a) Top view of the crystal structure of Ba$_2$CuGe$_2$O$_7$ produced by VESTA
\cite{VESTA}.
$a$ and $b$ axes indicate the crystallographic coordinates.
$X_A$ ($X_B$) and $Y_A$ ($Y_B$) axes are defined as the local coordinates at $A$ site ($B$ site).
$\pm \kappa$ indicates the tilting angle of CuO$_4$ tetrahedra about the c axis.
(b) Schematic view of the CuO$_4$ tetrahedron.
Vertical arrows indicate the deformation direction of CuO$_4$.
(c) Schematic view of the electric dipole $P_z$ and the conjugating external field of the electric field $E_z$.
(d) Schematic view of the electric quadrupole $O_{xy}$ and the conjugate external field of the crystal strain $\varepsilon_{xy}$.
}
\label{CrystalStructure}
\end{figure}

\begin{table*}[htbp]
\caption{
Characters and several basis functions of the point group $D_{2d}$.
Here, we do not consider the time-reversal symmetry.
$l_x = -i(y\partial_z - z\partial_y)$, $l_y = -i(z\partial_x - x\partial_z)$, and $l_z = -i(x\partial_y - y\partial_x)$ are the angular momentum operators.
$l_i$ ($i = x, y, z$) is related to the rotation of crystal lattice and electronic systems around the $i$-axis and the Zeeman effect.
$B_x$, $B_y$, and $B_z$ represent the components of the magnetic fields.
}
\begin{ruledtabular}
\label{table_character of D2d}
\begin{tabular}{cccccccccc}
\textrm{Irrep}
	& $E$
		& $2IC_4$
			& $C_2$
				& $2C_2'$
					& $2\sigma_d$
						& \textrm{Basis function: polar}
							& \textrm{axial}
								& \textrm{quadratic}
									& \textrm{producted}
\\
\hline
$A_1$
	& $1$
		& $1$
			& $1$
				& $1$
					& $1$
						& 
							&
								&$z^2$
									&$xyz$
									\\
	&	&	&	&	&	&	&	&	&$z \left(  l_x l_y + l_y l_x \right)$, $x l_x  - y l_y$, $yz l_x - zx l_y$										
\\
	&	&	&	&	&	&	&	&	&$z B_x B_y$, $x B_x  - y B_y$, $xy B_x B_y$, $yz B_x - zx B_y$									
\\
	&	&	&	&	&	&	&	&	&$l_x B_x  + l_y B_y$										
\\
$A_2$
	& $1$
		& $1$
			& $1$
				& $-1$
					& $-1$
						& 
							&$l_z$, $B_z$
								&
									&$z\left(x^2 - y^2 \right)$, $xy\left(x^2 - y^2 \right)$	
\\
$B_1$
	& $1$
		& $-1$
			& $1$
				& $1$
					& $-1$
						& 
							&
								&$x^2-y^2$
									&	
\\
$B_2$
	& $1$
		& $-1$
			& $1$
				& $-1$
					& $1$
						& $z$
							&
								& $xy$
									&$\left(  l_x l_y + l_y l_x \right)$	
\\
$E$
	& $2$
		& $0$
			& $-2$
				& $0$
					& $0$
						& $\{x, y\}$
							&$\{l_x, l_y\}$, $\{B_x, B_y \}$
								&$\{yz, zx\}$
									&	
\end{tabular}
\end{ruledtabular}
\end{table*}

Recently, the spin-dependent $d$-$p$ hybridization has been proposed in Ba$_2$$X$Ge$_2$O$_7$ ($X$ = Mn, Co, Cu) to describe the FIEP of  $P_c$ along the $c$ axis of the crystallographic orientation accompanied by a magnetization process
\cite{Murakawa_PRL105, Murakawa_PRB85}. 
Here, $P_c$ is the sum of the electric dipole $P_z$ in the solid, and the quantization axis $z$ is set along the $c$-axis.
Figure \ref{CrystalStructure}(a) shows the crystal structure of Ba$_2$CuGe$_2$O$_7$ with an \aa kermanite-type structure belonging to the tetragonal space group $P\overline{4}2_1m$.
Cu ions responsible for magnetic moments are at the center of deformed O$_4$ tetrahedra (see Fig. \ref{CrystalStructure}(b)).
Thus, a point group symmetry at Cu-site is $D_{2d}$ without the inversion operation ($I$)
\cite{Inui_group}.
The local coordinates of CuO$_4$ clusters, $X_i$ and $Y_i$, ($i = A$ and $B$), are tilted around $c$-axis with the angle $\pm \kappa = \pm 22^\circ$ (see Fig. \ref{CrystalStructure}(a)).
$3d$ and $2p$ orbitals in the CuO$_4$ clusters form valence and conduction bands
\cite{Corasaniti_PRB96}.
A magnetic structure below antiferromagnetic (AFM) transition at $T_\mathrm{N} = 3.2$ K shows a spiral structure 
\cite{Zheludev_PRB54}
because the DM interaction is active in such tilting crystals without the inversion operation.
In magnetic fields, the magnetic structure changes from an incommensurate spiral to a commensurate one
\cite{Zheludev_PRB57}.
The electric polarization $P_c$ for $B//[110]$ exhibits an anomaly at such a transition field in addition to the magnetization curve
\cite{Murakawa_PRB85}.

On the other hand, another mechanism describing the FIEP has been indicated.
In  \aa kermanite-type compounds Sr$_2$CoSi$_2$O$_7$ and Ca$_2$CoSi$_2$O$_7$, the FIEP has been observed even in spin saturation regions
\cite{Akaki_PRB86, Akaki_JPSJ83}.
Since the spin-dependent $d$-$p$ hybridization model is proportional to $(\boldsymbol{S} \cdot \boldsymbol{r}_{l} )^2 \boldsymbol{r}_{l}$, field-dependent electric polarization cannot be expected above the spin saturation fields.
Furthermore, a theoretical study has suggested that such a spin-dependent scenario based on the Kramers doublet with spin-1/2 is excluded in terms of the vanishment of single-site anisotropy
\cite{Ono_PRB102}.
Therefore, we deduce that other electronic degrees of freedom also contribute to the FIEP.

One of the candidate degrees of freedom can be electric multipoles.
A previous study in Ba$_2$CuGe$_2$O$_7$ has focused on a cross-correlation between the electric dipole, $P_z \propto z$, and the spin-nematic operator, $S_x S_y + S_y S_x$, that is equal to the electric quadrupole, $O_{xy} \propto xy$ (see Figs. \ref{CrystalStructure}(c) and \ref{CrystalStructure}(d))
\cite{Soda_PRL112}.
Because the spatial inversion symmetry is broken at Cu-sites centered at O$_4$ tetrahedra, both $P_z$ and $O_{xy}$ belong to the irreducible representation (irrep) $B_2$ of the point group $D_{2d}$ (see Tables \ref{table_character of D2d} and \ref{table_multipoles} and Appendix \ref{Appendix_BeforeBeforeA})
\cite{Inui_group, Luthi_Text}.
In other words, the matrix of $P_z$ should be proportional to the matrix of $O_{xy}$ for the quantum states under the $D_{2d}$. 
This fact indicates that the response of electric quadrupole $O_{xy}$ induces the electric polarization $P_c$ via the response of the electric dipole $P_z$.
Furthermore, due to the coupling between the electric quadrupole $O_{xy}$ and an elastic strain $\varepsilon_{xy}$ (see  Fig. \ref{CrystalStructure}(d)), we can expect the electric polarization mediated by the crystal distortion of the CuO$_4$ tetrahedra
\cite{Yamauchi_PRB84, Barone_PRB92}.
Above the spin saturation fields, we deduce that the field-dependent electric polarization is brought about by the cross-correlation described as $P_z \propto O_{xy}$.

Since the electric dipole and quadrupole responses are expected, the orbital part of wave functions plays a key role in the FIEP.
From the microscopic point of view, the contribution of $d$-$p$ hybridized states consisting of O-$2p$ and Cu-$3d$ orbitals is expected because parity-mixed wave functions lead the electric dipole degree of freedom
\cite{Hayami_PRB98}.
In addition, $3d$-$yz$ and $zx$ orbitals and $2p$-$x$ and $y$ orbitals in the $d$-$p$ hybridized states  also carry the electric quadrupole $O_{xy}$
\cite{Kurihara_JPSJ86}.
Therefore, we focused on the polarization and elastic constant measurements in high-magnetic fields to detect the electric dipole and electric quadrupole response in Ba$_2$CuGe$_2$O$_7$.
Furthermore, we discuss another possible mechanism of the FIEP based on the $d$-$p$ hybridization and spin-orbit coupling.

This paper is organized as follows.
In Sect. \ref{sect_exp}, the experimental procedures are described.
In Sect. \ref{sect_3}, we present the experimental results of the high-field magnetization, polarization, and ultrasonic measurements.
We show the field-induced elastic softening in addition to the field-dependent electric polarization above the spin saturation fields.
In Sect. \ref{sect_4}, we present quantum states in high fields and the electric dipole and quadrupole susceptibilities to describe our experimental results of polarization and elastic constants.
We conclude our results in Sect. \ref{conclusion}.

\begin{table*}[htb]
\caption{
Conjugate fields, electric multipoles, and response functions corresponding to the irreducible representations (irreps) of the point group $D_{2d}$ at Cu sites.
$E_i$ ($i = x$, $y$, $z$) is an electric field and $\varepsilon_\Gamma$ ($\Gamma = 3z^2-r^2$, $x^2-y^2$, $yz$, $zx$, $xy$) is a symmetry strain, respectively.
The electric dipole $P_i$ ($i = x$, $y$, $z$) and the electric quadrupole $O_\Gamma$ ($\Gamma = 3z^2-r^2$, $x^2-y^2$, $yz$, $zx$, $xy$) are written as the electric multipoles.
The sign in column $I$ indicates the spatial inversion property of conjugating fields and electric multipoles (even: $+$, odd: $-$).  
Extra suffixes g (gerade) and u (un-gerade) of the irreps describe the parity of conjugate fields, multipoles, and response functions.
}
\begin{ruledtabular}
\label{table_multipoles}
\begin{tabular}{cccccc}
\textrm{Irrep}
& \textrm{Conjugating field}
	& \textrm{Electric multipole}
		& \textrm{Response function}
			& $I$
\\
\hline
$A_\mathrm{1(g)}$
	& $\varepsilon_\mathrm{B} = \varepsilon_{xx} + \varepsilon_{yy} + \varepsilon_{zz}$
		& $O_\mathrm{B} = 1/r$
			& $ C_\mathrm{B} = \left( 2C_{11} + 2C_{12} + 4C_{13} + C_{33} \right)/9 $
				& $+$
\\

	& $\varepsilon_u = (2\varepsilon_{zz} - \varepsilon_{xx} - \varepsilon_{yy})/\sqrt{3}$
		& $O_{3z^2-r^2} = \left( 3z^2-r^2 \right)/r^2$
			& $C_u = \left( C_{11} + C_{12} - 4C_{13} + 2C_{33} \right)/6$
				& $+$
\\
$B_\mathrm{1(g)}$
	& $\varepsilon_{x^2-y^2} = (\varepsilon_{xx} - \varepsilon_{yy})/\sqrt{2}$
		& $O_{x^2-y^2} = \left( x^2 -y^2 \right)/( \sqrt{2} r^2)$
			& $C_\mathrm{T} = \left( C_{11} - C_{12} \right)/2$
				& $+$
\\
$B_\mathrm{2(g)}$
	& $\sqrt{2} \varepsilon_{xy}$
		& $O_{xy} =  \sqrt{2} xy/r^2$
			& $C_{66}$
				& $+$
\\
$B_\mathrm{2(u)}$
	& $E_z$
		& $P_z =  z/r$
			& $P_c$
				& $-$
\\
$E_\mathrm{(g)}$
	& $ \sqrt{2} \varepsilon_{yz}$
		& $O_{yz} = \sqrt{2} yz/r^2$
			& $C_{44}$
				& $+$
\\

	& $\sqrt{2} \varepsilon_{zx}$
		& $O_{zx} = \sqrt{2} zx/r^2$
			&$C_{44}$
				& $+$
\\
$E_\mathrm{(u)}$
	& $E_x$
		& $P_x = x/r$
			&$P_a$ 
				& $-$
\\

	& $E_y$
		& $P_y = y/r$
			&$P_b$  
				& $-$
\end{tabular}
\end{ruledtabular}
\end{table*}

\section{
\label{sect_exp}
Experimental details
}

Single crystals of Ba$_2$CuGe$_2$O$_7$ were grown by the floating zone method.
The Laue X-ray backscattering technique was used to cut samples with (100), ($\bar{1}$00), (001), and (00$\bar{1}$) faces and (110), (1$\bar{1}$0), (001), and (00$\bar{1}$) faces.
The magnetization $M$ in pulsed magnetic fields was measured by the induction method using coaxial pickup coils.
The FIEP of $P_c$ was obtained by integrating the polarization current as a function of time
\cite{Mitamura_JPSJ76}.
The ultrasonic pulse-echo method with a numerical vector-type phase detection technique was used for the ultrasonic velocity, $v$
\cite{Fujita_JPSJ80, Kohama_JAP132}.
Piezoelectric transducers using LiNbO$_3$ plates with a 36$^\circ$ Y-cut and an X-cut (Yamaju Co.) were employed to generate longitudinal ultrasonic waves with the fundamental frequency of approximately $f = 30$ MHz and the transverse waves with 16 MHz, respectively. 
Higher-harmonic frequencies of 68 MHz and 112 MHz were also employed.
The elastic constant, $C = \rho v^2$, was obtained from the ultrasonic velocity, $v$.
Here, the mass density, $\rho = 5.07$ g/cm$^3$, for Ba$_2$CuGe$_2$O$_7$ is calculated by the structural parameters
\cite{Zheludev_PRB54}.
The direction of ultrasonic propagation, $\boldsymbol{q}$, and the direction of polarization, $\boldsymbol{\xi}$, for the elastic constant, $C_{ij}$, are indicated in figures.
$g$ values of $2.04$ for the in-plane field direction and $2.44$ for the inter-plane field direction based on the Cu$^{2+}$ ions were estimated by the X-band electron spin resonance (ESR) measurements with the frequency of 9.12 GHz.

For high-field measurements up to 60 T, non-destructive pulse magnets with time durations of 25 and 36 ms installed at The Institute for Solid State Physics, The University of Tokyo were used. 

The energy scheme and the electric multipole susceptibility are calculated by the Julia language.

\section{
\label{sect_3}
Experimental results and discussions
}

\subsection{
\label{subsect_Magnetization}
High-field magnetization
}

\begin{figure}[htbp]
\begin{center}
\includegraphics[clip, width=0.5\textwidth, bb=0 0 600 360]{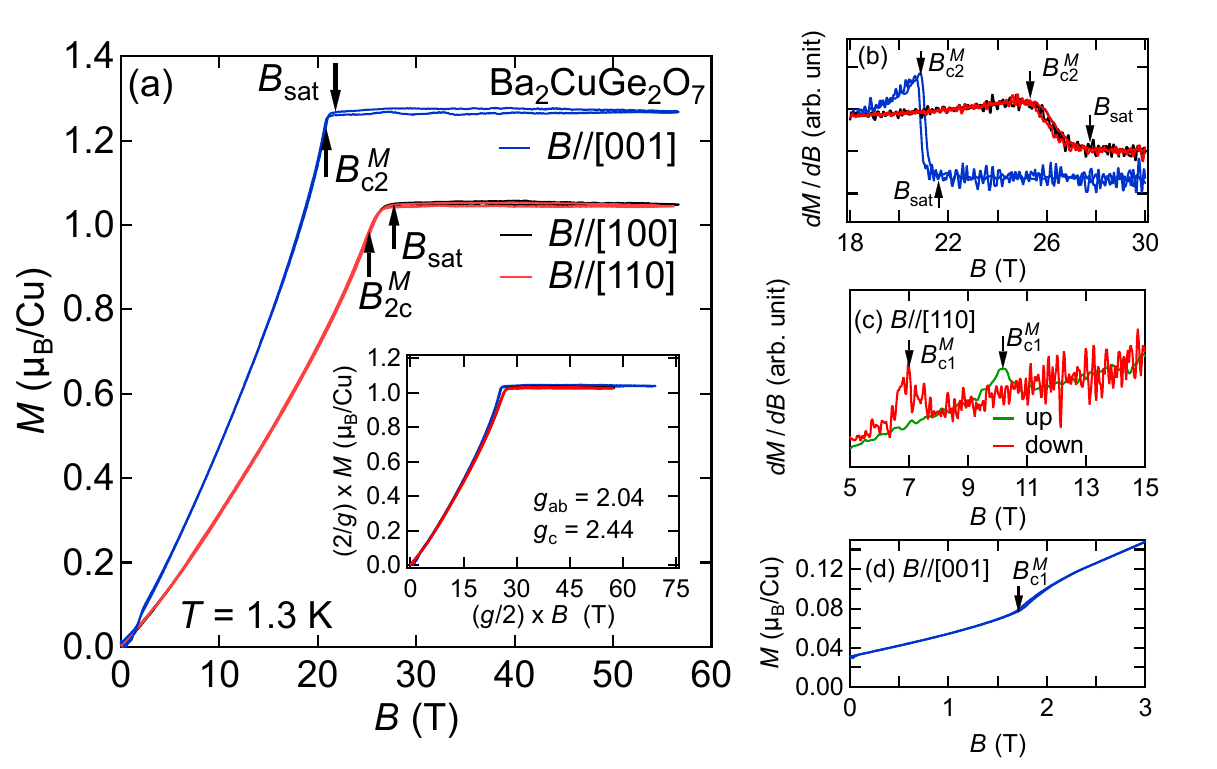}
\end{center}
\caption{
(a) Magnetization curve $M$ in Ba$_2$CuGe$_2$O$_7$ at 1.3 K for $B//[001]$, $B//[100]$, and $B//[110]$.
The vertical arrows indicate characteristic fields $B_\mathrm{c2}^M$ and $B_\mathrm{sat}$.
The inset in (a) shows the normalized magnetization curves, $(2/g)M$, as a function of normalized magnetic field, $(2/g)B$.
(b) First derivative of $M$ with respect to magnetic field $B$ for $B//[001]$, $B//[100]$, and $B//[110]$ in the range of 18 to 30 T.
The vertical arrows indicate transition fields $B_\mathrm{c2}^M$ and $B_\mathrm{sat}$.
The data sets are shifted consecutively along the $dM/dB$ axes for clarity. 
(c) $dM/dB$ for $B//[110]$ in the range of 5 to 15 T.
The vertical arrows indicate transition fields $B_\mathrm{c1}^M$ for the field up-sweep and down-sweep.
The data sets are shifted consecutively along the $dM/dB$ axes for clarity. 
(d) $M$ for $B//[001]$ below 3 T.
The vertical arrow indicates transition field $B_\mathrm{c1}^M$.
}
\label{M_1p3K}
\end{figure}

Figure \ref{M_1p3K}(a) shows the magnetization as a function of fields in Ba$_2$CuGe$_2$O$_7$ at $1.3$ K for various field directions.
We observed several characteristic anomalies and a magnetization plateau in high fields.
The magnetization curve $M$ for $B//[100]$ and $[110]$ show the increase in the field up to $B_\mathrm{c2}^M = 25.3$ T, which is determined by the first derivative of the magnetization with respect to the field, $dM/dB$, in Fig. \ref{M_1p3K}(b).
With further application of the fields, $M$ exhibits a plateau above $B_\mathrm{sat} = 27.8$ T up to 56 T.
The difference in $B_\mathrm{c2}^M$ and $B_\mathrm{sat}$ between the field up-sweep and down-dweep processes was hardly noticeable within our experimental resolution.
We also observed the anomaly at $B_\mathrm{c1}^M = 10.1$ T for field up-sweep and $7.1$ T for down-sweep in $dM/dB$ (Fig. \ref{M_1p3K}(c)). 

As discussed later, the anomaly $B_\mathrm{c1}^M$ corresponds to the transition field of the incommensurate spiral to the commensurate one.
For $B//[001]$, $M$ shows the increase in the field up to $B_\mathrm{c2}^M = 20.8$ T and a plateau above $B_\mathrm{sat} = 21.6$ T.
The anomaly at $B_\mathrm{c1}^M = 1.7 $ T is also observed as shown in Fig. \ref{M_1p3K}(d).
As shown in the inset of Fig. \ref{M_1p3K}(a), each magnetization curve is normalized using the $g$ value of $2.04$ for the in-plane field direction and $2.44$ for inter-plane field direction, which is comparable to the previous study
\cite{Zheludev_PRL78}.
Above $B_\mathrm{sat}$, normalized magnetizations $(2/g)M$ show almost 1, suggesting that the polarized paramagnetic (PPM) state is realized.

\begin{figure}[t]
\begin{center}
\includegraphics[clip, width=0.5\textwidth, bb=0 0 600 400]{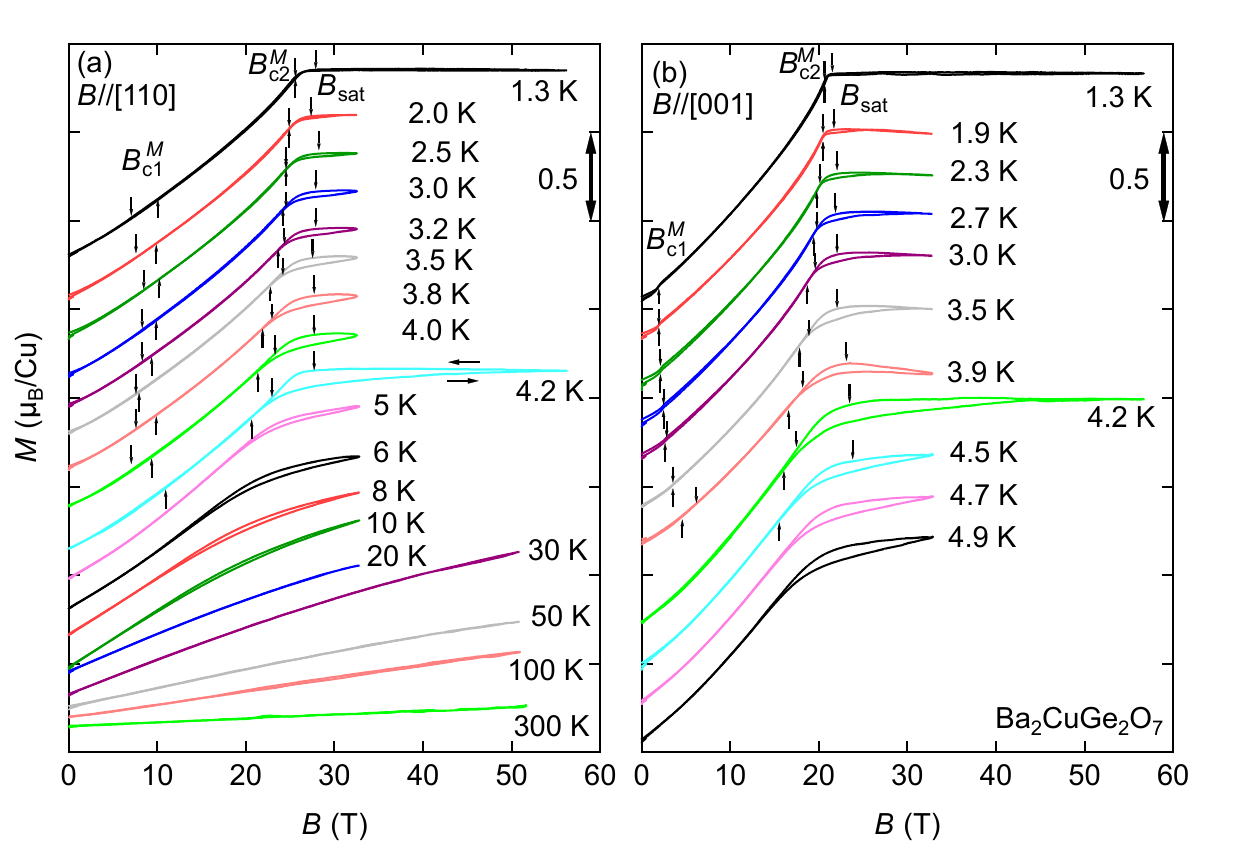}
\end{center}
\caption{
Magnetization curve $M$ in Ba$_2$CuGe$_2$O$_7$ at several temperatures for (a) $B//[110]$ and (b) $B//[001]$.
The vertical arrows indicate the characteristic fields $B_\mathrm{c1}^M$, $B_\mathrm{c2}^M$, and $B_\mathrm{sat}$.
Up-arrows (down-arrows) for $B_\mathrm{c1}^M$ and $B_\mathrm{c2}^M$ indicate the anomaly for field up-sweep (down-sweep).
The data sets are shifted consecutively along the $M$ axes for clarity. 
}
\label{M_All}
\end{figure}

Figures \ref{M_All} show the magnetization as a function of fields in Ba$_2$CuGe$_2$O$_7$ at several temperatures.
We investigated the evolution of anomalies with temperature changes.
As the temperature is raised, the anomaly at $B_\mathrm{c2}^M$ shifts to a lower field and the hysteresis loop opens for both field directions of $[110]$ and $[001]$.
The saturation field $B_\mathrm{sat}$ shifts to higher fields with increasing temperatures for both $B//[110]$ and $B//[001]$.
Here, $B_\mathrm{sat}$ was determined from the field down-sweep process because of the gradual change in $M$ for the up-sweep process.
The hysteresis behavior at $B_\mathrm{c1}^M$ appears below 4.0 K for both $B//[110]$ and $B//[001]$.
The difference of $B_{c1}^M$ for $B//[001]$ between the field up-sweep and field down-sweep is smaller than that for $B//[110]$.
Above 10 K, the loop closes for $B//[110]$.
$B_\mathrm{c2}^M$ also shifts to a lower field with increasing temperatures.
In addition, the bend in $M$ at $B_{c2}^M$ becomes more gradual, and hysteresis behavior becomes more pronounced.
In the magnetization curve for $B//[110]$ at 4.2 K, the shape for the field down-sweep seems to be similar to the $M$ at 1.3 K while the shape of $M$ for the field up-sweep can be similar to that at 5 K.
We deduce that the hysteresis behavior around $B_\mathrm{c2}$ and $B_\mathrm{sat}$ originates from temperature changes in the sample caused by magnetocaloric effects under quasi-adiabatic conditions
\cite{Kihara_RSI84, Miyake_JPSJ88, Miyake_JPSJ90}.
The magnetocaloric effect can lead to a lower sample temperature during the field down-sweep process compared to the field up-sweep process.


As shown above, we observed several anomalies in the magnetization curves.
In particular, the observation of spin saturation plays a key role in understanding the multiferroic mechanism in Ba$_2$CuGe$_2$O$_7$.
In the following Sect. \ref{subsect_PhaseDiagram}, we discuss the magnetic phase diagram and the origin of anomalies.

\subsection{
\label{subsect_PhaseDiagram}
Phase diagram
}

\begin{figure}[htbp]
\begin{center}
\includegraphics[clip, width=0.5\textwidth, bb=0 0 550 400]{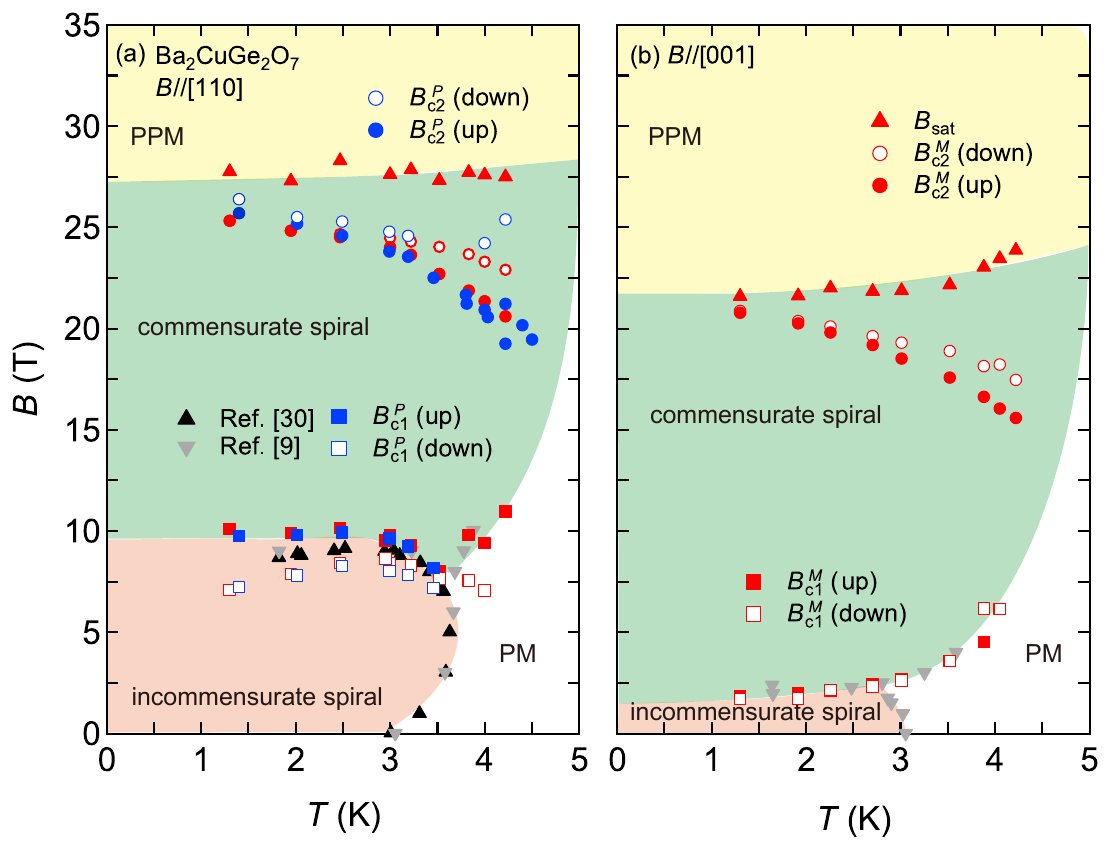}
\end{center}
\caption{
Temperature-field phase diagram of Ba$_2$CuGe$_2$O$_7$ for (a) $B//[110]$ and (b) $B//[001]$.
The spin saturation field $B_\mathrm{sat}$ determined by the magnetization is shown by the filled red triangles.
The characteristic field $B_\mathrm{c1}^M$ ($B_\mathrm{c1}^P$) determined by the magnetization (polarization) measurements is indicated by filled red (blue) squares for field up-seep and open red (blue) squares for down-sweep, respectively.
The characteristic field $B_\mathrm{c2}^M$ ($B_\mathrm{c2}^P$) determined by the magnetization (polarization) measurements is indicated by filled red (blue) circles for field up-seep and open red (blue) circles for down-sweep, respectively.
The filled black triangles and filled gray triangles show transition fields in Refs.
\cite{Muhlbauer_PRB86} and \cite{Murakawa_PRB85}.
The background colors are the guide to distinguish each phase.
}
\label{PhaseDiagram}
\end{figure}

To compare the characteristic anomalies to those of the previous reports, we summarized several characteristic fields of Ba$_2$CuGe$_2$O$_7$ in the temperature-magnetic field phase diagram in Fig. \ref{PhaseDiagram}.
For $B//[110]$, the phase boundary below 10 T seems to be consistent with the previous reports (see Fig. \ref{PhaseDiagram}(a))
\cite{Murakawa_PRB85, Muhlbauer_PRB86}.
Thus, we conclude that the anomaly at $B_\mathrm{c1}^M$ originates from the transition of the incommensurate spiral to the commensurate one.
The saturation field $B_\mathrm{sat}$ is our new observations.
$B_\mathrm{sat}$ seems to be almost independent of temperatures.
We also observed $B_\mathrm{c2}^M$ obtained from the peak structure of $dM/dB$.
$B_\mathrm{c2}^M$ decreases with increasing temperatures.

For $B//[001]$, the low-field phase boundary seems to be consistent with the previous results (see Fig. \ref{PhaseDiagram}(b))
\cite{Zheludev_PRB57, Muhlbauer_PRB86}.
Thus, $B_\mathrm{c1}^M$ for $B//[001]$ can correspond to the boundary of the incommensurate spiral to the commensurate one.
In contrast, the AF cone state
\cite{Muhlbauer_PRB86}
is hardly visible in our high-field measurements.
We also observed the anomaly at $B_\mathrm{c2}^M$ and $B_\mathrm{sat}$ for $B/[001]$.
$B_\mathrm{sat}$ increases with increasing temperature while $B_\mathrm{c2}^M$ decreases.
Our magnetization measurements confirmed the phase boundary between the incommensurate spiral to the commensurate one.
We observed the difference in $B_{c1}$ between the field up-sweep and down-sweep processes.

As shown in the phase diagram, our results in low fields are consistent with those in the previous studies
\cite{Murakawa_PRB85, Muhlbauer_PRB86}.
In addition to the previous studies, we demonstrate that the polarized paramagnetic state can be realized above $B_\mathrm{sat}$.

\subsection{
\label{subsect_ElectricPolarization}
Field-induced electric polarization
}

\begin{figure}[t]
\begin{center}
\includegraphics[clip, width=0.5\textwidth, bb=0 0 450 450]{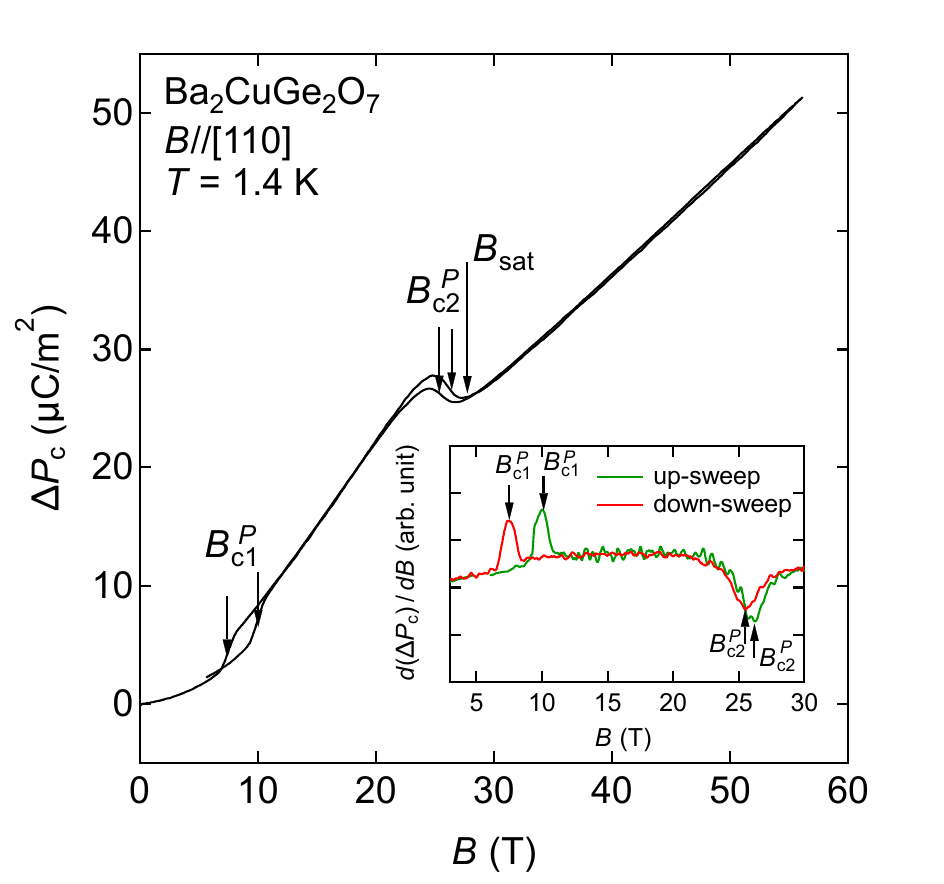}
\end{center}
\caption{
Magnetic field dependence of the electric polarization $\mathit{\Delta}P_c$ in Ba$_2$CuGe$_2$O$_7$ at 1.4 K for $B//[110]$.
The down arrows indicate the characteristic transition fields $B_\mathrm{c1}^P$ and $B_\mathrm{c2}^P$ determined by $P_c$ and $B_\mathrm{sat}$ by $M$.
The inset shows $d (\mathit{\Delta}P_c)/dB$.
}
\label{P_1p4K}
\end{figure}

To investigate the multiferroicity above the spin saturation regions, we measured the polarization in high fields.
Figure \ref{P_1p4K} shows the magnetic field dependence of the electric polarization $\mathit{\Delta}P_c$ in Ba$_2$CuGe$_2$O$_7$ for $B//[110]$ at 1.4 K.
We also exhibits the first derivative of $\mathit{\Delta}P_c$ with respect to $B$, $d(\mathit{\Delta}P_c)/dB$, in the inset of Fig. \ref{P_1p4K}.
We observed several anomalies in the magnetization.
$\mathit{\Delta}P_c$ and $d (\mathit{\Delta}P_c)/dB$ exhibit the anomalies at $B_\mathrm{c1}^P = 9.3$ T for field up-sweep and $7.2$ T for down-sweep.
The anomaly $B_\mathrm{c1}^P$ corresponds to the transition field of the incommensurate spiral to the commensurate one.
The hysteresis behavior around $B_\mathrm{c1}$ can be attributed to the field-induced magnetic structure change (see appendix \ref{Appendix_Before_A}).
For further increasing fields, $\mathit{\Delta}P_c$ shows the anomaly at $B_\mathrm{c2}^P = 25.7$ T (26.4 T) for field up-sweep (down-sweep).
$B_\mathrm{c2}^P$ in $\mathit{\Delta}P_c$ is consistent with $B_\mathrm{c2}^M$ in the magnetization for $B//[110]$.
Above the saturation field $B_\mathrm{sat}$, $\mathit{\Delta}P_c$ increases in the field up to 56 T.

\begin{figure}[htbp]
\begin{center}
\includegraphics[clip, width=0.45\textwidth, bb=0 0 450 520]{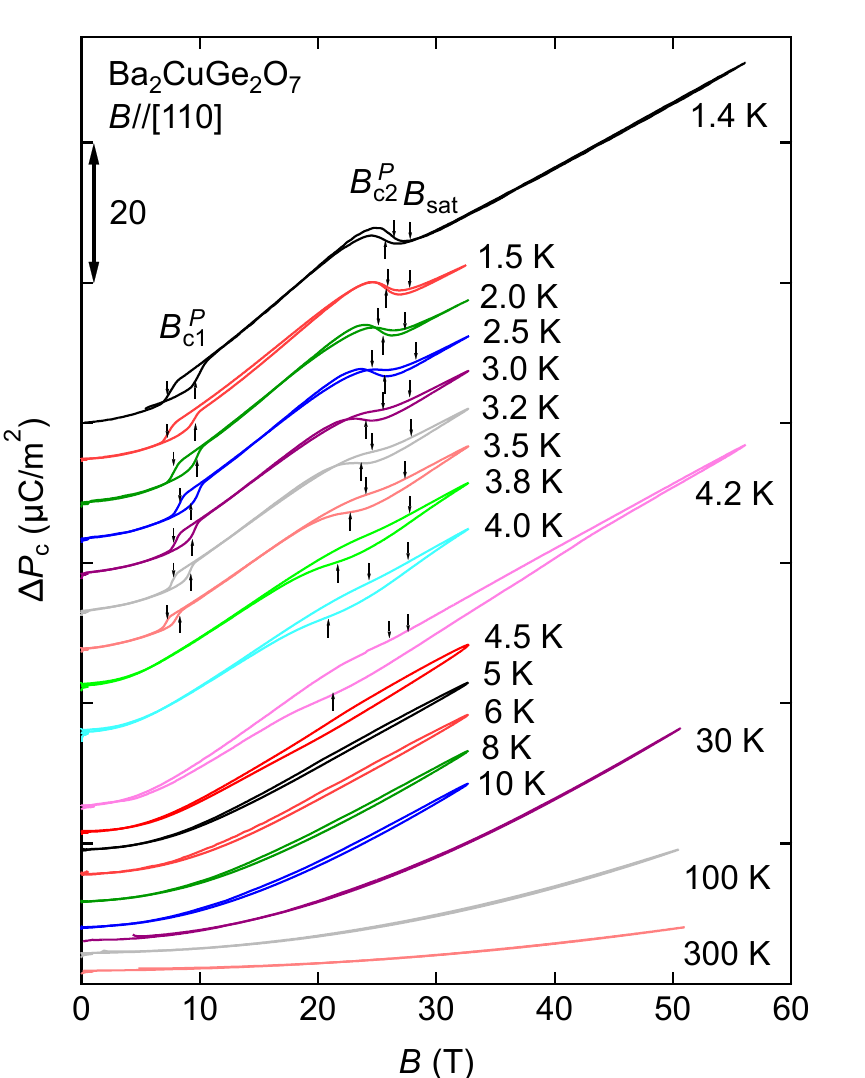}
\end{center}
\caption{
Magnetic-field dependence of the electric polarization $\mathit{\Delta}P_c$ in Ba$_2$CuGe$_2$O$_7$ at several temperatures for $B//[110]$.
The vertical arrows indicate the characteristic fields $B_\mathrm{c1}^P$ and $B_\mathrm{c2}^P$ determined by $P_c$ and $B_\mathrm{sat}$ by $M$.
In particular, up-arrows (down-arrows) for $B_\mathrm{c1}^P$ and $B_\mathrm{c2}^P$ indicate the anomaly for field up-sweep (down-sweep).
The data sets are shifted consecutively along the $\mathit{\Delta}P_c$ axes for clarity.
}
\label{P_All}
\end{figure}

To investigate the temperature dependence of these anomalies and the deviation of $\mathit{\Delta}P_c$ from 1.4 K, we also measured the magnetic-field dependence of $\mathit{\Delta}P_c$ in Ba$_2$CuGe$_2$O$_7$ at several temperatures (see Fig. \ref{P_All}). 
As the temperature is raised, $B_\mathrm{c1}^P$ shifts to lower fields and the hysteresis loop closes.
$B_\mathrm{c2}^P$ also shifts to a lower field with increasing temperatures.
In addition, the hump-shaped anomaly around $B_\mathrm{c2}^P$ disappears and hysteresis behavior becomes more pronounced.
Above 3.8 K, the anomaly at $B_\mathrm{c1}^P$ is hardly visible in the polarization measurements.
The hysteresis loop in high fields closes above 30 K.
We added $B_{c1}^P$ and $B_\mathrm{c2}^P$ appeared in $P_c$ to the phase diagram (see Fig. \ref{PhaseDiagram}).
These characteristic fields are consistent with those in $M$. 

Despite the spin saturation above $B_\mathrm{sat}$, our polarization measurements revealed that the polarization $P_\mathrm{c}$ showed still increase in the fields, indicating another mechanism of multiferroicity in Ba$_2$CuGe$_2$O$_7$.
Therefore, we focused on the cross-correlation described as $P_z \propto O_{xy}$ to understand the quantum states that carry the mechanism of the field-dependent electric polarization above $B\mathrm{sat}$.

\subsection{
\label{subsect_Elastic}
Elastic constants in high fields
}

\begin{figure}[htbp]
\begin{center}
\includegraphics[clip, width=0.5\textwidth, bb=0 0 450 670]{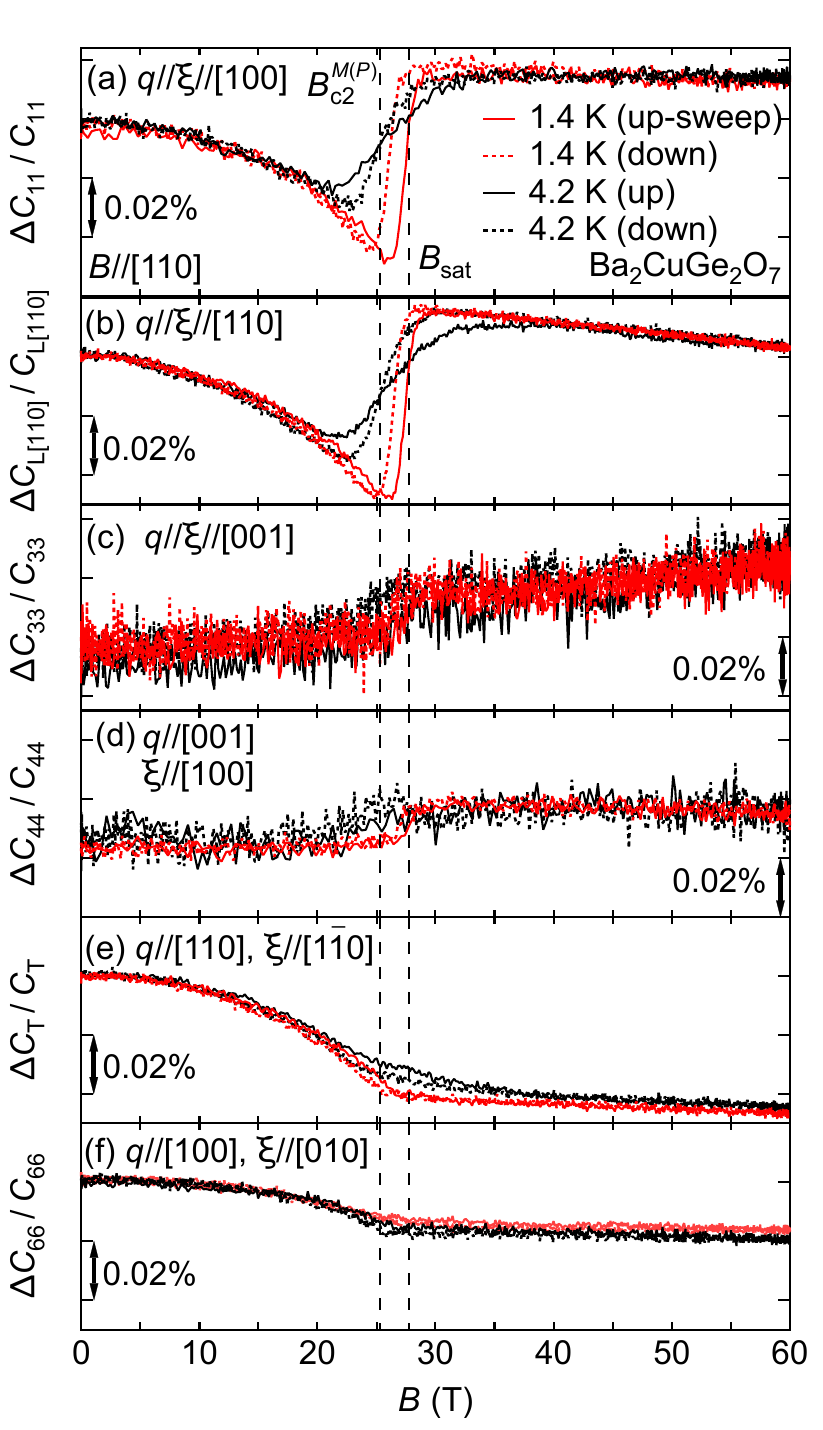}
\end{center}
\caption{
Magnetic-field dependence of the relative elastic constants
$\mathit{\Delta}C_{ij}/C_{ij} = [C_{ij}(B) - C_{ij}(B = 0)]/C_{ij}(B = 0)$
of longitudinal elastic constants (a) $C_{11}$, (b) $C_{\mathrm{L}[110]}$, and (c) $C_{33}$ and transverse elastic constants (d) $C_{44}$, (e) $C_\mathrm{T} = (C_{11} - C_{12})/2$, and (f) $C_{66}$ in Ba$_2$CuGe$_2$O$_7$ at $1.4$ and $4.2$ K for $B//[110]$.
The ultrasonic propagation direction $\boldsymbol{q}$ and polarization direction $\boldsymbol{\xi}$ are listed.
The solid (dotted) lines indicate the elastic constant for the field-upsweep (down-dweep) process.
The dashed lines in each panel indicate $B_\mathrm{c2}^{M(P)}$ and $B_\mathrm{sat}$ determined by the magnetization curve.
}
\label{C_All}
\end{figure}

To investigate the cross-correlation between the electric dipole and electric quadrupole, we measured elastic constants in Ba$_2$CuGe$_2$O$_7$.
Figure \ref{C_All} shows the magnetic-field dependence of the relative elastic constants $\mathit{\Delta} C_{ij}/C_{ij}$ at 1.4 and 4.2 K for $B//[110]$.
In the AFM phase at 1.4 K, we observed the elastic softening with increasing fields and characteristic anomalies.
At 1.4 K, the longitudinal elastic constants $C_{11}$ in Fig. \ref{C_All}(a) and $C_\mathrm{L[110]} = (C_{11} + C_{12} + 2C_{66})/2$ in Fig. \ref{C_All}(b) with the in-plane-type ultrasonic propagating direction show softening with the increase in the field up to around $B_\mathrm{c2}^{M(P)}$ accompanying hysteresis loop.
With the further application of the fields, each elastic constant shows a rapid increase with the bending point around $B_\mathrm{sat}$, then elastic constants exhibit slight softening up to 60 T.
In contrast to the $C_{11}$ and $C_\mathrm{L[110]}$, the longitudinal elastic constant $C_{33}$ with the inter-plane-type ultrasonic propagating direction in Fig. \ref{C_All}(c) shows hardening with the increase in the fields accompanying the anomaly at around $B_\mathrm{sat}$.
The transverse elastic constant $C_{44}$ with inter-plane-type ultrasonic propagation in Fig. \ref{C_All}(d) shows a step-like anomaly around $B_\mathrm{sat}$ while the field dependence can be almost field independent.
The transverse elastic constants $C_\mathrm{T} = (C_{11} - C_{12})/2$ in Fig. \ref{C_All}(e) and $C_{66}$ in Fig. \ref{C_All}(f) show softening with the increasig fields up to $B_\mathrm{sat}$.
With the further application of the fields, each elastic constant shows slight softening up to 60 T.
We could not observe the elastic anomaly at $B_\mathrm{c1}^{M(P)}$ in all elastic constants within our experimental resolution, suggesting that the contribution of the magnetoelastic coupling around $B_\mathrm{c1}^{M(P)}$ is small.
To study the contribution of the magnetoelastic coupling to the transition of the incommensurate spiral to the commensurate one, we need high-precision ultrasonic measurements.

At 4.2 K, each elastic constant shows a similar field dependence to that at 1.4 K except around 27 T, where the anomaly appears.
Below 20 T and above 30 T, the field dependence of each elastic constant is almost the same as that at 4.2 K.
The anomaly in $C_{11}$ and $C_\mathrm{L[110]}$ changes from a rapid increase to a gradual slope.
$C_{33}$ can be almost temperature independent.
The step-like anomaly of $C_{44}$ shifts to the lower field side.
The anomaly in $C_\mathrm{T}$ and $C_{66}$ changes from the bending-like form to a gradual slope.

We also observed elastic softening above $B_\mathrm{sat}$ indicating the electric quadrupole response.
This fact demonstrates that the quantum states in high fields provide the finite diagonal elements of multipole matrices and the expectation value of multipoles.

\subsection{
\label{subsect_Discussion}
Discussions
}

\begin{figure}[t]
\begin{center}
\includegraphics[clip, width=0.4\textwidth, bb=0 0 420 690]{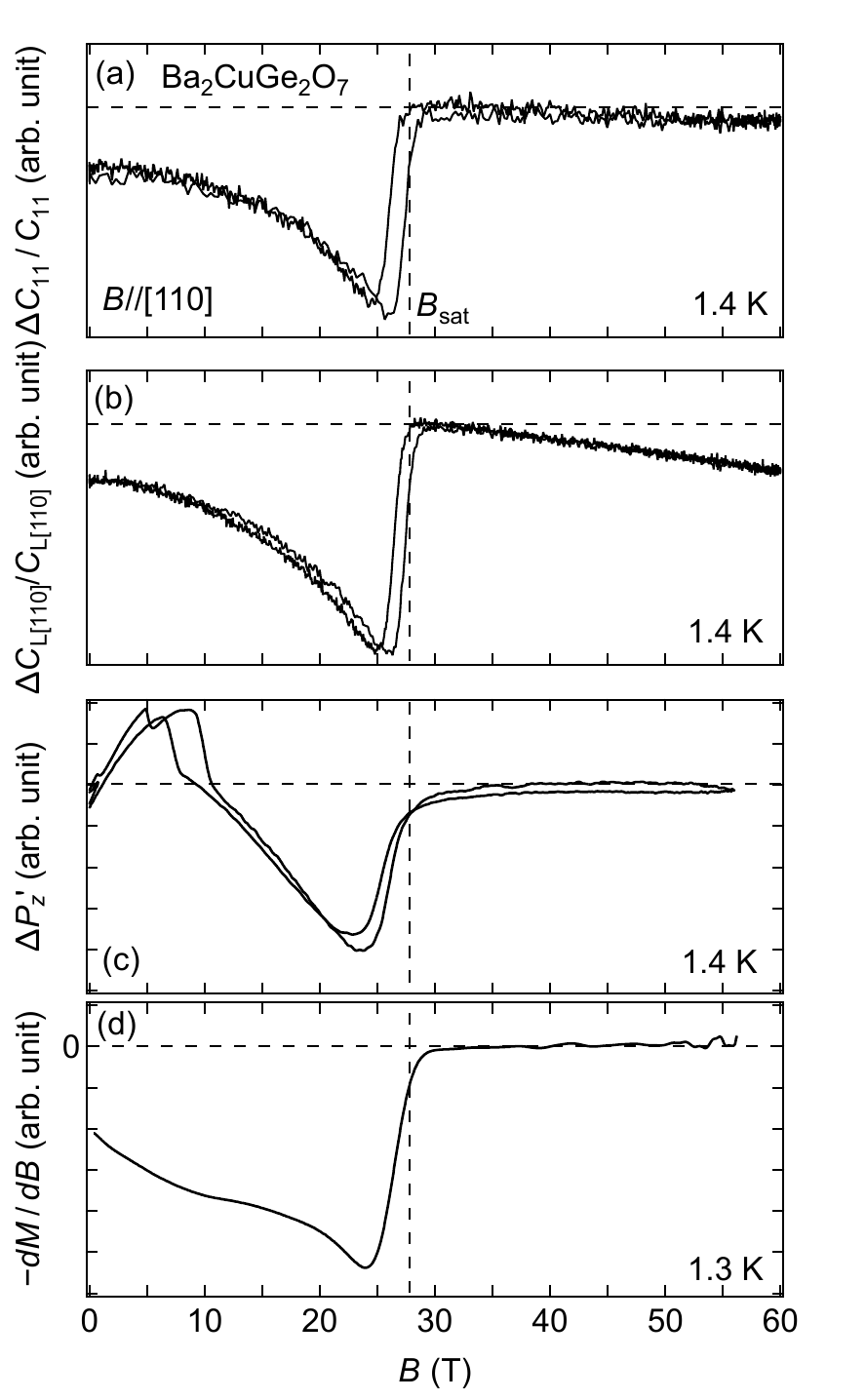}
\end{center}
\caption{
Magnetic-field dependence of the relative elastic constants (a) $C_{11}$ and (b) $C_{\mathrm{L}[110]}$, (c) the electric polarization $\mathit{\Delta}P_c'$, and (d) the first derivative of the magnetization with respect to magnetic fields $-dM/dB$ for the field down-sweep process in Ba$_2$CuGe$_2$O$_7$.
$dM/dB$ is calculated after applying binominal smoothing to $M$.
The horizontal broken lines indicating the field-independent factor are shown in panels (a) - (c) as guides for the eyes.
The horizontal broken line in panel (d) represents the zero line of $dM/dB$ indicating saturation magnetization.
The vertical broken lines indicate $B_\mathrm{sat}$.
Because the bend in $dM/dB$ becomes gradual due to the smoothing, $B_\mathrm{sat}$ may appear different compared to $dM/dB$ before smoothing in Fig. \ref{M_1p3K}(b).
}
\label{Compare}
\end{figure}

We observed the spin saturation above $B_\mathrm{sat}$ and the characteristic anomaly at $B_\mathrm{c2}^{M(P)}$ in Ba$_2$CuGe$_2$O$_7$ for $B//[110]$.
Around $B_\mathrm{sat}$ and  $B_\mathrm{c2}^{M(P)}$, both the polarization and the elastic constants show step-wise change.
We discuss the origin of this step-wise anomaly in terms of the thermodynamic relation between magnetization and elastic constants.
Figure \ref{Compare} shows the magnetic-field dependence of the relative elastic constants, the subtracted polarization $\mathit{\Delta}P_c'$, and the first derivative of the magnetization with respect to magnetic fields, $dM/dB$.
Using the linear slope $f(\mathrm{T})$ of $\mathit{\Delta}P_c$ above 40 T, we calculated $\mathit{\Delta}P_c'$ as $f(\mathrm{T}) - \mathit{\Delta}P_c$ to demonstrate the step-wise change in the electric polarization.
The form of the step-wise anomaly in $\mathit{\Delta}C_{11}/C_{11}$, $\mathit{\Delta}C_\mathrm{L[110]}/C_\mathrm{L[110]}$, and $\mathit{\Delta}P_c$ seems to be reproduced by $-dM/dB$, suggesting that the elastic constants and the electric polarization are proportional to $dM/dB$ around the spin saturation fields.
In the low-dimensional spin magnetic compound NH$_4$CuCl$_3$, experimental studies have proposed that the elastic anomaly is proportional to $dM/dB$ around the magnetization plateaus
\cite{Schmidt_EPL53}.
A theoretical study has confirmed $\mathit{\Delta}C/C \propto dM/dB$ based on the Ehrenfest relation, albeit under special circumstances
\cite{Matsumoto_JPSJ74}.
Since $\mathit{\Delta}P_c'$ can also be proportional to $dM/dB$ around $B_\mathrm{sat}$, we deduce that this Ehrenfest relation is applicable to polarization.
Therefore, we conclude that the step-like behavior in the elastic constants and the polarization originates from the spin saturation of $M$ in Ba$_2$CuGe$_2$O$_7$.
As shown in Fig. \ref{C_All}, the little step-like anomaly in $C_{44}$ can also be attributed to $dM/dB$.

In contrast, our experimental results suggest that the alternative mechanism of the FIEP can be realized in Ba$_2$CuGe$_2$O$_7$.
As shown in Fig. \ref{M_1p3K}, the spin saturation for $B//[110]$ above $B_\mathrm{sat}$ suggests that magnetic moments can be aligned parallel to the magnetic field direction. 
Since the electric dipole $P_z$ is proportional to the square of magnetic moments in the spin-dependent $d$-$p$ hybridization model, the polarization should be independent of fields in the spin saturation regions.
However, the polarization for $B//[110]$ increases in the fields above $B_\mathrm{sat}$.
This fact indicates that the quantum states describing the polarization still depend on the magnetic fields in the saturation magnetic moment region.
In addition, our ultrasonic measurements revealed the response of electric quadrupole in magnetic fields.
Therefore, we need to consider new quantum states and the mechanism of the FIEP accompanying the elastic response in magnetic fields.

\begin{table}[t]
\caption{
Longitudinal strains, that of reduction by the symmetry strains, and decomposition for the point group $D_{2d}$.
}
\begin{ruledtabular}
\label{table2}
\begin{tabular}{ccc}
\textrm{Strain}
	& \textrm{Reduction}
		& \textrm{Decomposition}
		\\
\hline
$\varepsilon_{xx} $
	&$\varepsilon_\mathrm{B}/3 - \varepsilon_{u}/(2\sqrt{3}) + \varepsilon_{x^2-y^2}/\sqrt{2}$
		& $2A_1 \oplus B_1$
\\
$\varepsilon_{yy} $
	&$\varepsilon_\mathrm{B}/3 - \varepsilon_{u}/(2\sqrt{3}) - \varepsilon_{x^2-y^2}/\sqrt{2}$
		& $2A_1 \oplus B_1$
\\
 $\varepsilon_\mathrm{L[110]}$
	& $\varepsilon_\mathrm{B}/3 - \varepsilon_{u}/(2\sqrt{3}) + \sqrt{2} \varepsilon_{xy}$
		& $2A_1 \oplus B_2$
\\
$\varepsilon_{zz} $
	& $\varepsilon_\mathrm{B}/3 +  \varepsilon_{u}/\sqrt{3}$
		& $2A_1$
\end{tabular}
\end{ruledtabular}
\end{table}

To describe another possible multiferroic mechanism, we focus on active electric multipoles based on the point group $D_{2d}$.
The irrep of elastic constants and polarization provides information on possible quantum states in terms of group theory.
As shown in the Table \ref{table_multipoles},  the transverse elastic constant $C_{44}$ with the irrep $E$ reflects the response of the electric quadrupoles $O_{yz}$ and $O_{zx}$ because the ultrasonic waves for $C_{44}$ induces the strains $\varepsilon_{yz}$ and $\varepsilon_{zx}$.
The monotonic hardening of $C_{44}$ with increasing fields indicates that the contribution of the quadrupole-strain coupling between $O_{yz}$ ($O_{zx}$) and $\varepsilon_{yz}$ ($\varepsilon_{zx}$) to elasticity is negligibly small.
The elastic softening of $C_\mathrm{T}$ with the irrep $B_\mathrm{1}$ originates from the coupling of the quadrupole $O_{x^2-y^2}$ with the strain $\varepsilon_{x^2-y^2}$.
The softening of $C_{66}$ with the irrep $B_\mathrm{2}$ is attributed to the $O_{xy}$ and $\varepsilon_{xy}$.
The elastic constant $C_{11}$ is measured by the longitudinal ultrasonic waves inducing the strain $\varepsilon_{xx}$ ($\varepsilon_{yy}$).
As listed in Table \ref{table2}, since $\varepsilon_{xx}$ ($\varepsilon_{yy}$) includes $\varepsilon_{x^2-y^2}$, the softening of $C_{11}$ originates from the coupling of $O_{x^2-y^2}$ to $\varepsilon_{x^2-y^2}$ because little contribution of the coupling between $O_\mathrm{B}$ ($O_{3z^2-r^2}$) and $\varepsilon_\mathrm{B}$ ($\varepsilon_{3z^2-r^2}$) with the irrep $A_1$ is indicated by the hardening of $C_{33}$.
This fact for the inter-plane type strain $\varepsilon_{3z^2-r^2}$ is in contrast to magnetostriction around the spin saturation in Ba$_2$FeSi$_2$O$_7$
\cite{Watanabe_JPSJ92}.
Also, the elastic softening of $C_\mathrm{L[110]}$ is attributed to the coupling of $O_{xy}$ to $\varepsilon_{xy}$.
In addition to the elastic constants, the increase of polarization $P_c$ indicates the response of the electric dipole $P_z$ with the irrep $B_2$.
Thus, our experiments suggest that the active multipoles in high fields are the electric dipole $P_z$ with the irrep $B_2$ and the electric quadrupoles $O_{x^2-y^2}$ with $B_1$ and $O_{xy}$ with $B_2$.

Since the above group-theoretical analysis is based on the global coordinates of the crystal lattice, we should discuss the active multipoles for the local coordinates.
Considering the tilting of CuO$_4$ tetrahedra, we deduce that both the response of the electric quadrupole $O_{x^2-y^2}$ with the irrep $B_1$ and $O_{xy}$ with $B_2$ for the global coordinates originate from the response of an electric quadrupole $O_{XY}$ with $B_2$ for the local coordinates.
As a result of this response, the elastic softening of $C_\mathrm{T}$ with the irrep $B_1$ for the global coordinates is induced by the elastic softening of $C_{66}$ with the irrep $B_2$ for the local coordinates. 
The response of the electric dipole $P_z$ for the global coordinates is the same as that for the local coordinates because the $z$ component is invariant for the tilting of CuO$_4$ tetrahedra around the $c$-axis.
Therefore, we conclude that the active multipoles are $P_z$ and $O_{XY}$ with the irrep $B_2$ of the point group $D_{2d}$.

Focusing on the active representation $B_2$ obtained by the group-theoretical analysis, we can describe the free energy based on the Landau phenomenological theory for the $D_{2d}$ point group.
Considering the basis functions with the irrep $A_1$ in Table \ref{table_character of D2d}, the conjugate fields and electric multipoles in Table \ref{table_multipoles}, and the product table of the $D_{2d}$ point group in Table \ref{table_product}, we can describe the free energy $F$ composed of the minimal elements as a function of the electric dipole $P_z$, the electric quadrupole $O_{XY}$, the strain $\varepsilon_{XY}$, the electric field $E_z$, and the in-plane magnetic field $\left( B_X, B_Y, 0 \right)$ for the local coordinate as below:
\begin{align}
\label{Landau F}
F
= & \frac{1}{2}\alpha_z P_z^2 + \frac{1}{2}\alpha_{XY}O_{XY}^2 + \frac{1}{2}C_{66}^{'0} \varepsilon_{XY}^2
\nonumber	\\
&- g_z P_z E_z - g_{XY}O_{XY}\varepsilon_{XY} - \mu_\mathrm{Z} \left( l_X B_X + l_Y B_Y \right) 
\nonumber	\\
& - \alpha_{A_1}P_z O_{XY}  - g_{A_1} P_z B_X B_Y 
.
\end{align}
Here, $\alpha_z$, $\alpha_{XY}$, $C_{66}^{'0}$, $g_z$, $g_{XY}$, $\mu_\mathrm{Z}$, $\alpha_{A_1}$, and $g_{A_1}$ are the coefficients.
The terms from the first to the sixth on the right-hand side in Eq. (\ref{Landau F}) indicate the conventional terms incorporated in the free energy.
While the angular momentums $l_x$ and $l_y$ are bases of the irrep $E$,  we include the Zeeman effect for the in-plane magnetic fields as the seventh term.
Although the eighth and ninth terms are constructed from the products of the basis functions with different properties under spatial inversion, each is also invariant for the symmetry operations of $D_{2d}$.
We emphasize that the eighth term can be the characteristic term describing the cross-correlation between the electric dipole and electric quadrupole.
We can also expect field-induced electric polarization because the contribution of the tenth term is allowed under the $D_{2d}$.
Based on the character table of Table \ref{table_character of D2d},  we can also include other terms, such as the sum of the possible permutations of $l_XO_{YZ}$, $l_Y O_{ZX}$, $l_X l_Y P_z$, and $P_X B_X - P_Y B_Y$,  in the free energy of Eq. (\ref{Landau F}).
However, the contributions of $O_{YZ}$ and $O_{ZX}$ with the irrep $E$ can be negligible because the elastic hardening of $C_{44}$ indicates no-contribution of $O_{yz}$ and $O_{zx}$.
We also deduce that $\left( l_X l_Y + l_Y l_X \right) P_z$ is included in the eighth term of $\alpha_{A_1} P_z O_{XY}$ because $l_Xl_Y + l_Y l_X$ is equivalent to $O_{XY}$ in terms of the Stevens' operator
\cite{Kusunose_JPSJ77}.
Since $P_X B_X - P_Y B_Y$ is not constructed by the electric dipole $P_z$ with the irrep $B_2$, we do not include this term in the free energy. 

Based on the free energy of Eq. (\ref{Landau F}), we can qualitatively describe the cross-correlation between $P_z$ and $O_{XY}$ and the FIEP and elastic softening.
From the equilibrium conditions, denoted by $\partial F / \partial P_z = 0$ and $\partial F / \partial O_{XY} = 0$, the following relationships are obtained:
\begin{align}
\label{Cross-corelation of Oxy and Pz}
P_z
&= \frac{1}{\alpha_z} \left( g_z E_z + \alpha_{A_1} O_{XY} + g_{A_1} B_X B_Y  \right)
,	\\
\label{Cross-corelation of Pz and Oxy}
O_{XY}
&= \frac{1}{\alpha_{XY}} \left( \alpha_{A_1} P_z + g_{XY} \varepsilon_{XY}  \right)
.
\end{align}
These results indicate the cross-correlation between the electric dipole $P_z$ and the electric quadrupole $O_{xy}$.
In addition, we can describe $P_z$ and $O_{XY}$ as a function of $E_z$, $\varepsilon_{XY}$, and $\left( B_X, B_Y \right)$ as 
\begin{align}
\label{Piezo and FIEP and ES}
\left(
\begin{array}{c}
P_z	\\
O_{XY}	\\
\end{array}
\right)
&=\frac{g_z}{ \alpha_z \alpha_{XY} - \alpha_{A_1}^2 }
\left(
\begin{array}{c}
\alpha_{XY} 	\\
\alpha_{A_1} 	\\
\end{array}
\right) E_z
\nonumber	\\
&+ \frac{g_{XY}}{ \alpha_z \alpha_{XY} - \alpha_{A_1}^2 } 
\left(
\begin{array}{c}
\alpha_{A_1} 	\\
\alpha_z 	\\
\end{array}
\right) \varepsilon_{XY}
\nonumber	\\
&+\frac{g_{A_1}}{ \alpha_z \alpha_{XY} - \alpha_{A_1}^2 } 
\left(
\begin{array}{c}
\alpha_{XY}  	\\
\alpha_{A_1}  	\\
\end{array}
\right) B_X B_Y
.
\end{align}
This result indicates the FIEP and elastic softening and piezoelectric.
However, we cannot explain the microscopic origin of the multipoles and the mechanism of the FIEP and elastic softening.
Therefore, we discuss the possible quantum states so as not to contradict our group-theoretical analysis.

Since $P_z$ and $O_{XY}$ with the irrep $B_2$ can be the active multipoles, the direct product of the irrep of the quantum states should contain the decomposition of
$B_\mathrm{2(u)} (P_{z}) \oplus  B_\mathrm{2(g)} (O_{xy})$.
Here, we use the irrep with the extra suffix g (gerade) and u (un-gerade) to distinguish between even and odd parity.
As shown in Table \ref{table_product}, the irrep of candidate orbitals constracting the quantum states is $E$.
This result indicates that the quantum states are formed by the $yz$ and $zx$ orbitals of Cu-$3d$ electrons and $x$ and $y$ orbitals of O-$2p$ electrons.
Furthermore, because the spatial inversion property of these electric multipoles is different from each other as listed in Table \ref{table_multipoles}, we should describe parity-hybridized wave functions.
This fact indicates that the hybridization between $yz$ and $x$, $zx$ and $y$, and $xy$ and $z$ orbitals can play a key role in the quantum states.
The possibility of $d$-$p$ hybridization as an origin of such parity-hybridized states is indicated by the basis function of $xyz$ with the irrep $A_1$ of $D_{2d}$.

\begin{table}[t]
\caption{
Product table of the point group $D_{2d}$ calculated by the characters in Table \ref{table_character of D2d}.
The basis functions of the irreps $A_1$, $A_2$, $B_1$, $B_2$, and $E$ are $z^2$ and $xyz$, $l_z$, $x^2 - y^2$, $z$ and $xy$, and $\{x,y\}$, $\{yz, zx\}$, $\{l_x, l_y\}$, and $\{B_x, B_y \}$, respectively.
}
\begin{ruledtabular}
\label{table_product}
\begin{tabular}{c|ccccc}
$D_{2d}$
	& $A_1$
		& $A_2$
			& $B_1$
				& $B_2$
					& $E$
\\
\hline
$A_1$
	& $A_1$
		& $A_2$
			& $B_1$
				& $B_2$
					& $E$
\\
$A_2$
	& $A_2$
		& $A_1$
			& $B_2$
				& $B_1$
					& $E$
\\
$B_1$
	& $B_1$
		& $B_2$
			& $A_1$
				& $A_2$
					& $E$
\\
$B_2$
	& $B_2$
		& $B_1$
			& $A_2$
				& $A_1$
					& $E$
\\
$E$
	& $E$
		& $E$
			& $E$
				& $E$
					& $A_1 \oplus A_2 \oplus B_1 \oplus B_2$
\end{tabular}
\end{ruledtabular}
\end{table}

In addition to the multipole effects, we should consider the magnetic field effects to describe the quantum states.
Since the AFM ordering has been observed in Ba$_2$CuGe$_2$O$_7$, the spin degrees of freedom can be necessary to describe the quantum states.
Thus, we consider the spin-orbit coupling.
Furthermore, we deduce that the field dependence of the electric multipole response is dominated by the Zeeman effect for the orbital part of the high-field quantum states because the FIEP and elastic softening appear above $B_\mathrm{sat}$.
Therefore, we focus on the angular momentum operators $l_X$ and $l_Y$ for the local coordinates with the irrep $E$ of the $D_{2d}$ point group for the in-plane magnetic field direction. 
To take into account the Zeeman effect, the direct product of the irrep of the quantum states should contain the irrep $E$.
This fact indicates that the quantum states should contain $E$ and $B_2$ orbitals (see Table \ref{table_product}). 
We would like to mention that if the magnetic field is applied along the $c$ axis, $l_z$ with the irrep $A_2$ contributes to the Zeeman effect.
Then, we consider the tilting crystal structure of CuO$_4$ tetrahedra.
In our measurements, the magnetic field was applied along the $[110]$ direction for the global coordinates.
Thus, based on the tilting angle $\pm \kappa$ of CuO$_4$ tetrahedra, we should calculate the high-field quantum states and the magnetic field dependence of the multipole response.

In the above discussions, we assumed that the crystal symmetry and the point group remained conserved above $B_\mathrm{sat}$.
Before delving into the analysis of quantum states, it is necessary to evaluate the validity of this assumption. 
Below $B_\mathrm{sat}$, we also observed the FIEP and elastic softening, indicating that the crystal symmetry breaking characterized by the irrep $B_2$ was already induced by the magnetic field.
Due to the crystal symmetry breaking, the symmetry lowering of the point group from $D_{2d}$ to $C_2$ can be realized above $B_\mathrm{sat}$.
Furthermore, the field-independent behavior of the electric polarization and the transverse elastic constants are expected above $B_\mathrm{sat}$.
However, this contradicts our experimental results.
Hence, it can be reasonably assumed that the group-theoretical analysis based on $D_{2d}$ is approximately applicable above $B_\mathrm{sat}$.

Based on the above discussions, we describe the quantum states based on the Cu-$3d$ and O-$2p$ orbitals with the $d$-$p$ hybridization, the spin-orbit coupling, and the Zeeman effect, whose contributions are indicated by our group-theoretical analysis.
We show several theoretical studies of the wave functions in high fields, the electric multipole susceptibility, and group theoretical analysis in the following Sect. \ref{sect_4}.

\section{
\label{sect_4}
Quantum states in high fields and electric multipole susceptibility
}

\subsection{
\label{subsect_WaveFunctions}
Wave functions and field dependence of eigenenergies
}

In this section, we discuss the origin of the electric dipole and the electric quadrupole in Ba$_2$CuGe$_2$O$_7$ to calculate quantum states based on the crystalline electric field (CEF) $H_\mathrm{CEF} $, the $d$-$p$ hybridization $H_\mathrm{d-p}$, the spin-orbit coupling $H_\mathrm{SO} $, and the Zeeman effect $H_\mathrm{Zeeman}$.
Taking into account the electric dipole-electric field interaction $H_\mathrm{DE}^\mathrm{L}$ (electric quadrupole-strain interaction $H_\mathrm{QS}^\mathrm{L}$), we calculate the electric dipole (electric quadrupole) susceptibility for the local coordinate.
Furthermore, we discuss the possible contribution of inter-site quadrupole interaction $H_\mathrm{QQ}^\mathrm{G}$ to the high-field quantum states.
Therefore, total Hamiltonian $H_\mathrm{total}$ is described as below:
\begin{align}
H_\mathrm{total}
&= H_0 + H_\mathrm{ext}  + H_\mathrm{QQ}^\mathrm{G}
,	\\
H_0
&= H_\mathrm{CEF} + H_\mathrm{d-p} + H_\mathrm{SO} + H_\mathrm{Zeeman}
,	\\
H_\mathrm{ext}
&= H_\mathrm{DE}^\mathrm{L} +  H_\mathrm{QS}^\mathrm{L}
.
\end{align}
Based on this Hamiltonian, we demonstrate that the orbital part of the wave functions, in addition to the spins, contributes to the field-induced multiferroicity.

\begin{figure*}[htbp]
\begin{center}
\includegraphics[clip, width=1\textwidth, bb=0 0 600 320]{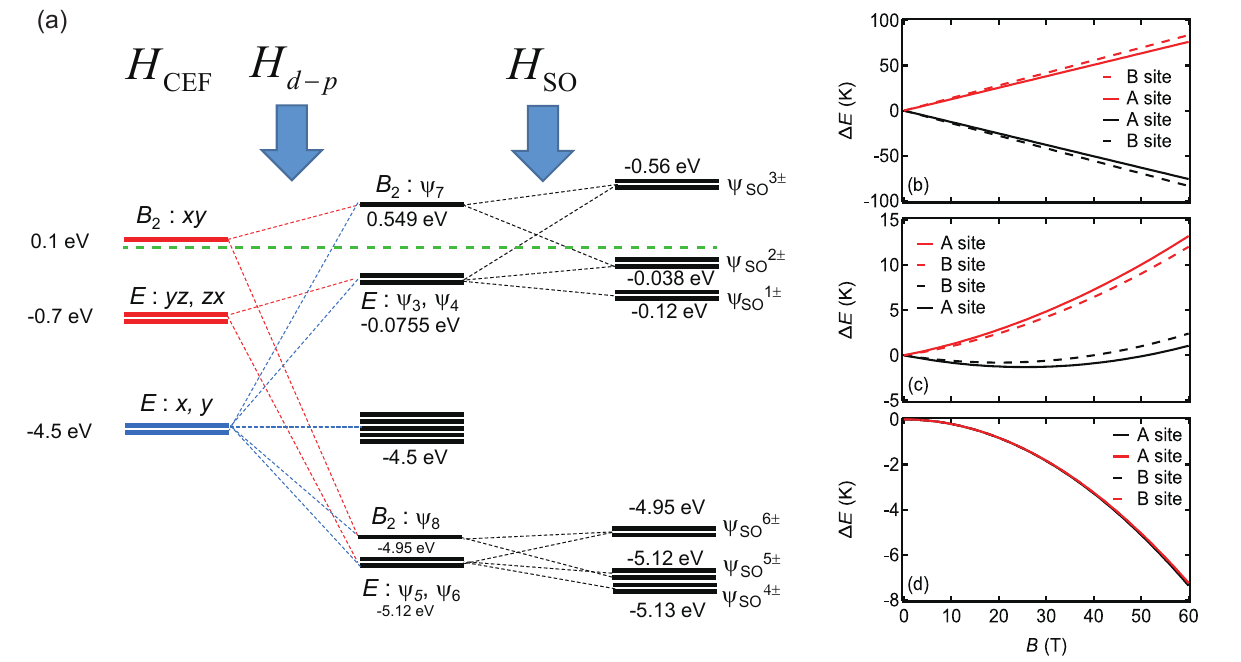}
\end{center}
\caption{
(a) Energy scheme of the quantum state for the crystalline electric field $H_\mathrm{CEF}$ with $D_{2d}$ point group symmetry, $d$-$p$ hybridization $H_\mathrm{CEF} + H_{d-p}$, and spin-orbit coupling $H_\mathrm{CEF} + H_{d-p} + H_\mathrm{SO}$.
The red (blue) broken lines indicate the 3$d$-orbital (2$p$-orbital) contributions for $d$-$p$ hybridized states.
The black broken lines exhibit the contributions of $d$-$p$ hybridized states to the spin-orbit coupled states.
The green broken line indicates the origin of the energy (0 eV).
Zeeman splitting of the spin-orbit coupled states (b) $\psi_\mathrm{SO}^{3\pm}$, (c)$\psi_\mathrm{SO}^{2\pm}$, and (d)$\psi_\mathrm{SO}^{1\pm}$.
The solid (dashed) lines indicate the energy of the quantum state at the A (B) site. 
}
\label{EnergyScheme}
\end{figure*}

At first, we describe the wave functions of Cu-$3d$ electrons and the molecular orbitals constructed by the $2p$ electrons of O$_4$ tetrahedra.
In Ba$_2$CuGe$_2$O$_7$, Cu ions are centered at O$_4$ tetrahedron with the point group symmetry $D_{2d}$.
Thus, considering the CEF for Cu-$3d$ electrons described as
\cite{Kurihara_JPSJ86}
\begin{align}
\label{H_CEF}
H_\mathrm{CEF} 
= A_2^0\left( \frac{3z^2 - r^2}{2r^2} \right) + A_4^0 \left( \frac{35z^4-30z^2r^2 + 3r^4}{8r^4} \right)
\\ \nonumber
+ A_4^4 \left( \frac{\sqrt{35} }{8} \frac{x^4-6x^2y^2+y^4}{r^4} \right)
,
\end{align}
we obtain doubly degenerate $3d$-orbitals $yz$ and $zx$ and an excited singlet orbital $xy$ as shown in Fig. \ref{EnergyScheme}(a).
Here, $A_2^0$, $A_4^0$, and $A_4^4$ denote the CEF parameters.
The wave functions of $yz$, $zx$, and $xy$ orbitals and the relationship between the energy levels and the CEF parameters are written in Appendix \ref{Appendix_A}.
Based on the previous studies in the \aa kermanite-type compound
\cite{Yamauchi_PRB84},
we set the energy levels of $xy$ and degenerate $yz$ and $zx$ orbitals to $0.1$ and $-0.7$ eV, respectively (see Fig. \ref{EnergyScheme}(a)).
We deduce that $3z^2-r^2$ and $x^2-y^2$ orbitals of $3d$ electrons do not contribute to the polarization and elasticity because they exist deep in the energy levels
\cite{Corasaniti_PRB96}.
The $2p$ orbitals on O$_4$ tetrahedra are also transformed by the symmetry operations of $D_{2d}$.
As shown in Appendix \ref{Appendix_B}, a group theoretical analysis provides eight molecular orbitals consisting of $x$ and $y$ orbitals of O-$2p$ electrons.
We ignore the molecular orbitals consisting of $z$ orbital of $2p$ electrons because $z$ orbital with the irrep $B_2$ does not carry $P_z$, $O_{x^2-y^2}$, and $O_{xy}$ (see Table \ref{table_product}).
Considering the previous studies in the \aa kermanite-type compound
\cite{Yamauchi_PRB84},
we set the energy level of degenerate $x$ and $y$ orbitals to $-4.5$ eV.
As we will discuss late, our calculations using these values can reproduce the electronic structure near the Fermi level proposed in the previous study
\cite{Corasaniti_PRB96}.

We stress that the electric dipole and electric quadrupole degrees of freedom are attributed to the $3d$-$yz$ and $zx$ and $2p$-$x$ and $y$ orbitals with the irrep $E$.
The matrix elements of the electric dipole and electric quadrupoles for the $yz$, $zx$, and $xy$ orbitals of $3d$ electrons and the $x$ and $y$ orbitals of $2p$ electrons are described as
\begin{align}
\label{Ov}
\boldsymbol{O}_{x^2-y^2}
&= \bordermatrix{
& yz	& zx	& xy & x	& y	\cr
& -\frac{\sqrt{2}}{7}r_d^2	 &0& 0	& 0	&0	\cr
& 0	& \frac{\sqrt{2}}{7}r_d^2 &0	& 0	&0	\cr
& 0 & 0 & 0 & 0 & 0	\cr
& 0	& 0 &0	& \frac{\sqrt{2}}{5}r_p^2	&0	\cr
& 0	& 0 &0	& 0	& -\frac{\sqrt{2}}{5}r_p^2	\cr
}, \\
\label{Oxy}
\boldsymbol{O}_{xy}
&=\left(
\begin{array}{ccccc}
0	&\frac{\sqrt{2}}{7}r_d^2 &0	&0	&0	\cr
\frac{\sqrt{2}}{7}r_d^2	&0 &0	&0	&0	\cr
0  &0  &0  &0  &0	\cr
0	&0	 &0&0	&\frac{\sqrt{2}}{5}r_p^2	\cr
0	&0	 &0&\frac{\sqrt{2}}{5}r_p^2	&0	\cr
\end{array}
\right)
, \\
\label{Pz}
\boldsymbol{P}_{z}
&=\left(
\begin{array}{ccccc}
0	&0	 &0	&0	&p	\cr
0	&0	 &0	&p	&0	\cr
0	&0	 &0	&0	&0	\cr
0	&p	 &0	&0	&0	\cr
p	&0	 &0	&0	&0	\cr
\end{array}
\right)
.
\end{align}
Here, we set $r_d^2 = \int_0^\infty dr \left[ r f_{3d}(r) \right]^2$ and $r_p^2 = \int_0^\infty dr \left[r f_{2p}(r) \right]^2$ for the convenience.
$p$ in $\boldsymbol{P}_z$ of Eq. (\ref{Pz}) indicates the radial integration of the wave functions for the $3d$-$yz$ and $zx$ and $2p$-$x$ and $y$ orbitals. 
Using the matrices of Eqs. (\ref{Ov}) - (\ref{Pz}), we can describe the matrices of these electric multipoles for the molecular orbitals with the irrep $E$ (see Eqs. (\ref{Ov_Molecular}) - (\ref{Pz_Molecular})  in Appendix \ref{Appendix_B}).
This matrix analysis is consistent with the group-theoretical analysis (see Table \ref{table_product}).

The group-theoretical analysis also indicates the contributions of the spin-orbit coupling and the Zeemam effect for the quantum states in CuO$_4$ clusters.
$yz$ and $zx$ orbitals with the irrep $E$ and $xy$ orbital with the irrep $B_2$ carry the angular momentums $l_x$ and $l_y$ with the irrep $E$ to reduce $(E \oplus B_2) \otimes (E \oplus B_2)$ (see Table \ref{table_product}).
The matrix elements of $l_x$ and $l_y$ are written as
\begin{align}
\boldsymbol{l}_{x}
&=\left(
\begin{array}{ccccc}
0	&0	&0	&0	&0	\cr
0	&0	&i	&0	&0	\cr
0 	&-i 	&0 	&0 	&0	\cr
0	&0	&0	&0	&0	\cr
0	&0	&0	&0	&0	\cr
\end{array}
\right)
,	\\
\boldsymbol{l}_{y}
&=\left(
\begin{array}{ccccc}
0	&0	&-i	&0	&0	\cr
0	&0	&0	&0	&0	\cr
i 	&0 	&0 	&0 	&0	\cr
0	&0	&0	&0	&0	\cr
0	&0	&0	&0	&0	\cr
\end{array}
\right)
.
\end{align}
Therefore, both the spin-orbit coupling proportional to $l_x s_x + l_y s_y$ and the Zeeman effect for $B // [110]$ including $l_x$ and $l_y$ cause the orbital mixing in quantum states.

To obtain the finite expectation value of the electric dipole $P_z$, we consider the parity hybridization between the Cu-$3d$ and O-$2p$ orbitals.
The hybridization between $3d$ orbitals of Cu ions and the $2p$ molecular orbitals of O$_4$ clusters is described by the following  Hamiltonian
\cite{Yamauchi_PRB84}:
\begin{eqnarray}
\label{H_dp}
H_{d - p}
= \sum_{l'} \sum_{l} \sum_{\sigma , \sigma'}
t_{l', l, \sigma, \sigma'} \left( p_{l', \sigma'}^\dagger d_{l, \sigma} + \mathrm{h.c.} \right)
.
\end{eqnarray}
Here, $l = yz, zx, xy$ and $l' = x, y$ are the indices of the $3d$- and $2p$-electron orbitals, $\sigma$ and $\sigma' = \uparrow, \downarrow$ are the indices of spin, $t_{l, l', \sigma, \sigma'}$ is a transfer energy between Cu-$3d$ and O-$2p$ electrons
\cite{Slater_PR94},
$d_{l, \sigma}$ and $d_{l, \sigma}^\dagger$ are an annihilation operator and a creation operator of $3d$ electrons, and $p_{l', \sigma'}$ and $p_{l', \sigma'}^\dagger$ are an annihilation operator and a creation operator of $2p$ electrons, respectively.
We used $t_{x, yz} = -0.73$ eV, $t_{x, xy} = -0.53$ eV, and $t_{x, zx} = -0.41$ eV to estimate the energy scheme (see Appendix \ref{Appendix_B}).
We obtain eleven $d$-$p$ hybridized states and that of eigenenergies to diagonalize $H_\mathrm{CEF} +  H_{d-p}$ shown in Appendix \ref{Appendix_B}.
The energy scheme and the irrep of hybridized states are depicted in Fig. \ref{EnergyScheme}(a).
The energy levels of $\psi_3$, $\psi_4$, and $\psi_7$ can reproduce the electronic structure around the Fermi level in the previous study
\cite{Corasaniti_PRB96}.
As shown in Table. \ref{Coefficients of Wave Functions}, the higher-level hybridized states $\psi_3$ and $\psi_4$ with the irrep $E$ and $\psi_7$ with $B_2$ are mainly contributed by $3d$ orbitals. 
The difference of the energy levels between $\psi_3$ and $\psi_4$ states and $\psi_7$ state is attributed to the energy scheme of $yz$, $zx$, and $xy$ orbitals for $H_\mathrm{CEF}$.
In contrast, the low-energy states $\psi_5$ and $\psi_6$ with $E$ and $\psi_8$ with $B_2$ are mainly constructed by the $2p$ orbitals.
The other five states consist of the $2p$ orbitals.

\begin{table}[t]
\caption{
Coefficients of $\psi_{yz}$, $\psi_{zx}$, $\psi_{E(1+)}$, $\psi_{E(1-)}$, $\psi_{E(2+)}$, $\psi_{E(2-)}$, $\psi_{xy}$, and $\psi_{B_2}$ that constitute the wave function $\psi_i$ ($i = 3$ - $8$).
}
\begin{ruledtabular}
\label{Coefficients of Wave Functions}
\begin{tabular}{c|cccccccc}

	&$\psi_{yz}$ 
		& $\psi_{zx}$ 
			& $\psi_{E(1+)}$ 
				& $\psi_{E(1-)}$
					& $\psi_{E(2+)}$ 
						& $\psi_{E(2-)}$ 
							&$\psi_{xy}$
								&$\psi_{B_2}$
\\
\hline
$\psi_3$
	&$ -0.66$
		&$-0.66$
			& 0.31
				&0
					& 0.17
						& 0
							&0
								&0
\\
$\psi_4$
	&$-0.66$
		&0.66
			&0
				&0.31
					&0
						&0.17
							&0
								&0
\\
$\psi_5$
	&0.25
		&0.25
			&0
				&0.82
					&0
						&0.46
							&0
								&0
\\
$\psi_6$
	&0.25
		&$-0.25$
			&0.82
				&
					&0.46
						&0
							&0
								&0
\\
$\psi_7$
	&0
		&0
			&0
				&0
					&0
						&0
							&$-0.96$
								&0.29
\\
$\psi_8$
	&0
		&0
			&0
				&0
					&0
						&0
							&$-0.29$
								&$-0.96$
\end{tabular}
\end{ruledtabular}
\end{table}

\begin{table*}[t]
\caption{
Coefficients of $\psi_{3, \uparrow}$, $\psi_{4, \uparrow}$, $\psi_{7, \downarrow}$, $\psi_{3, \downarrow}$, $\psi_{4, \downarrow}$, and $\psi_{7, \uparrow}$ that constitute the wave function $\psi_\mathrm{SO}^{i\pm}$ ($i = 1$ - $3$) at a zero magnetic field.
}
\begin{ruledtabular}
\label{Coefficients of Wave Functions SO}
\begin{tabular}{c|cccccc}

	&$\psi_{3, \uparrow}$ 
		& $\psi_{4, \uparrow}$ 
			& $\psi_{7, \downarrow}$ 
				& $\psi_{3, \downarrow}$
					& $\psi_{4, \downarrow}$ 
						& $\psi_{7, \uparrow}$ 
\\
\hline
\\
$\psi_\mathrm{SO}^{1+}$
	&$-0.71$
		&$0.71i$
			& 0
				&0
					& 0
						& 0
\\
$\psi_\mathrm{SO}^{1-}$
	&0
		&0
			& 0
				&$-0.71$
					& $- 0.71i$
						& 0
\\
$\psi_\mathrm{SO}^{2+}$
	&$-0.50 + 0.50i$
		& $-0.50 - 0.50i$
			&$- 0.11$
				&0
					&0
						&0
\\
$\psi_\mathrm{SO}^{2-}$
	&0
		& 0
			&0
				&$-0.50 - 0.50i$
					&$-0.50 + 0.50i$
						&$ 0.11$
\\
$\psi_\mathrm{SO}^{3+}$
	&$- 0.05 + 0.05i$
		&$- 0.05 - 0.05i$
			&$0.99$
				&0
					&0
						&0
\\
$\psi_\mathrm{SO}^{3-}$
	&0
		&0
			&0
				&$ 0.05 + 0.05i$
					&$0.05 - 0.05i$
						&$0.99$
\end{tabular}
\end{ruledtabular}
\end{table*}

We emphasize that the multipoles are active for the $d$-$p$ hybridized states.
As shown in Eqs. (\ref{Ov_Molecular_2}) - (\ref{Pz_Molecular_2}) in Appendix \ref{Appendix_B}, $\psi_3$ and $\psi_4$ states with the irrep $E$ provide the multipole matrices with non-zero elements.
This result is also confirmed in terms of the group theoretical analysis for the $\psi_3$ and $\psi_4$ states that are constructed by $yz$ and $zx$ orbitals with the irrep $E_{(g)}$ and $x$ and $y$ orbitals with the irrep $E_{(u)}$ because the decomposition of the direct product of $(E_{(\mathrm{g})} \oplus E_{(\mathrm{u})}) \otimes (E_{(\mathrm{g})} \oplus E_{(\mathrm{u})})$ includes $B_{1(\mathrm{u})} (P_{z}) \oplus B_{1(\mathrm{g})}(O_{x^2-y^2}) \oplus B_{2(\mathrm{g})} (O_{xy})$.
We also stress that both the matrix $\boldsymbol{P}_z$ and $\boldsymbol{O}_{xy}$ are proportional to the $z$-component of Pauli matrix, $\boldsymbol{\sigma}_z$ (see Eqs. (\ref{Oxy_Molecular_2}) - (\ref{sigmaz})).
Therefore, we can conclude that $\boldsymbol{P}_z$ is proportional to $\boldsymbol{O}_{xy}$ in a zero field.

We also need to reproduce the magnetic-field dependence of the polarization and elastic constants as experimentally observed.
Thus, we introduce the spin-orbit coupling of Cu-$3d$ electrons written as
\begin{align}
\label{H_SO}
H_\mathrm{SO} 
= \lambda_\mathrm{SO} \left( \boldsymbol{l} \cdot \boldsymbol{s} \right)
.
\end{align}
Here, $\lambda_\mathrm{SO}$ is the coupling constant for $3d$ electrons, $\boldsymbol{l} = (l_x, l_y, l_z)$ is the vector form of the azimuthal angular momentums, and $\boldsymbol{s} = (s_x, s_y, s_z)$ is the vector form of spin angular momentums.
We used $\lambda_\mathrm{SO} = -0.1$ eV for calculations
\cite{Hayn_PRB66}.
Using the $d$-$p$ hybridized states of $H = H_\mathrm{CEF} + H_{d-p}$, the matrix elements of $H + H_\mathrm{SO}$ are written in Appendix \ref{Appendix_C}.
Diagonalizing $H + H_\mathrm{SO}$, we obtain the wave functions describing the spin-dependent quantum states.
The energy scheme of each state is illustrated in Fig. \ref{EnergyScheme}(a).
As shown in Tables \ref{Coefficients of Wave Functions SO}, $\psi_\mathrm{SO}^{1+}$ ($\psi_\mathrm{SO}^{1-}$) consists of $\psi_{3, \uparrow}$ and $\psi_{4, \uparrow}$ ($\psi_{3, \downarrow}$ and $\psi_{4, \downarrow}$).
$\psi_\mathrm{SO}^{2+}$ ($\psi_\mathrm{SO}^{2-}$) is dominated by the contribution of  $\psi_{3, \uparrow}$ and $\psi_{4, \uparrow}$ ($\psi_{3, \downarrow}$ and $\psi_{4, \downarrow}$) rather than $\psi_{7, \downarrow}$ ($\psi_{7, \uparrow}$).
In contrast, $\psi_\mathrm{SO}^{3+}$($\psi_\mathrm{SO}^{3-}$)  is mainly constructed by $\psi_{7, \downarrow}$ ($\psi_{7, \uparrow}$).
We also show that the low-energy states $\psi_\mathrm{SO}^{1\pm}$ consist of the $3d$-$yz$ and $zx$ orbitals and the molecular orbitals with the irrep $E$ (see Table \ref{Coefficients of Wave Functions from SO to yz zx at 0 T} in Appendix \ref{Appendix_C}).
The $\psi_\mathrm{SO}^{2\pm}$ states contain the $xy$ and the molecular orbital with the irrep $B_2$ in addition to these $E$ orbitals.
In contrast, the high-energy states $\psi_\mathrm{SO}^{3\pm}$ consist of these $B_2$ orbitals orbitals.
We deduce that the reason why the $\psi_\mathrm{SO}^{1+}$ states have the lowest energy is attributed to the energy scheme of $yz$, $zx$, and $xy$ orbitals for $H_\mathrm{CEF}$.

Each degenerate state formed by $H_\mathrm{SO}$ may seem to carry the electric multipole degrees of freedom in terms of the contribution of $\psi_{3, \uparrow (3, \downarrow)}$ and $\psi_{4, \uparrow (4, \downarrow)}$, however, the diagonal elements are absent because $\psi_\mathrm{SO}^{1\pm}$, $\psi_\mathrm{SO}^{2\pm}$, and $\psi_\mathrm{SO}^{3\pm}$ are each Kramers doublet.
From a different perspective, the diagonal elements of the multipole matrices become zero because the absolute values of the coefficients of $\psi_{3, \uparrow (3, \downarrow)}$ and $\psi_{4, \uparrow (4, \downarrow)}$ that constitute these three states are equal.
On the other hand, as shown in Eqs. (\ref{Ov for SO calc}) - (\ref{Pz for SO calc}) in Appendix \ref{Appendix_C}, $\psi_{3, \uparrow (3, \downarrow)}$ and $\psi_{4, \uparrow (4, \downarrow)}$ in these states bring about the off-diagonal elements between these states.
This fact indicates that mixing between these three states provides multipole contributions to polarization and elasticity in fields.
Therefore, we focus on the Zeeman effect described below:
\begin{align}
\label{H_Zeeman}
H_\mathrm{Zeeman} 
= - \mu_\mathrm{B} \left( \boldsymbol{l} + 2 \boldsymbol{s} \right) \cdot \boldsymbol{B}
.
\end{align}
Here, $\mu_\mathrm{B}$ is the Bohr magneton and $\boldsymbol{B} = (B_x, B_y, B_z)$ is the vector form of magnetic fields.
We show the matrix of $H + H_\mathrm{SO} + H_\mathrm{Zeeman}$ in Appendix \ref{Appendix_D}.
Diagonalizing $H_\mathrm{SO} + H_\mathrm{Zeeman} $ for the $d$-$p$ hybridized wave functions, we obtain the field dependence of the eigenenergies for the magnetic fields along the crystallographic orientation of $[110]$ as shown in Figs. \ref{EnergyScheme}(b) - \ref{EnergyScheme}(d).
The energy levels depend on magnetic fields, but no level crossings or hybridized gaps are observed below 60 T, suggesting that such anomalous states cannot be the origin of multipole response in high fields. 

\begin{table*}[t]
\caption{
Coefficients of $\psi_{3, \uparrow}$, $\psi_{4, \uparrow}$, $\psi_{7, \downarrow}$, $\psi_{3, \downarrow}$, $\psi_{4, \downarrow}$, and $\psi_{7, \uparrow}$ that constitute the wave function $\psi_\mathrm{SO}^{i\pm'}$ ($i = 1$ - $3$) at 50 T for $A$ site.
}
\begin{ruledtabular}
\label{Coefficients of Wave Functions SO_50T}
\begin{tabular}{c|cccccc}

	&$\psi_{3, \uparrow}$ 
		& $\psi_{4, \uparrow}$ 
			& $\psi_{7, \downarrow}$ 
				& $\psi_{3, \downarrow}$
					& $\psi_{4, \downarrow}$ 
						& $\psi_{7, \uparrow}$ 
\\
\hline
\\
$\psi_\mathrm{SO}^{1+'}$
	&$0.39 + 0.33i$
		&$0.26 - 0.42i$
			&$-0.01$
				&$-0.31 - 0.40i$
					& $-0.34 - 0.35i$
						& $-0.01$
\\
$\psi_\mathrm{SO}^{1-'}$
	&$0.42 + 0.26i$
		&$0.33 - 0.39i$
			& $0.01$
				&$0.35 - 0.34i$
					& $0.40 - 0.31i$
						&$ -0.01$
\\
$\psi_\mathrm{SO}^{2+'}$
	&$-0.23 + 0.44i$
		& $-0.47 - 0.16i$
			&$-0.07 + 0.03i$
				&$0.38 + 0.32i$
					&$0.38 - 0.33i$
						&$-0.08$
\\
$\psi_\mathrm{SO}^{2-'}$
	&$-0.16 + 0.47i$
		& $-0.44 - 0.23i$
			&$-0.07 + 0.03i$
				&$-0.33 - 0.38i$
					&$-0.33 + 0.38i$
						&$ 0.01$
\\
$\psi_\mathrm{SO}^{3+'}$
	&$- 0.02 + 0.05i$
		&$-0.05 - 0.02i$
			&$0.65 - 0.26i$
				&$0.04 + 0.04i$
					&$0.04 - 0.04i$
						&$0.70$
\\
$\psi_\mathrm{SO}^{3-'}$
	&$0.02 - 0.05i$
		&$0.05 + 0.02i$
			&$-0.65 + 0.26i$
				&$ 0.04 + 0.04i$
					&$0.04 - 0.04i$
						&$0.70$
\end{tabular}
\end{ruledtabular}
\end{table*}

To discuss the origin of the multipole response, for instance, we show the coefficients of the wave functions at 50 T for $A$ site (see Table \ref{Coefficients of Wave Functions SO_50T}). 
The proportion of $\psi_{3, \uparrow}$ ($\psi_{4, \downarrow}$) in the lowest energy state $\psi_\mathrm{SO}^{1+'}$ (second-lowest energy state $\psi_\mathrm{SO}^{1-'}$) in Fig. \ref{EnergyScheme}(d) increases with the applied fields.
Due to this imbalance of the proportion of $\psi_{3, \uparrow (3, \downarrow)}$ and $\psi_{4, \uparrow (4, \downarrow)}$, the diagonal elements of the multipole matrices become non-zero value (see Table \ref{Diagonal elements at 50T} in Appendix \ref{Appendix_D}). 
In other words, the quantum states in high fields obtain the multipole degrees of freedom.
Such imbalance of the proportion of $\psi_{3, \uparrow (3, \downarrow)}$ and $\psi_{4, \uparrow (4, \downarrow)}$ causes the field-induced proportion change of the $3d$-$yz$ and $zx$ orbitals and $2p$-$x$ and $y$ orbitals.

\begin{figure}[t]
\begin{center}
\includegraphics[clip, width=0.50\textwidth, bb=0 0 470 280]{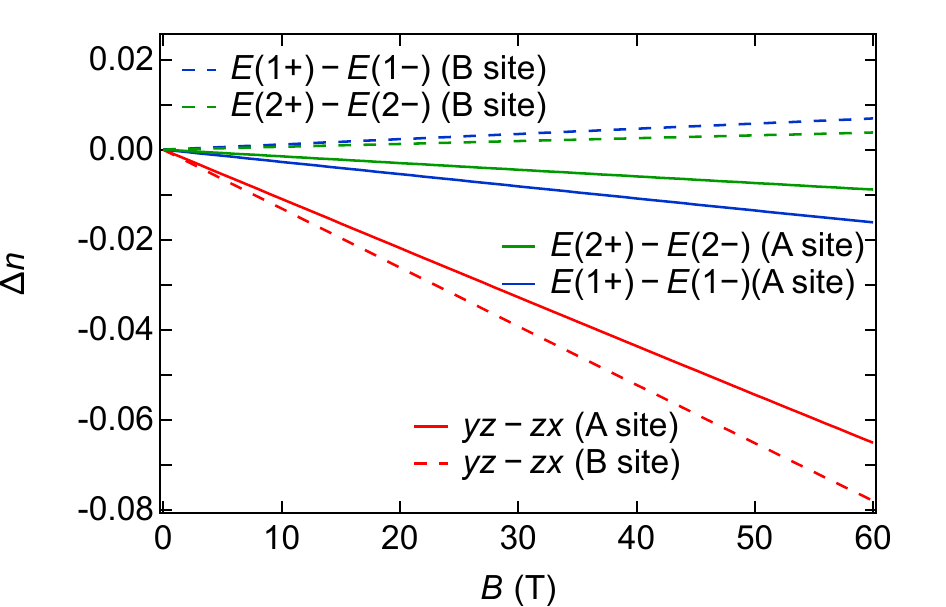}
\end{center}
\caption{
Magnetic-field dependence of the difference in the coefficients of the wave functions $\psi_{yz, \uparrow}$, $\psi_{zx \uparrow}$, $\psi_{E(1+) \uparrow}$, $\psi_{E(1-) \uparrow}$, $\psi_{E(2+) \uparrow}$, and $\psi_{E(2-) \uparrow}$ that constitute the lowest-energy states $\psi_\mathrm{SO}^{1+}$.
The solid (broken) lines indicate the $\mathit{\Delta} n$ at $A$ site ($B$ site).
}
\label{Delta_n}
\end{figure}

To confirm the field-induced proportion change, we calculated the field-dependence of the difference in the coefficients of the wave functions, which was denoted by $\mathit{\Delta} n$ (see Fig. \ref{Delta_n}).
In zero magnetic fields, the values of $|\mathit{\Delta} n|$ for $\psi_{yz}$ and $\psi_{zx}$, $\psi_{E(1+)}$ and $\psi_{E(1-)}$, and $\psi_{E(2+)}$ and $\psi_{E(2-)}$ are zero.
On the other hand, $|\mathit{\Delta} n|$ for each wave function show finite values, indicating that magnetic fields induce an anisotropic charge distribution breaking spatial inversion and rotational operations of the $D_{2d}$ point group.
This anisotropy of charge distribution can be the microscopic origin of the electric dipole and quadrupoles in high fields.

\subsection{
\label{subsect_Multipolesusceptibility}
Electric multipole susceptibility without inter-site interaction
}

Here, we show the field dependence of the multipole susceptibility of $P_z$, $O_{x^2-y^2}$, and $O_{xy}$ and the thermal average of $P_z$ without inter-site interaction.
Our analysis can explain the field-induced electric polarization and elastic softening.

To describe multipole properties, we focus on the characteristic crystal structure of Ba$_2$CuGe$_2$O$_7$.
We need to identify CuO$_4$ on the corner ($A$ site) and the center ($B$ site) of the crystal shown in Fig. \ref{CrystalStructure}(a).
Relationship between the crystallographic coordinate $\boldsymbol{r} = (x, y, z)$ and the local coordinate $\boldsymbol{R}_A = (X_A,  Y_A, Z_A)$ for A sites is described as
\begin{align}
\label{r to R_A}
\boldsymbol{R}_A = \boldsymbol{R} \left( \kappa \right) \boldsymbol{r}
\end{align}
Here, $\kappa$ is a tilting angle of CuO$_4$ tetrahedron at $A$ sites and $\boldsymbol{R} \left( \kappa \right)$ is described as
\begin{align}
\label{Define R(theta)}
\boldsymbol{R} \left( \kappa \right)
=
\left(
\begin{array}{ccc}
\cos\kappa	&-\sin\kappa	&0	\cr
\sin\kappa	&\cos\kappa	&0	\cr
0	&0	&1
\end{array}
\right).
\end{align}
Relation between $\boldsymbol{r}$ and $\boldsymbol{R}_B = (X_B, Y_B, Z_B)$ for B sites is also described as
\begin{align}
\label{r to R_B}
\boldsymbol{R}_B = \boldsymbol{R} \left( -\kappa \right) \boldsymbol{r}
\end{align}
As shown in Eqs. (\ref{r to R_A}) and (\ref{r to R_B}), $z$ component is invariant for this tilting structure.

Due to the relation of Eqs. (\ref{r to R_A}) and (\ref{r to R_B}), the multipoles in local coordinates are also described by those of the global coordinates.
The electric quadrupoles $O_{X^2-Y^2}^i$ and $O_{XY}^i$ and electric dipole $P_Z^i$ on the local sites $i = (A, B)$ are written by a vector-type coordinate of multipoles
$\boldsymbol{O} \left(\boldsymbol{r} \right) = \left( O_{x^2-y^2}, O_{xy}, P_z \right)$
for the global coordination and
$\boldsymbol{O} \left(\boldsymbol{R}_i \right) = \left( O_{X^2-Y^2}^i, O_{XY}^i, P_Z^i \right)$ for the local coordinates
as 
\begin{align}
\label{O to O_A}
\boldsymbol{O} \left(\boldsymbol{R}_A \right)
&= \boldsymbol{R} \left(2 \kappa \right) \boldsymbol{O} \left(\boldsymbol{r} \right) 
, \\
\boldsymbol{O} \left(\boldsymbol{R}_B \right)
&= \boldsymbol{R} \left(-2 \kappa \right) \boldsymbol{O} \left(\boldsymbol{r} \right)
.
\end{align}
Here, the quadrupoles for each coordinate are defined as follows:
\begin{align}
\label{Define Quadrupole1}
O_{x^2-y^2} &=  \frac{1}{\sqrt{2}}\frac{ x^2 -y^2 }{r^2} , \\
\label{Define Quadrupole2}
O_{xy} &= \sqrt{2} \frac{xy}{r^2}, \\
\label{Define Quadrupole3}
O_{X^2-Y^2}^i &= \frac{1}{ \sqrt{2}} \frac{ X_i^2 - Y_i^2 }{R_i^2}, \\
\label{Define Quadrupole4}
O_{XY}^i &= \sqrt{2} \frac{X_i Y_i} {R_i^2}.
\end{align}
These quadrupoles for the local coordinates are written by the linear combination of $O_{x^2-y^2}$ and $O_{xy}$.
Because the $z$ component is invariant for the in-plane tilting of CuO$_4$, the electric dipole $P_z$ is also invariant for the rotational operation $\boldsymbol{R} (\pm 2\kappa)$.
We stress that $O_{x^2-y^2}$ with the irrep $B_1$ for the global coordinates includes $O_{XY}^i$ with $B_2$ for the local coordinates.

To describe the electric quadrupole and dipole susceptibility, we also need to consider the quadrupole-strain interaction and the dipole-electric field interaction for both local and global coordinates. 
The rotational transformation of Eqs. (\ref{r to R_A}) and (\ref{r to R_B}) provides the strains for local coordinate (see Appendix \ref{Appendix_E}).
The strains $\varepsilon_{X^2-Y^2}^i$ and $\varepsilon_{XY}^i$ and electric field $E_z$ for the local coordinates are written by the global coordinate as
\begin{align}
\label{CoordinateTransformation_1}
\boldsymbol{\varepsilon} \left(\boldsymbol{R}_A \right)
&= \boldsymbol{R} \left(2 \kappa \right) \boldsymbol{\varepsilon} \left(\boldsymbol{r} \right)
, \\
\label{CoordinateTransformation_2}
\boldsymbol{\varepsilon} \left(\boldsymbol{R}_B \right)
&= \boldsymbol{R} \left(-2 \kappa \right) \boldsymbol{\varepsilon} \left(\boldsymbol{r} \right)
.
\end{align}
Here,
$\boldsymbol{\varepsilon} \left(\boldsymbol{r} \right) = \left( \varepsilon_{x^2-y^2},\varepsilon _{xy}, E_z \right)$
and
$\boldsymbol{\varepsilon} \left(\boldsymbol{R}_i \right) = \left( \varepsilon_{X^2-X^2}^i,\varepsilon _{XY}^i, E_Z^i \right)$
are the vector-type coordinates of external fields for the global and local coordinates, respectively.
Thus, each strain for the local coordinates is written by the linear combination of $\varepsilon_{x^2-y^2}$ and $\varepsilon_{xy}$.
The electric field $E_z$ is invariant for the rotational operation $\boldsymbol{R} (\pm 2\kappa)$.

As discussed above, we obtained the multipoles and external fields for local coordinates.
Thus, the quadrupole-strain interaction for the local coordinates,
\begin{align}
\label{HQS_v_local}
H_\mathrm{QS}^\mathrm{L}
= -\sum_{i = A, B} \sum_{\Gamma = X^2-Y^2, XY} g_{\Gamma}O_{\Gamma}^i \varepsilon_{\Gamma}^i
\end{align}
also depends on $O_{x^2-y^2}$, $O_{xy}$, $\varepsilon_{x^2-y^2}$, and $\varepsilon_{xy}$ as written in Eqs. (\ref{H_QS_L_v}) and (\ref{H_QS_L_xy}) (see Appendix \ref{Appendix_E}).
The partition function for electric quadrupoles for the site $i$ is described as
\begin{widetext}
\begin{align}
\label{func_Z_quad}
Z_i \left( \varepsilon_{X^2-Y^2}^i, \varepsilon_{XY}^i, T, B \right)
=\sum_{\Gamma = X^2-Y^2, XY} \sum_{l} \exp \left[  - \frac{E_l^i \left( \varepsilon_{\Gamma}^i, B \right) }{ k_\mathrm{B}T } \right]
\end{align}
Here, $E_l^i \left( \varepsilon_{\Gamma}^i, B \right)$ is the second perturbation energy for the quantum state $l$ in the magnetic fields $B$ at $i$ site of CuO$_4$ clusters based on $H_\mathrm{QS}^\mathrm{L}$ described below:
 \begin{align}
\label{2nd_perturbation}
E_l^i \left( \varepsilon_{\Gamma}^i, B \right) 
= E_l^i \left( \varepsilon_{\Gamma}^i = 0, B \right)
- \left\langle l, B| H_\mathrm{QS}^\mathrm{L}| l, B  \right\rangle
+\sum_{l' (\neq l)}\frac{
\left\langle l, B| H_\mathrm{QS}^\mathrm{L}| l', B  \right\rangle  \left\langle l', B| H_\mathrm{QS}^\mathrm{L}| l, B  \right\rangle
}{E_{l}^i \left( \varepsilon_{\Gamma}^i = 0, B \right) - E_{l'}^i \left( \varepsilon_{\Gamma}^i = 0, B \right)
}
\end{align}
Here, $E_l^i \left( \varepsilon_{\Gamma}^i = 0, B \right)$ is the eigen energy of the non-purturbation Hamiltonian, $H_\mathrm{CEF} + H_{d-p} + H_\mathrm{SO} + H_\mathrm{Zeeman}$.
Considering the free energy of quadrupoles for the local sites,
\begin{align}
\label{F_total}
F &= F_\mathrm{lattice} + F_\mathrm{electronic} \\ \nonumber
&= \sum_\Gamma \frac{1}{2}C_\Gamma^0 \varepsilon_{\Gamma}^{2}
- \sum_i N_i k_\mathrm{B}T \ln Z_i \left( \varepsilon_{X^2-Y^2}^i, \varepsilon_{XY}^i, T, B \right)
,
\end{align}
which include the elastic part $F_\mathrm{lattice}$ for the global strain $\varepsilon_\Gamma$ and electronic part $F_\mathrm{electronic}$,
we show that the elastic constants are written as
\begin{align}
\label{Elastic_xy}
C_{66} \left( T, B \right)
&= \frac{\partial^2 F}{\partial \varepsilon_{xy}^{2} }	\\ \nonumber
&= C_{66}^0 - N \left\{
g_{XY}^2 \left[ \chi_{XY}^A \left(T, B \right) + \chi_{XY}^B \left(T, B \right)  \right] \cos^2 2\kappa \right.
\left. + g_{X^2-Y^2}^2 \left[ \chi_{X^2-Y^2}^A \left(T, B \right) + \chi_{X^2-Y^2}^B \left(T, B \right)  \right] \sin^2 2\kappa \right.
\\ \nonumber
&\left. - \frac{ g_{XY} g_{X^2-Y^2} }{k_\mathrm{B}T} \left( 
\left\langle O_{XY}^A \left( T, B \right) \right\rangle \left\langle O_{X^2-Y^2}^A \left( T, B \right) \right\rangle
- \left\langle O_{XY}^B \left( T, B \right) \right\rangle \left\langle O_{X^2-Y^2}^B \left( T, B \right) \right\rangle
\right) \sin 4\kappa \right\}
,\\ 
\label{Elastic_v}
C_\mathrm{T} \left( T, B \right)
&= \frac{\partial^2 F}{\partial \varepsilon_{x^2-y^2}^{2} }	\\ \nonumber
&= C_\mathrm{T}^0 - N \left\{
g_{XY}^2 \left[ \chi_{XY}^A \left(T, B \right) + \chi_{XY}^B \left(T, B \right)  \right] \sin^2 2\kappa \right.
\left. + g_{X^2-Y^2}^2 \left[ \chi_{X^2-Y^2}^A \left(T, B \right) + \chi_{X^2-Y^2}^B \left(T, B \right)  \right] \cos^2 2\kappa \right.
\\ \nonumber
&\left. - \frac{ g_{XY} g_{X^2-Y^2} }{k_\mathrm{B}T} \left( 
-\left\langle O_{XY}^A \left( T, B \right) \right\rangle \left\langle O_{X^2-Y^2}^A \left( T, B \right) \right\rangle
+ \left\langle O_{XY}^B \left( T, B \right) \right\rangle \left\langle O_{X^2-Y^2}^B \left( T, B \right) \right\rangle
\right) \sin 4\kappa \right\}
.
\end{align}
Here, $N$ is half the number of CuO$_4$ clusters in the unit cell and
$\left\langle O_\Gamma^i \left( T, B \right) \right\rangle$
denotes the thermal average for the Boltzmann statistics as
$\sum_l \left\langle l, B | O_\Gamma^i | l, B \right\rangle \exp \left[ -E_l^i \left( \varepsilon_{\Gamma}^i, B \right) / k_\mathrm{B}T   \right]
/Z_i \left( \varepsilon_{X^2-Y^2}^i, \varepsilon_{XY}^i, T, B \right) \left|_{\varepsilon_\Gamma^i \rightarrow 0}  \right. $.
To calculate $C_{66}$ and $C_\mathrm{T}$, we used the transformation of the derivative with respect to the strains in Eqs. (\ref{DerivativeTransformation_1}) and (\ref{DerivativeTransformation_2}) (see Appendix. \ref{Appendix_E}).
Because 
$\left\langle O_{XY}^A \left( T, B \right) \right\rangle \left\langle O_{X^2-Y^2}^A \left( T, B \right) \right\rangle
- \left\langle O_{XY}^B \left( T, B \right) \right\rangle \left\langle O_{X^2-Y^2}^B \left( T, B \right) \right\rangle$
in Eqs. (\ref{Elastic_xy}) and (\ref{Elastic_v}) is on the oder of $10^{-6}$ K$^{-1}$ at 60 T, we ignore this term.
$\chi_\Gamma^i \left( T, B \right)$ is the susceptibility of quadrupole $O_\Gamma$ at $i$ site described below:
\begin{align}
\label{chi_Gamma}
-g_\Gamma^2 \chi_\Gamma^i \left( T, B \right)
&= \left\langle \frac{\partial^2 E_l\left( \varepsilon_{X^2-Y^2}^i, \varepsilon_{XY}^i, B \right) }{\partial \varepsilon_\Gamma^{i2} } \left|_{\varepsilon_\Gamma^i \rightarrow 0}  \right. \right \rangle
\\ \nonumber
& - \frac{1}{k_\mathrm{B}T}\left\{
\left\langle \left( \frac{\partial E_l\left( \varepsilon_{X^2-Y^2}^i, \varepsilon_{XY}^i, B \right) }{\partial \varepsilon_\Gamma^i} \left|_{\varepsilon_\Gamma^i \rightarrow 0}\right. \right)^2 \right \rangle
- \left\langle \frac{\partial E_l\left( \varepsilon_{X^2-Y^2}^i, \varepsilon_{XY}^i, B \right) }{\partial \varepsilon_\Gamma^i} \left|_{\varepsilon_\Gamma^i \rightarrow 0} \right. \right \rangle^2
\right\}
.
\end{align}

We also describe the electric dipole susceptibility.
The electric dipole-electric field interaction for the local coordinates is written as
\begin{align}
\label{HDE_v_local}
H_\mathrm{DE}^\mathrm{L}
= -\sum_{i = \mathrm{A}, \mathrm{B}} g_z P_z^i E_z^i
.
\end{align}
The partition function for electric dipole for the site $i$ is described as
\begin{align}
\label{func_Z_dipole}
Z_i \left( E_z^i, T, B \right)
= \sum_{l} \exp \left[  - \frac{E_l^i \left( E_z^i, B \right) }{ k_\mathrm{B}T } \right]
\end{align}
Considering the second perturbation energy of Eq. (\ref{2nd_perturbation}) for $H_\mathrm{DE}$ and the free energy for the  electric dipole,
\begin{align}
\label{F_total_dipole}
F = - \sum_i N_i k_\mathrm{B}T \ln Z_i \left( E_z^i, T, B \right)
,
\end{align}
we can describe the thermal average of the electric dipole $P_z$ for the Boltzmann statistics as below:
\begin{align}
\left\langle P_z^i \left( T, B \right) \right\rangle
= \frac{ \sum_l \left\langle l, B | P_z^i | l, B \right\rangle \exp \left[ -E_l^i \left( E_z^i, B \right) / k_\mathrm{B}T   \right] }
{ Z_i \left( E_z^i, T, B \right) } \left|_{ E_z^i \rightarrow 0}  \right.
.
\end{align}
Here, $E_l^i \left( E_z^i, B \right)$ is the second perturbation energy for $H_\mathrm{DE}^\mathrm{L}$.
The susceptibility of $P_z$ is written as
\begin{align}
\label{chi_Pz}
-g_z^2 \chi_z^i \left( T, B \right)
= \left\langle \frac{\partial^2 E_l\left( E_z^i, B \right) }{\partial E_z^{i2} } \left|_{E_z^i \rightarrow 0}  \right. \right \rangle
- \frac{1}{k_\mathrm{B}T}\left\{
\left\langle \left( \frac{\partial E_l\left(E_z^i, B \right) }{\partial E_z^i} \left|_{E_z^i \rightarrow 0}\right. \right)^2 \right \rangle
- \left\langle \frac{\partial E_l\left( E_z^i, B \right) }{\partial E_z^i} \left|_{E_z^i \rightarrow 0} \right. \right \rangle^2
\right\}
.
\end{align}
\end{widetext}

\begin{figure}[t]
\begin{center}
\includegraphics[clip, width=0.50\textwidth, bb=0 0 490 480]{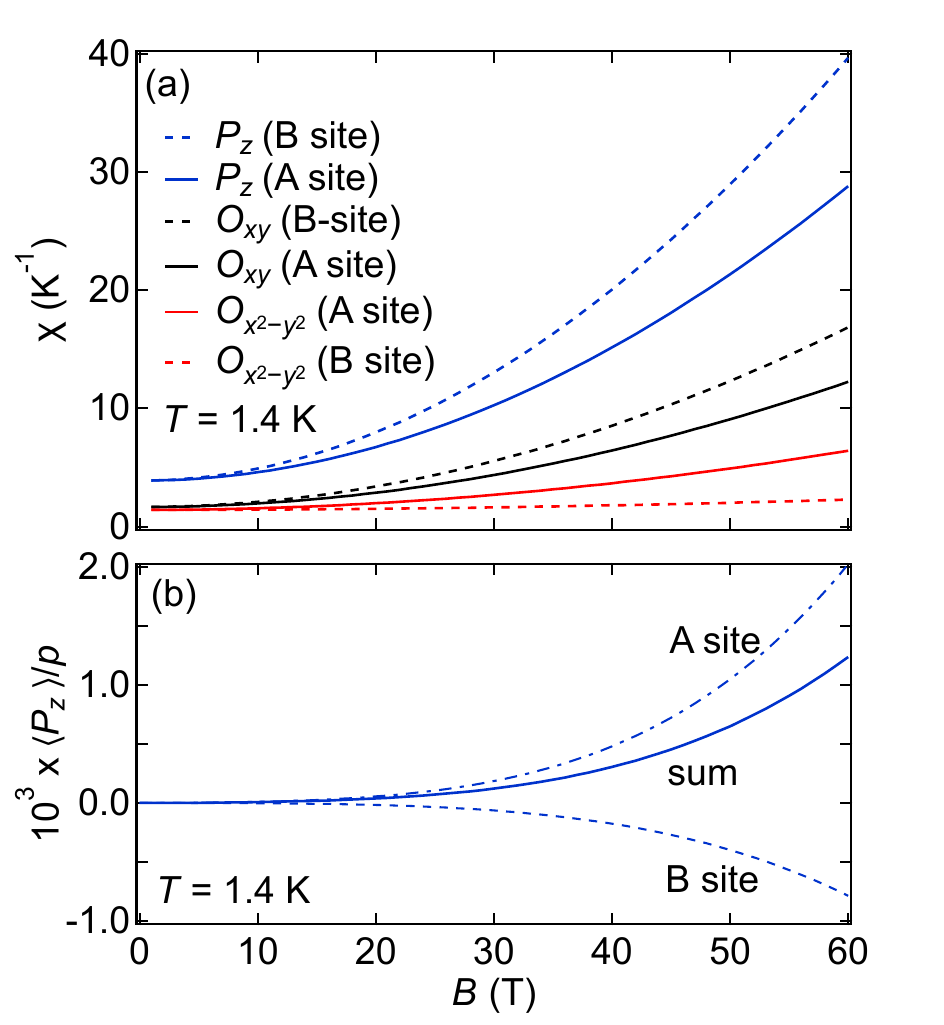}
\end{center}
\caption{
(a) Magnetic-field dependence of the electric multipole susceptibility at $A$ and $B$ sites of CuO$_4$ clusters at 1.4 K.
The solid (broken) lines indicate the susceptibilities at the $A$ site ($B$ site).
(b) Magnetic-field dependence of the thermal average of the electric dipole $P_z$ for the Boltzmann statistics divided by $p$ in Eq. (\ref{Pz}) at 1.4 K.
The dashed (broken) lines indicate the thermal average at the $A$ site ($B$ site).
The solid line shows the sum of the thermal averages for the $A$ and $B$ sites. 
}
\label{Suscept}
\end{figure}

Our calculations demonstrate that the multipoles are active for the quantum states in high fields.
We show the field dependence of multipole susceptibility for $A$ and $B$ sites at 1.4 K (see Fig. \ref{Suscept}(a)).
These susceptibilities of the electric dipole $P_z$ and the electric quadrupoles $O_{X^2-Y^2}$ and $O_{XY}$ for the local coordinates demonstrate the increase in the fields.
Furthermore, we show that the thermal average of the electric dipole denoted by $\left\langle P_z \right\rangle$ for $A$ site ($B$ site) at 1.4 K exhibits the increase (decrease) in the fields.
The sum of $\left\langle P_z \right\rangle$ also shows the increase in the field.
These results are qualitatively consistent with the field-induced electric polarization and elastic softening in Ba$_2$CuGe$_2$O$_7$.

\begin{figure}[t]
\begin{center}
\includegraphics[clip, width=0.50\textwidth, bb=0 0 490 610]{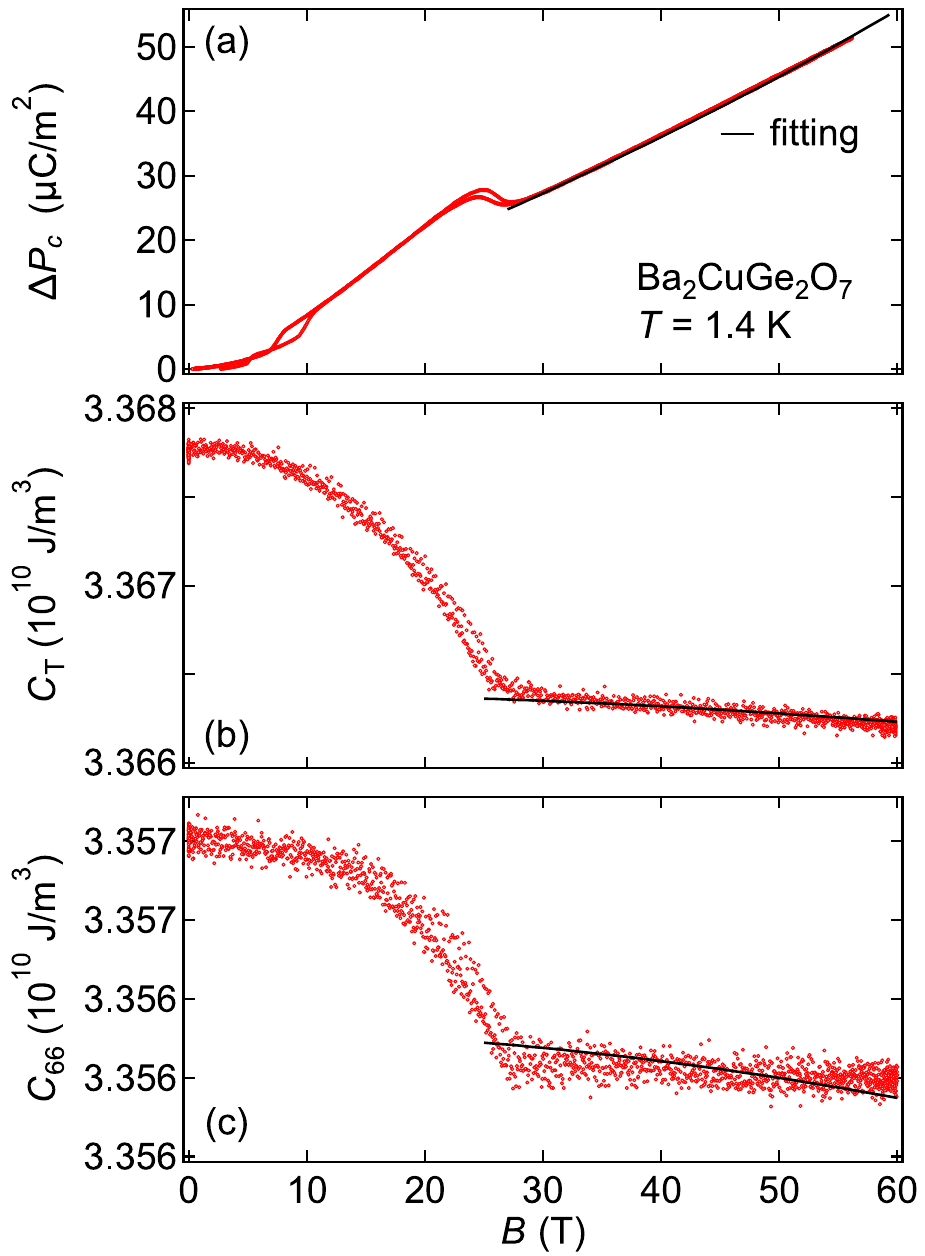}
\end{center}
\caption{
Analytical results of the field dependence of (a) the electric polarization $\mathit{\Delta} P_c$ and (b) the elastic constant $C_\mathrm{T}$ and (c) $C_{66}$.
The black lines indicate the fit of each physical quantity above $B_\mathrm{sat}$.
}
\label{fitting}
\end{figure}

Using these susceptibilities and expectation values, we analyzed our experimental results.
Figure $\ref{fitting}$ shows the results of the analysis.
The FIEP above $B_\mathrm{sat}$ is described by $P_c^0 + P_c^1 \times B + p \times \left \langle P_z \right \rangle$ where $P_c^0$ is the constant, $P_c^1$ is the coefficient of the first-order term of $B$, and $p$ is the matrix element of $P_z$ in Eq. (\ref{Pz}).
We obtaind $P_c^0 = 2.8$ $\mathrm{\mu}$C/m$^2$, $P_c^1 = 0.80$ $\mathrm{\mu}$C/m$^2 \cdot$T, and $p = 4.0 \times 10^{3}$ $\mathrm{\mu}$C/m$^2$.
Furthermore, the transverse elastic constants $C_\mathrm{T}$ and $C_{66}$ above $B_\mathrm{sat}$ are described by Eqs. (\ref{Elastic_xy}) and (\ref{Elastic_v}).
We obtained $C_\mathrm{T}^0 = 3.3664 \times 10^{10}$ J/m$^3$, $C_{66}^0 = 3.3563 \times 10^{10}$ J/m$^3$, $g_{X^2-Y^2} = 0$ K, and $g_{XY} = 1.9$ K.
This result indicates that the elastic softening of $C_{66}$ is attributed to the electric quadrupole $O_{XY}$ with the irrep $B_{2g}$ for the local coordinates.
The softening of $C_\mathrm{T}$ originates from the response of $O_{XY}$ and the tilting crystal structure of CuO$_4$ tetrahedra.
Since $p$ and $g_{XY}$ are attributed to the radial integration of the wave functions, estimated values can reflect the spatial extent of these wave functions (see Appendix \ref{Appendix_C}.)
We stress that this field-induced phenomenon is attributed to the proportion change of the $3d$-$yz$ and $zx$ orbitals and $2p$-$x$ and $y$ orbitals in the fields.

While our model successfully provides a qualitative explanation of the experimental results, several aspects could be further improved.
One is the field dependence of $\left\langle P_z \right\rangle$.
Our experimental and analytical results of FIEP show the contribution of $B$-linear term in addition to our calculated $\left\langle P_z \right\rangle$.
For quantitative explanations, it may be necessary to incorporate other interactions such as dipole-dipole and spin interactions to calculate the quantum state in high fields.
Another is the qualitative explanation of $g$ value.
The models with the CEF, the $d$-$p$ hybridization, and the spin-orbit coupling provide a qualitative description of the anisotropy of magnetization.
Based on these wave functions and energy schemes, we can estimate an effective $g$ value 
\cite{Pryce_PRSA63, Tanaka_JPSJ68}
for the $[110]$ ($[001]$) crystallographic orientation to be $2.13$ ($2$).
While we can explain the enhancement of the $g$ value greater than $2$ for the in-plane field direction, there is room for improvement to reproduce our experimental result of $g = 2.44$.
By adjusting the energy levels of each orbital under the CEF as follows: $\epsilon_{dB_2} = 0.1$ eV, $\epsilon_{dE} = -0.25$ eV, and $\epsilon_{pE} = -2.5$ eV, we can reproduce $g \sim 2.44$.
We can also reproduce the $g$ value by increasing the magnitude of the $d$-$p$ hybridization.
However, the energy levels of the hybridized orbitals calculated from these parameters are narrower compared to the previous study
\cite{Corasaniti_PRB96}.
Another approach to reproduce $g = 2.44$ is to enhance the strength of the spin-orbit coupling from $\lambda_\mathrm{SO} = -0.1$ eV to $-0.35$ eV.
Nevertheless, such an enhancement cannot be considered realistic since the magnitude of the spin-orbit coupling is determined by the atomic orbitals.
Therefore, we deduce that other interactions, such as magnetic and anisotropic exchange interactions, contribute to the observed magnetic anisotropy in Ba$_2$CuGe$_2$O$_7$.
Although a quantitative explanation for the energy levels cannot be achieved, it is possible to provide a quantitative explanation for the multipole degrees of freedom originating from the quantum states in Fig. \ref{EnergyScheme}(a).

\subsection{
\label{subsect_Possible}
Possible contribution of inter-site interaction to multipole susceptibility
}

In addition to the quadrupole-strain interaction, quadrupole-quadrupole interaction is also needed for the quadrupole susceptibility.
The quadrupole-quadrupole interaction for the global coordinate is written as
\begin{align}
\label{H_QQ_grobal}
H_\mathrm{QQ}^\mathrm{G}
= -\sum_{i \in A, j \in B} \sum_{\Gamma = x^2-y^2, xy} G_\Gamma^{ij} O_\Gamma^i O_\Gamma^j
.
\end{align}
Here, $G_\Gamma^{ij}$ is an interaction coefficient. 
If we define the quadrupole-quadrupole interaction for local coordinates using the interaction coefficient $G_{\Gamma'}^{ij}$ as
\begin{align}
\label{H_QQ_local}
H_\mathrm{QQ}^\mathrm{L}
= -\sum_{i \in A, j \in B} \sum_{\Gamma' = X^2-Y^2, XY} G_{\Gamma'}^{ij} O_{\Gamma'}^i O_{\Gamma'}^j
,
\end{align}
$H_\mathrm{QQ}^\mathrm{G}$ is decomposed as below:
\begin{align}
\label{H_QQ_local}
H_\mathrm{QQ}^\mathrm{G}
= H_\mathrm{QQ}^\mathrm{L} +H_\mathrm{cross}
.
\end{align}
Here, $H_\mathrm{cross}$ is additional Hamiltonian, which is similar to the DM interaction described as
\begin{align}
\label{Hcross}
H_\mathrm{cross}
&= -\sum_{i \in A, j \in B} G_\mathrm{cross}^{ij} \left( O_{X^2-Y^2}^i O_{XY}^j - O_{XY}^i O_{X^2-Y^2}^j \right)
\nonumber \\
&= -\sum_{i \in A, j \in B} G_\mathrm{cross}^{ij} \left[  \boldsymbol{O} \left(\boldsymbol{R}_i \right) \times \boldsymbol{O} \left(\boldsymbol{R}_j \right) \right]_z
.
\end{align}
The interaction coefficients $G_{X^2-Y^2}^{ij}$, $G_{XY}^{ij}$, and $G_\mathrm{cross}^{ij}$ are described as 
\begin{align}
G_{X^2-Y^2}^{ij} &= G_{x^2-y^2}^{ij} \cos^2 2\kappa - G_{xy}^{ij} \sin^2 2\kappa
,\\
G_{XY}^{ij} &= -G_{x^2-y^2}^{ij} \sin^2 2\kappa + G_{xy}^{ij} \cos^2 2\kappa
, \\
G_\mathrm{cross}^{ij} &= - \left( G_{x^2-y^2}^{ij} + G_{xy}^{ij} \right) \sin 4\kappa
.
\end{align}
$H_\mathrm{cross}$ originates from the tilting crystal structure.
The additional term of the dipole-dipole interaction for $P_z$ does not appear because $P_z$ is invariant for the rotational operation $\boldsymbol{R} ({\kappa})$.
Considering the tilting angle $\kappa$ in Ba$_2$CuGe$_2$O$_7$, we deduce that $\sin4\kappa$ in $G_\mathrm{cross}^{ij}$ is not negligible.
Therefore, if the sum of the interaction coefficients, $G_{x^2-y^2}^{ij} + G_{xy}^{ij}$, has a finite value, we should take $H_\mathrm{cross}$ into account for calculating the multipole susceptibility.

We could calculate the quantum states and the quadrupole susceptibility with $H_\mathrm{cross}$, however, it is hard to obtain analytical results.
We desire the theoretical calculations of the elastic constants and the polarization with the inter-site quadrupole-quadrupole interaction in Eq. (\ref{H_QQ_local}) and the DM-like quadrupole-quadrupole interaction in Eq. (\ref{Hcross}). 
Considering these interactions may provide a more quantitative explanation of the experimental results.
Although we ignore magnetic interactions for calculations, we believe that our calculations successfully demonstrate the importance of the orbital contributions to the field-induced multiferroicity.
Taking into account the magnetic interactions, our model can also explain the FIEP and elastic softening below $B_\mathrm{sat}$.


\section{
\label{conclusion}
Conclusion}

We investigated magnetization, polarization, and elastic constants in Ba$_2$CuGe$_2$O$_7$ under high-magnetic fields to elucidate the contribution of electric dipoles and electric quadrupoles to the multiferroicity.
Above the spin saturation fields, we found that the electric polarization $P_c$ exhibited increasing up to 56 T and the elastic constants showed softening with the increase in the fields up to 60 T.
Our theoretical calculation revealed that orbital degrees of freedom of O-2$p$ and Cu-3$d$ electrons can play a key role in the multiferroicity between the electric polarization and elastic constants in high-magnetic fields.
This characteristic phenomenon originates from the crystalline electric field, $d$-$p$ hybridization between Cu-$3d$ and O-$2p$ electrons, and the spin-orbit coupling of $3d$ electrons.
Cross-correlation between the electric dipole $P_z$ and the electric quadrupole $O_{xy}$ satisfying $P_z \propto O_{xy}$ can be another possible multiferroic mechanism in such quantum systems.

\section*{Acknowledgment}
The authors thank Yuichi Nemoto and Mitsuhiro Akatsu for supplying the LiNbO$_3$ piezoelectric transducers.
The authors also thank Kunihiko Yamauchi and Hiroshi Yaguchi for valuable discussions.
This work was partly supported by JSPS Grants-in-Aid for early-career scientists (KAKENHI JP20K14404, JP22K13999).

\begin{table*}[htbp]
\label{Representation matrices of D2d}
\caption{
Representation matrix of the symmetry operation of the point group $D_{2d}$ for the polar bases $x$, $y$, and $z$, quadratic bases $yz$, $zx$, and $xy$, the axial bases $l_x$, $l_y$, and $l_z$, and the producted bases.
The lines drawn within the matrix are used to distinguish between the block matrices of the two- and one-dimensional representations.
}
\centering
\begin{tabular}{ccccccccc}
\hline  \hline
Basis	&$E$	&$IC_4$   &$IC_4^{-1}$     &$C_2$    &$C_2^y$  &$C_2^x$    &$\sigma_d^{\overline{y}}$   &$\sigma_d^{\overline{x}}$    \\  \hline
$\left( \begin{array}{c}
                    x     \\
                    y     \\
                    z     \\
                \end{array}\right)$        
&$\left( \begin{array}{cc|c}
                    1   &0  &0  \\
                    0   &1  &0  \\  \hline
                    0   &0  &1  \\
                \end{array}\right)$
&$\left( \begin{array}{cc|c}
                    0   &\overline{1}  &0  \\
                    1   &0  &0  \\ \hline
                    0   &0  &\overline{1}  \\
                \end{array}\right)$
&$\left( \begin{array}{cc|c}
                    0   &1  &0  \\
                    \overline{1}   &0  &0  \\  \hline
                    0   &0  &\overline{1}  \\
                \end{array}\right)$   
&$\left( \begin{array}{cc|c}
                    \overline{1}   &0  &0  \\
                    0   &\overline{1}  &0  \\ \hline
                    0   &0  &1  \\
                \end{array}\right)$
&$\left( \begin{array}{cc|c}
                    \overline{1}   &0  &0  \\
                    0   &1  &0  \\  \hline
                    0   &0  &\overline{1}  \\
                \end{array}\right)$
&$\left( \begin{array}{cc|c}
                    1   &0  &0  \\
                    0   &\overline{1}  &0  \\ \hline
                    0   &0  &\overline{1}  \\
                \end{array}\right)$
&$\left( \begin{array}{cc|c}
                    0   &\overline{1}  &0  \\
                    \overline{1}   &0  &0  \\ \hline
                    0   &0  &1  \\
                \end{array}\right)$
&$\left( \begin{array}{cc|c}
                    0   &1  &0  \\
                    1   &0  &0  \\  \hline
                    0   &0  &1  \\
                \end{array}\right)$   
\\	\\
$\left( \begin{array}{c}
                    yz     \\
                    zx     \\
                    xy     \\
                \end{array}\right)$
&$\left( \begin{array}{cc|c}
1   &0	&0  \\
0   &1	&0  \\	\hline
0	&0	&1	\\
\end{array} \right)$
&$\left( \begin{array}{cc|c}
0   &\overline{1}  &0	\\
1   &0  &0	\\	\hline
0	&0	&\overline{1}	\\
\end{array} \right)$
&$\left( \begin{array}{cc|c}
0   &1  &0	\\
\overline{1}   &0  &0	 \\	\hline
0	&0	&\overline{1}	\\
\end{array} \right)$
&$\left( \begin{array}{cc|c}
\overline{1}   &0  &0	\\
0   &\overline{1}  &0	\\	\hline
0	&0	&1	\\
\end{array} \right)$
&$\left( \begin{array}{cc|c}
\overline{1}   &0	&0  \\
0   &1	&0  \\	\hline
0	&0	&\overline{1}	\\
\end{array} \right)$
&$\left( \begin{array}{cc|c}
1   &0	&0 \\
0   &\overline{1}  &0	\\	\hline
0	&0	&\overline{1}	\\
\end{array} \right)$
&$\left( \begin{array}{cc|c}
0   &\overline{1}  &0	\\
\overline{1}   &0  &0	\\	\hline
0	&0	&1	\\
\end{array} \right)$
&$\left( \begin{array}{cc|c}
0   &1  &0	\\
1   &0  &0	\\	\hline
0	&0	&1	\\
\end{array} \right)$   
\\	\\
$\left( \begin{array}{c}
                    l_x     \\
                    l_y     \\ 
                    l_z     \\
                \end{array}\right)$
&$\left( \begin{array}{cc|c}
1   &0	&0  \\
0   &1	&0  \\	\hline
0	&0	&1	\\
\end{array} \right)$
&$\left( \begin{array}{cc|c}
0   &1  &0	\\
\overline{1}   &0  &0	\\	\hline
0	&0	&1	\\
\end{array} \right)$
&$\left( \begin{array}{cc|c}
0   &\overline{1}  &0	\\
1   &0 &0	 \\	\hline
0	&0	&1	\\
\end{array} \right)$
&$\left( \begin{array}{cc|c}
\overline{1}   &0  &0	\\
0   &\overline{1}  &0	\\	\hline
0	&0	&1	\\
\end{array} \right)$
&$\left( \begin{array}{cc|c}
\overline{1}   &0	&0  \\
0   &1	&0  \\	\hline
0	&0	&\overline{1}	\\
\end{array} \right)$
&$\left( \begin{array}{cc|c}
1   &0	&0 \\
0   &\overline{1}  &0	\\	\hline
0	&0	&\overline{1}	\\
\end{array} \right)$
&$\left( \begin{array}{cc|c}
0   &1  &0	\\
1   &0  &0	\\	\hline
0	&0	&\overline{1}	\\
\end{array} \right)$
&$\left( \begin{array}{cc|c}
0   &\overline{1}  &0	\\
\overline{1}   &0  &0	\\	\hline
0	&0	&\overline{1}	\\
\end{array} \right)$   \\
\hline  \hline
\end{tabular}
\end{table*}

\appendix

\section{
\label{Appendix_BeforeBeforeA}
Symmetry operations of the $D_{2d}$ point group for several basis functions
}

We show the representation matrices of the symmetry operations of the $D_{2d}$ point group for several basis functions in Table \ref{Representation matrices of D2d}.
Considering the transformation of basis functions, we can reproduce the character table of Table \ref{table_character of D2d}.

\section{
\label{Appendix_Before_A}
Field-sweep dependence of the anomaly around $B_\mathrm{c1}$
}

\begin{figure}[htbp]
\begin{center}
\includegraphics[clip, width=0.50\textwidth, bb=0 0 350 270]{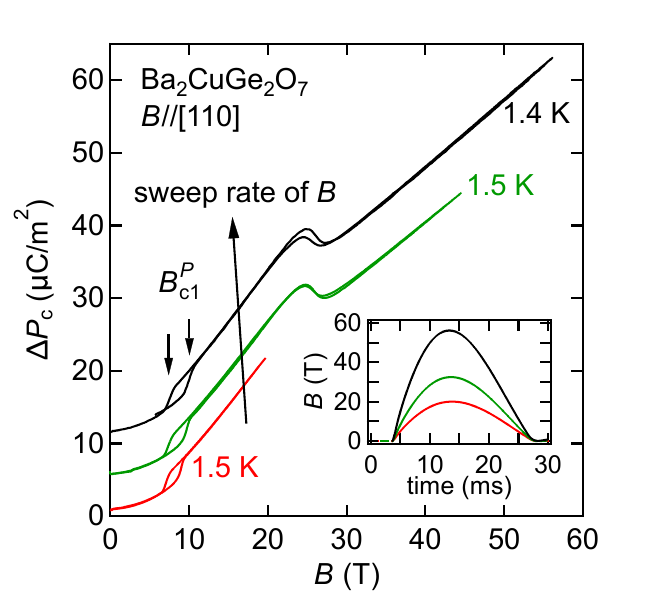}
\end{center}
\caption{
Magnetic-field dependence of the electric polarization $\mathit{\Delta}P_c$ in Ba$_2$CuGe$_2$O$_7$ at several temperatures for $B//[110]$ measured at several sweep rates of magnetic fields.
The data sets are shifted consecutively along the $\mathit{\Delta}P_c$ axes for clarity.
The inset shows the time dependence of magnetic fields.
}
\label{SweepRateDep}
\end{figure}

In this section, we discuss the origin of the hysteresis behavior around $B_\mathrm{c1}$ appearing in the magnetization and polarization.
Figure \ref{SweepRateDep} shows the magnetic-field dependence of the electric polarization $\mathit{\Delta}P_c$ in Ba$_2$CuGe$_2$O$_7$ at several temperatures measured at several sweep rates of magnetic fields.
In high-field measurements using the pulsed magnet, the sweep time of the magnetic field is kept constant while varying the maximum value of the magnetic field (see the inset of Fig. \ref{SweepRateDep}).
As a result, a higher maximum magnetic field corresponds to a faster sweep rate of magnetic fields.
In these measurements, the hysteresis behavior around $B_\mathrm{c1}^P$ seems to be independent of the sweep rate of magnetic fields.
If the hysteresis behavior is due to temperature changes induced by the magnetic fields, we can expect that $B_\mathrm{c1}$ depends on the time for the temperature to relax.
In other words, $B_\mathrm{c1}$ can depend on the sweep rate of the magnetic fields.
Therefore, we concluded that the origin of this hysteresis behavior is attributed to the field-induced magnetic structure change from the incommensurate spiral to the commensurate one.

\section{
\label{Appendix_A}
Wave functions of Cu-$3d$ and O-$2p$ electrons and CEF parameters
}

We show the wave functions describing Cu-$3d$ electrons and O-$2p$ electrons as follows:
\begin{align}
\psi_{yz} (\boldsymbol{r})
&= \sqrt{ \frac{15}{4\pi} } f_{3d}(r) \frac{yz}{r^2}
, \\
\psi_{zx}(\boldsymbol{r})
&= \sqrt{ \frac{15}{4\pi} } f_{3d}(r) \frac{zx}{r^2}
, \\
\psi_{xy}(\boldsymbol{r})
&= \sqrt{ \frac{15}{4\pi} } f_{3d}(r) \frac{xy}{r^2}
, \\
\psi_{x}(\boldsymbol{r})
&= \sqrt{ \frac{3}{4\pi} } f_{2p}(r) \frac{x}{r}
, \\
\psi_{y}(\boldsymbol{r})
&= \sqrt{ \frac{3}{4\pi} } f_{2p}(r) \frac{y}{r}
.
\end{align}
Here, $f_{3d}(r)$ and $f_{2d}(r)$ are radial distribution functions for $3d$ and $2p$ electrons, respectively.
Based on the wave functions of $3d$ electrons, we can describe the relationship between the energy levels $\epsilon_E$ ($\epsilon_{B_2}$) for the $yz$ and $zx$ ($xy$) orbitals and the CEF parameters $A_2^0$ $A_4^0$ and $A_4^4$ in Eq. (\ref{H_CEF}) as below:
\begin{align}
\epsilon_E
&= \frac{1}{7}A_2^0 - \frac{4}{21}A_4^0
,	\\
\epsilon_{B_2}
&= -\frac{2}{7}A_2^0 + \frac{4}{21}A_4^0 - \frac{\sqrt{35} }{21}A_4^4 
.
\end{align}

\section{
\label{Appendix_B}
Molecular orbitals of CuO$_4$ tetrahedra
}

\begin{table*}[htbp]
\caption{
Transformations of the O-$2p$ orbitals $(p_x^i, p_y^i, p_z^i)$ at O$_i$ $(i = 1,2, 3, 4)$ by the eight operations $R $ ($= E$, $C_2$, $C_2^y$, $C_2^x$, $IC_4$, $IC_4^{-1}$, $\sigma_d^{\overline{y}}$, $\sigma_d^{\overline{x}}$) of the point group $D_{2d}$ and the characters $\chi(R)$.
}
\begin{ruledtabular}
\label{table_p orbitals}
\begin{tabular}{cccccccccccccc}
$R$
& $p_x^1$
	& $p_x^2$
		& $p_x^3$
			& $p_x^4$
				& $p_y^1$
					& $p_y^2$
						& $p_y^3$
							& $p_y^4$
								& $p_z^1$
									& $p_z^2$
										& $p_z^3$
											& $p_z^4$
												& $\chi(R)$
\\
\hline
$E$
& $p_x^1$
	& $p_x^2$
		& $p_x^3$
			& $p_x^4$
				& $p_y^1$
					& $p_y^2$
						& $p_y^3$
							& $p_y^4$
								& $p_z^1$
									& $p_z^2$
										& $p_z^3$
											& $p_z^4$
												& $12$
\\
$IC_4$
& $-p_y^4$
	& $-p_y^1$
		& $-p_y^2$
			& $-p_y^3$
				& $p_x^4$
					& $p_x^1$
						& $p_x^2$
							& $p_x^3$
								& $-p_z^4$
									& $-p_z^1$
										& $-p_z^2$
											& $-p_z^3$
												& $0$
\\
$IC_4^{-1}$
& $p_y^2$
	& $p_y^3$
		& $p_y^4$
			& $p_y^1$
				& $-p_x^2$
					& $-p_x^3$
						& $-p_x^4$
							& $-p_x^1$
								& $-p_z^2$
									& $-p_z^3$
										& $-p_z^4$
											& $-p_z^1$
												& $0$
\\
$C_2$
& $-p_x^3$
	& $-p_x^4$
		& $p_x^1$
			& $-p_y^2$
				& $-p_y^3$
					& $-p_y^4$
						& $-p_y^1$
							& $-p_y^2$
								& $p_z^3$
									& $p_z^4$
										& $p_z^1$
											& $p_z^2$
												& $0$
\\
$C_2^y$
& $-p_x^2$
	& $-p_x^1$
		& $-p_x^4$
			& $-p_x^3$
				& $p_y^2$
					& $p_y^1$
						& $p_y^4$
							& $p_y^3$
								& $-p_z^2$
									& $-p_z^1$
										& $-p_z^4$
											& $-p_z^3$
												& $0$
\\
$C_2^x$
& $p_x^4$
	& $p_x^3$
		& $p_x^2$
			& $p_x^1$
				& $-p_y^4$
					& $-p_y^3$
						& $-p_y^2$
							& $-p_y^1$
								& $-p_z^4$
									& $-p_z^3$
										& $-p_z^2$
											& $-p_z^1$
												& $0$
\\
$\sigma_d^{\overline{y}}$
& $-p_y^3$
	& $-p_y^2$
		& $-p_y^1$
			& $-p_y^4$
				& $-p_x^3$
					& $-p_x^2$
						& $-p_x^1$
							& $-p_x^4$
								& $p_z^3$
									& $p_z^2$
										& $p_z^1$
											& $p_z^4$
												& $2$
\\
$\sigma_d^{\overline{x}}$
& $p_y^1$
	& $p_y^4$
		& $p_y^3$
			& $p_y^2$
				& $p_x^1$
					& $p_x^4$
						& $p_x^3$
							& $p_x^2$
								& $p_z^1$
									& $p_z^4$
										& $p_z^3$
											& $p_z^2$
												& $2$
\end{tabular}
\end{ruledtabular}
\end{table*}

In this section, we describe the molecular orbitals of CuO$_4$ tetrahedra consisting of Cu-$3d$ and O-$2p$ electrons.
At first, we describe the molecular orbitals of O$_4$ tetrahedron consisting of O-$2p$ electrons.
We consider the O-$2p$ orbitals $(p_x^i, p_y^i, p_z^i)$ at the O$_i$ for $i = 1, 2, 3, 4$.
The symmetry transformations of the point group $D_{2d}$ for these orbitals are summarized in Table \ref{table_p orbitals}.
The characters $\chi(R)$ in Table \ref{table_p orbitals} leads the decomposition 
$2A_1 \oplus A_2 \oplus B_1 \oplus 2B_2 \oplus 3E$.
Considering the projection operators for the irrep $\Gamma$ of $D_{2d}$, we obtain the twelve molecular orbitals of O$_4$ tetrahedron. 
Because $2p$-$z$ orbital does not contribute to the response of the electric multipoles with the irreps $B_1$ and $B_2$, we focus on the eight molecular orbitals consisting of $x$ and $y$ orbitals described as follows:
\begin{widetext}
\begin{align}
\label{molecular orbitals}
\left(
\begin{array}{c}
\psi_{A_1} \left( \boldsymbol{r}_1, \boldsymbol{r}_2, \boldsymbol{r}_3, \boldsymbol{r}_4 \right)	\cr
\psi_{A_2} \left( \boldsymbol{r}_1, \boldsymbol{r}_2, \boldsymbol{r}_3, \boldsymbol{r}_4 \right)	\cr
\psi_{B_1} \left( \boldsymbol{r}_1, \boldsymbol{r}_2, \boldsymbol{r}_3, \boldsymbol{r}_4 \right)	\cr
\psi_{B_2} \left( \boldsymbol{r}_1, \boldsymbol{r}_2, \boldsymbol{r}_3, \boldsymbol{r}_4 \right)	\cr
\psi_{E_{(1+)}} \left( \boldsymbol{r}_1, \boldsymbol{r}_2, \boldsymbol{r}_3, \boldsymbol{r}_4 \right)	\cr
\psi_{E_{(1-)}} \left( \boldsymbol{r}_1, \boldsymbol{r}_2, \boldsymbol{r}_3, \boldsymbol{r}_4 \right)	\cr
\psi_{E_{(2+)}} \left( \boldsymbol{r}_1, \boldsymbol{r}_2, \boldsymbol{r}_3, \boldsymbol{r}_4 \right)	\cr
\psi_{E_{(2-)}} \left( \boldsymbol{r}_1, \boldsymbol{r}_2, \boldsymbol{r}_3, \boldsymbol{r}_4 \right)	\cr
\end{array}
\right)
=\frac{1}{2\sqrt{2}}\left(
\begin{array}{cccccccc}
1	&1	&-1	&1	&-1	&-1	&1	&-1	\cr
1	&-1	&1	&1	&-1	&1	&-1	&-1	\cr
1	&-1	&-1	&-1	&-1	&1	&1	&1	\cr
1	&1	&1	&-1	&-1	&-1	&-1	&1	\cr
1	&1	&1	&1	&1	&1	&1	&1	\cr
1	&-1	&1	&-1	&1	&-1	&1	&-1	\cr
1	&1	&-1	&-1	&1	&1	&-1	&-1	\cr
-1	&1	&1	&-1	&-1	&1	&1	&-1	\cr
\end{array}
\right) \left(
\begin{array}{c}
\psi_x \left( \boldsymbol{r}_1 \right)	\cr
\psi_y \left( \boldsymbol{r}_1 \right) 	\cr
\psi_x \left( \boldsymbol{r}_2 \right)	\cr
\psi_y \left( \boldsymbol{r}_2 \right)	\cr
\psi_x \left( \boldsymbol{r}_3 \right)	\cr
\psi_y \left( \boldsymbol{r}_3 \right)	\cr
\psi_x \left( \boldsymbol{r}_4 \right)	\cr
\psi_y \left( \boldsymbol{r}_4 \right)	\cr
\end{array}
\right)
\end{align}
We stress that the molecular orbitals with the irrep $E$ in Eq. (\ref{molecular orbitals}) and 3$d$-$yz$ and $zx$ orbitals with the irrep $E$ provide the finite value of the matrix elements of the electric multipoles described as follows:
\begin{align}
\label{Ov_Molecular}
\boldsymbol{O}_{x^2-y^2}
&= \bordermatrix{
&\psi_{yz}	&\psi_{zx}	&\psi_{E(1+)}	&\psi_{E(1-)}	&\psi_{E(2+)}	&\psi_{E(2-)}	\cr
&-\sqrt{2}/7 r_d^2	&0	&0	&0	&0	&0 \cr
	&0	&\sqrt{2}/7 r_d^2	&0	&0	&0	&0	 \cr
			&0	&0	& \sqrt{2}/5 r_p^2	&0	&0	&0 \cr
				&0	&0	& 0	&-\sqrt{2}/5r_p^2	&0	&0	\cr
					&0	&0	&0	&0	&\sqrt{2}/5r_p^2	&0	\cr
						&0	&0	&0	&0	&0	&-\sqrt{2}/5r_p^2
},
\end{align}
\begin{align}
\label{Oxy_Molecular}
\boldsymbol{O}_{xy}
&=\left(
\begin{array}{cccccc}
0	&\sqrt{2}/7 r_d^2	&0	&0	&0	&0	\cr
\sqrt{2}/7 r_d^2	&0	&0	&0	&0	&0	\cr
0	&0	&0	&\sqrt{2}/5r_p^2	&0	&0	\cr
0	&0	&\sqrt{2}/5r_p^2	&0	&0	&0	\cr
0	&0	&0	&0	&0	&-\sqrt{2}/5r_p^2	\cr
0	&0	&0	&0	&-\sqrt{2}/5r_p^2	&0	\cr
\end{array}
\right),
\end{align}
\begin{align}
\label{Pz_Molecular}
\boldsymbol{P}_{z}
&=\left(
\begin{array}{cccccc}
0	&0	&\sqrt{2}p	&-\sqrt{2}p	&0	&0	\cr
0	&0	&\sqrt{2}p	&\sqrt{2}p	&0	&0	\cr
\sqrt{2}p	&\sqrt{2}p	&0	&0	&0	&0	\cr
-\sqrt{2}p	&\sqrt{2}p	&0	&0	&0	&0	\cr
0	&0	&0	&0	&0	&0	\cr
0	&0	&0	&0	&0	&0	\cr
\end{array}
\right).
\end{align}
The $d$-$p$ hybridization between $3d$-$yz$, $zx$, and $xy$ orbitals and molecular orbitals in Eq. (\ref{molecular orbitals}) leads following matrix, $H_\mathrm{CEF} + H_{d-p}$, as 
\begin{align}
\label{Matrix of CEF + d-p}
&H_\mathrm{CEF} + H_{d-p}	\nonumber	\\
&= \bordermatrix{
& \psi_{yz}	&\psi_{zx}	&\psi_{E(1+)}	&\psi_{E(1-)}	&\psi_{E(2+)}	&\psi_{E(2-)}	&\psi_{xy}	&\psi_{B_2}	&\psi_{A_1}	&\psi_{A_2}	&\psi_{B_1} \cr
& \epsilon_{dE}
	& 0
		& \sqrt{2} t_{x,yz}
			& \sqrt{2} t_{x,yz}
				& \sqrt{2} t_{x,zx}
					& \sqrt{2} t_{x,zx}
						&0	&0	&0	&0	&0 \cr
& 0
	& \epsilon_{dE}
		& \sqrt{2} t_{x,yz}
			& -\sqrt{2} t_{x,yz}
				& \sqrt{2} t_{x,zx}
					& \sqrt{2} t_{x,zx}
						& 0	&0	&0	&0	&0 \cr
& \sqrt{2} t_{x,yz}
	& \sqrt{2} t_{x,yz}
		& \epsilon_{pE}
			& 0
				& 0
					& 0
						&0	&0	&0	&0	&0 \cr
& \sqrt{2} t_{x,yz}
	& -\sqrt{2} t_{x,yz}
		& 0
			& \epsilon_{pE}
				& 0
					& 0
						&0	&0	&0	&0	&0 \cr
& \sqrt{2} t_{x,zx}
	& \sqrt{2} t_{x,zx}
		& 0
			& 0
				& \epsilon_{pE}
					& 0
						&0	&0	&0	&0	&0 \cr
& \sqrt{2} t_{x,zx}
	& -\sqrt{2} t_{x,zx}
		& 0
			& 0
				& 0
					& \epsilon_{pE}
						&0	&0	&0	&0	&0 \cr
&0	&0	&0	&0	&0	&0 
						& \epsilon_{dB_2}
							& 2\sqrt{2} t_{x,xy}
								&0	&0	&0 \cr
&0	&0	&0	&0	&0	&0 
						& 2\sqrt{2} t_{x,xy}
							& \epsilon_{pB_2}
								&0	&0	&0 \cr
&0	&0	&0	&0	&0	&0	&0	&0
								&  \epsilon_{pA_1}
									&0 	&0 \cr
&0	&0	&0	&0	&0	&0	&0	&0	&0
									& \epsilon_{pA_2}
										&0 \cr
&0	&0	&0	&0	&0	&0	&0	&0	&0	&0
										& \epsilon_{pB_1}
}
.
\end{align}
Here, the diagonal elements $\varepsilon_{d\Gamma}$ and $\varepsilon_{p\Gamma}$ are the eigenenergies of CEF states, respectively.
$t_{x,yz}$, $t_{x,zx}$, and $t_{x,xy}$ in the matrix of Eq. (\ref{Matrix of CEF + d-p}) are the transfer energy between $2p$ orbitals and $3d$ orbitals written as
\cite{Slater_PR94}
\begin{align}
\label{t_xyz}
t_{x,yz} &= s^2c \left(\sqrt{3} V_{pd\sigma} - 2 V_{pd\pi} \right), \\
\label{t_xzx}
t_{x,zx} &= s^2c \left(\sqrt{3} V_{pd\sigma} - 2 V_{pd\pi} \right) + c V_{pd\pi}, \\
\label{t_xxy}
t_{x,xy} &= s^3 \left(\sqrt{3} V_{pd\sigma} - 2 V_{pd\pi} \right) + s V_{pd\pi}
.
\end{align}
Here,
$s = \sin \left( \kappa_0 + \delta\theta \right)/\sqrt{2}$, $c = \cos \left( \kappa_0 + \delta\theta \right)$, and $\cos^2 \kappa_0 = 1/3$ are the structural parameters describing the distortion of CuO$_4$ tetrahedra.
$\sin ( \kappa_0 + \delta\theta ) = 0.880$ and $\cos ( \kappa_0 + \delta\theta ) = 0.474$ are estimated by the structural parameters in Ba$_2$CuGe$_2$O$_7$
\cite{Zheludev_PRB54}.
We used $V_{pd\sigma} = -1.5$ eV and $V_{pd\pi} = -0.45 V_{pd\sigma}$
\cite{Maiti_EPL37, Mizokawa_JESRP92}
in Eqs. (\ref{t_xyz}) - (\ref{t_xxy}) to estimate the energy scheme.
Diagonalizing the matrix of Eq. (\ref{Matrix of CEF + d-p}), we obtains the eigenenergies as
\begin{align}
\label{Matrix of CEF + d-p diag}
H_\mathrm{CEF} + H_{d-p}
= \bordermatrix{
& \psi_1	&\psi_2	&\psi_3	&\psi_4	&\psi_5	&\psi_6	&\psi_7	&\psi_8	&\psi_{A_1}	&\psi_{A_2}	&\psi_{B_1} \cr
& \epsilon_{pE}	&	&	&	&	&	&	&	&	&	&	\cr
&	& \epsilon_{pE}	&	&	&	&	&	&	&	&	&	\cr
&	&	& \epsilon_E^+	&	&	&	&	&	&	&	&	\cr
&	&	&	& \epsilon_E^+	&	&	&	&	&	&	&	\cr
&	&	&	& 	& \epsilon_E^-	&	&	&	&	&	&	\cr
&	&	&	& 	&	& \epsilon_E^-	&	&	&	&	&	\cr
&	&	&	&	&	& 	& \epsilon_{B_2}^+	&	&	&	&	\cr
&	&	&	&	&	& 	&	& \epsilon_{B_2}^-	&	&	&	\cr
&	&	&	&	&	&	&	&	&  \epsilon_{pA_1}	& 	& \cr
&	&	&	&	&	&	&	&	&	& \epsilon_{pA_2}		& \cr
&	&	&	&	&	&	&	&	&	&	& \epsilon_{pB_1}
}.
\end{align}
Here, zeros in the off-diagonal elements are omitted for simplicity.
The eigenenergies are described as follows:
\begin{align}
\epsilon_E^\pm = \frac{  \epsilon_{dE} + \epsilon_{pE} }{2} \pm \delta \epsilon_E,
\\
\delta \epsilon_E = \sqrt{ \left( \frac{ \Delta\epsilon_{E} }{2} \right)^2 + 4 \left[ \left( t_{x,yz} \right)^2 + \left( t_{x,zx} \right)^2 \right] },
\\
\Delta\epsilon_{E} = \epsilon_{pE} - \epsilon_{dE}, 
\\
\epsilon_{B_2}^\pm = \frac{  \epsilon_{dB_2} + \epsilon_{pB_2} }{2} \pm \delta \epsilon_{B_2},
\\
\delta \epsilon_{B_2} = \sqrt{ \left( \frac{ \Delta\epsilon_{B_2} }{2} \right)^2 + 8 \left( t_{x,xy} \right)^2 },
\\
\Delta\epsilon_{B_2} = \epsilon_{pB_2} - \epsilon_{dB_2}
.
\end{align}
Thus, the energy of six orbitals is shifted due to $H_{d-p}$.
The eigenfunctions describing CuO$_4$ molecular orbitals and those coefficients are written as 
\begin{align}
\label{psi1}
\psi_1 \left( \boldsymbol{r}_1, \boldsymbol{r}_2, \boldsymbol{r}_3, \boldsymbol{r}_4 \right)
= \frac{ t_{x,zx}}{ \sqrt{ \left( t_{x,yz}\right)^2 + \left( t_{x,zx}\right)^2 } } \psi_{E(1+)} \left( \boldsymbol{r}_1, \boldsymbol{r}_2, \boldsymbol{r}_3, \boldsymbol{r}_4 \right)
 -  \frac{ t_{x,yz}}{ \sqrt{ \left( t_{x,yz}\right)^2 + \left( t_{x,zx}\right)^2 } } \psi_{E(2+)} \left( \boldsymbol{r}_1, \boldsymbol{r}_2, \boldsymbol{r}_3, \boldsymbol{r}_4 \right),
\\
\label{psi2}
\psi_2
= \frac{ t_{x,zx}}{ \sqrt{ \left( t_{x,yz}\right)^2 + \left( t_{x,zx}\right)^2 } } \psi_{E(1-)}
 -  \frac{ t_{x,yz}}{ \sqrt{ \left( t_{x,yz}\right)^2 + \left( t_{x,zx}\right)^2 } } \psi_{E(2-)} 
\\
\label{psi3}
\psi_3
= \frac{1}{ 2C_{\epsilon_E^+} } \left( \frac{ \Delta \epsilon_E }{2} - \delta\epsilon_E \right) \psi_{yz}
+ \frac{1}{ 2C_{\epsilon_E^+} } \left( \frac{ \Delta \epsilon_E }{2} - \delta\epsilon_E \right) \psi_{zx}
- \frac{ \sqrt{2}t_{x,yz} }{ C_{\epsilon_E^+} } \psi_{E(1+)}
- \frac{ \sqrt{2}t_{x,zx} }{ C_{\epsilon_E^+} } \psi_{E(2+)}
\\
\label{psi4}
\psi_4
= \frac{1}{ 2C_{\epsilon_E^+} } \left( \frac{ \Delta \epsilon_E }{2} - \delta\epsilon_E \right) \psi_{yz}
- \frac{1}{ 2C_{\epsilon_E^+} } \left( \frac{ \Delta \epsilon_E }{2} - \delta\epsilon_E \right) \psi_{zx}
- \frac{ \sqrt{2}t_{x,yz} }{ C_{\epsilon_E^+} } \psi_{E(1-)}
- \frac{ \sqrt{2}t_{x,zx} }{ C_{\epsilon_E^+} } \psi_{E(2-)}
\\
\label{psi5}
\psi_5
= \frac{1}{ 2C_{\epsilon_E^-} } \left( \frac{ \Delta \epsilon_E }{2} + \delta\epsilon_E \right) \psi_{yz}
+ \frac{1}{ 2C_{\epsilon_E^-} } \left( \frac{ \Delta \epsilon_E }{2} + \delta\epsilon_E \right) \psi_{zx}
- \frac{ \sqrt{2}t_{x,yz} }{ C_{\epsilon_E^-} } \psi_{E(1+)}
- \frac{ \sqrt{2}t_{x,zx} }{ C_{\epsilon_E^-} } \psi_{E(2+)}
\\
\label{psi6}
\psi_6
= \frac{1}{ 2C_{\epsilon_E^-} } \left( \frac{ \Delta \epsilon_E }{2} + \delta\epsilon_E \right) \psi_{yz}
- \frac{1}{ 2C_{\epsilon_E^-} } \left( \frac{ \Delta \epsilon_E }{2} + \delta\epsilon_E \right) \psi_{zx}
- \frac{ \sqrt{2}t_{x,yz} }{ C_{\epsilon_E^-} } \psi_{E(1-)}
- \frac{ \sqrt{2}t_{x,zx} }{ C_{\epsilon_E^-} } \psi_{E(2-)}
\\
\label{psi7}
\psi_7
= \frac{1}{ C_{\epsilon_{B_2}} } \left( \frac{ \Delta \epsilon_{B_2} }{2} - \delta\epsilon_{B_2} \right) \psi_{xy}
- \frac{ 2\sqrt{2}t_{x,xy} }{ C_{\epsilon_{B_2}} } \psi_{B_2}
\\
\label{psi8}
\psi_8
= \frac{ 2\sqrt{2}t_{x,xy} }{ C_{\epsilon_{B_2}} } \psi_{xy}
+ \frac{1}{ C_{\epsilon_{B_2}} } \left( \frac{ \Delta \epsilon_{B_2} }{2} - \delta\epsilon_{B_2} \right) \psi_{B_2}
\\
\label{psiE}
C_{\epsilon_E^\pm} 
= \sqrt{
\frac{1}{2} \left( \frac{ \Delta\epsilon_E }{2} \mp \delta\epsilon_E \right)^2
+ \sqrt{2} \left( t_{x,yz} \right)^2 + \sqrt{2} \left( t_{x,zx} \right)^2
}
\\
\label{psiB2}
C_{\epsilon_{B_2}} 
= \sqrt{
\left( \frac{ \Delta\epsilon_{B_2} }{2} - \delta\epsilon_{B_2} \right)^2
+ 8 \left( t_{x,xy} \right)^2
}
.
\end{align}
Here, $\left( \boldsymbol{r}_1, \boldsymbol{r}_2, \boldsymbol{r}_3, \boldsymbol{r}_4 \right)$ in Eqs. (\ref{psi2}) - (\ref{psiB2}) is ommited for simplicity.

Considering these molecular orbitals with the irrep $E$, we can describe the matrices of multipoles as follows:
\begin{align}
\label{Ov_Molecular_2}
\boldsymbol{O}_{x^2-y^2}
&= C_{\epsilon_E^+ }^{-2}  \bordermatrix{
&\psi_3	&\psi_4	\cr
&0	&\frac{\sqrt{2} }{14} r_d^2 \left( \frac{\Delta \epsilon_E }{2} - \delta \epsilon_E \right)^2
		+ \frac{2\sqrt{2} }{5} r_p^2  \left( t_{x, yz}^2 - t_{x, zx}^2 \right) \cr
	&\frac{\sqrt{2} }{14} r_d^2 \left( \frac{\Delta \epsilon_E }{2} - \delta \epsilon_E \right)^2
		+ \frac{2\sqrt{2} }{5} r_p^2\left( t_{x, yz}^2 - t_{x, zx}^2 \right)	&0	\cr
}, \\
\label{Oxy_Molecular_2}
\boldsymbol{O}_{xy}
&= C_{\epsilon_E^+ }^{-2} \left(
\begin{array}{cc}
\frac{\sqrt{2} }{14} r_d^2 \left( \frac{\Delta \epsilon_E }{2} - \delta \epsilon_E \right)^2
+ \frac{2\sqrt{2} }{5} r_p^2 \left( t_{x, yz}^2 + t_{x, zx}^2 \right)	&0	\cr
0	&- \frac{\sqrt{2} }{14} r_d^2 \left( \frac{\Delta \epsilon_E }{2} - \delta \epsilon_E \right)^2
			- \frac{2\sqrt{2} }{5} r_p^2 \left( t_{x, yz}^2 + t_{x, zx}^2 \right)	\cr
\end{array}
\right), \\
\label{Pz_Molecular_2}
\boldsymbol{P}_{z}
&=C_{\epsilon_E^+ }^{-2} \left(
\begin{array}{cc}
-4pt_{x,yz} \left( \frac{\Delta \epsilon_E }{2} - \delta \epsilon_E \right)	&0	\cr
0	&4pt_{x,yz} \left( \frac{\Delta \epsilon_E }{2} - \delta \epsilon_E \right)	\cr
\end{array}
\right).
\end{align}
Using the z-component of the Pauli matrix,
\begin{align}
\label{sigmaz}
\boldsymbol{\sigma}_z
= \left(
\begin{array}{cc}
1	&0	\cr
0	&-1	\cr
\end{array}
\right),
\end{align}
we obtain the relations between the electric multipoles and the Paul matrix, $\boldsymbol{O}_{xy} \propto \boldsymbol{\sigma}_z$ and $\boldsymbol{P}_{z} \propto \boldsymbol{\sigma}_z$.
Therefore, anomalous elasticity as a result of electric quadrupoles in addition to the dipole responses is expected in the CuO$_4$ molecular orbitals.
Because CuO$_4$ tetrahedra are compressed along the $c$-axis, the eigenenergies of the $yz$ and $zx$ orbitals for $H_\mathrm{CEF}$ are lower than that of the $xy$ orbital.
On the other hand, $\psi_3$ and $\psi_4$ wave functions are mainly constructed by $yz$ and $zx$ orbitals.
We deduce that these energy levels for $H_\mathrm{CEF}$ can be the reason why the Hilbert space spanned by $\psi_3$ and $\psi_4$ becomes important both in zero and high-magnetic fields.

\section{
\label{Appendix_C}
Spin-orbit coupling and eigenfunctions
}

In this section, we describe the spin-orbit coupling for the molecular orbitals.
First, we write the matrix of spin-orbit coupling for the $yz$, $zx$, and $xy$ orbitals as
\begin{align}
\label{HSO_for 3d}
H_\mathrm{SO}
= \frac{\lambda_\mathrm{SO} }{2}
\bordermatrix{
&\psi_{yz, \uparrow}	&\psi_{zx, \uparrow}	&\psi_{xy, \downarrow}	&\psi_{yz, \downarrow}	&\psi_{zx, \downarrow}	&\psi_{xy, \uparrow}	\cr
&0	&i	&-1	&0	&0	&0	\cr
&-i	&0	&i	&0	&0	&0	\cr
&-1	&-i	&0	&0	&0	&0	\cr
&0	&0	&0	&0	&-i	&1	\cr
&0	&0	&0	&i	&0	&i	\cr
&0	&0	&0	&1	&-i	&0	\cr
}
\end{align}
Thus, we can write the matrix of $H (= H_\mathrm{CEF} + H_{d-p}) + H_\mathrm{SO}$ as below:
\begin{align}
\label{Matrix of HSO}
H + H_\mathrm{SO}
= \bordermatrix{
& \psi_{3, \uparrow}	&\psi_{4, \uparrow}	&\psi_{5, \uparrow}	&\psi_{6, \uparrow}	&\psi_{7, \downarrow}	&\psi_{8, \downarrow} \cr
& \epsilon_E^+	& -i\alpha_E^+	& 0	& -i\beta_E	& e^{i \frac{3\pi}{4} } \gamma^+	& e^{i \frac{3\pi}{4} } \rho^+	\cr
& i\alpha_E^+	& \epsilon_E^+	& i\beta_E	& 0	& e^{-i \frac{3\pi}{4} } \gamma^+	& e^{-i \frac{3\pi}{4} } \rho^+	\cr
& 0	& -i\beta_E	& \epsilon_E^-	& -i\alpha_E^-	& e^{i \frac{3\pi}{4} } \gamma^-	& e^{i \frac{3\pi}{4} } \rho^-	\cr
& i\beta_E	& 0	& i\alpha_E^-	& \epsilon_E^-	& e^{-i \frac{3\pi}{4} } \gamma^-	& e^{-i \frac{3\pi}{4} } \rho^-	\cr
&  e^{-i \frac{3\pi}{4} } \gamma^+	&  e^{i \frac{3\pi}{4} } \gamma^+	&  e^{-i \frac{3\pi}{4} } \gamma^-	&  e^{i \frac{3\pi}{4} } \gamma^- 	& \epsilon_{B_2}^+	& 0	\cr
& e^{-i \frac{3\pi}{4} } \rho^+	& e^{i \frac{3\pi}{4} } \rho^+	& e^{-i \frac{3\pi}{4} } \rho^-	& e^{i \frac{3\pi}{4} } \rho^-	& 0	& \epsilon_{B_2}^-
}
\oplus (\mathrm{C.C.}).
\end{align}
Here, we set matrix elements for the convenience as below:
\begin{align}
\alpha_E^\pm &= \frac{ \lambda_\mathrm{SO} }{4\left( C_{\epsilon_E^\pm} \right)^2 }
	\left( \frac{ \Delta\epsilon_E }{2} \mp \delta\epsilon_E \right)^2, 
\\
\beta_E &= \frac{ \lambda_\mathrm{SO} }{4C_{\epsilon_E^+}C_{\epsilon_E^-} } \left[ \left( \frac{ \Delta\epsilon_E }{2} \right)^2 - \delta\epsilon_E^2 \right],
\\
\gamma^\pm &=  \frac{ \lambda_\mathrm{SO} }{2\sqrt{2} C_{\epsilon_E^\pm}C_{\epsilon_{B_2}} } 
	\left( \frac{ \Delta\epsilon_E }{2} \mp \delta\epsilon_E \right) \left( \frac{ \Delta\epsilon_{B_2} }{2} - \delta\epsilon_{B_2} \right),
\\
\rho^\pm &=  \frac{ \lambda_\mathrm{SO} }{ C_{\epsilon_E^\pm}C_{\epsilon_{B_2}} } 
	\left( \frac{ \Delta\epsilon_E }{2} \mp \delta\epsilon_E \right)t_{x,xy}^1
.
\end{align}
We ignore $\psi_1$ and $\psi_2$ because of the absence of $3d$ orbitals in these wave functions.
We deduce that the matrix elements between the high-energy states of $\psi_3$, $\psi_4$, and $\psi_7$ and low-energy states of $\psi_5$, $\psi_6$, and $\psi_8$ are negligible because the spin-orbit coupling constant $\lambda_\mathrm{SO} \sim 0.1$ eV is much smaller than the energy gap of $\sim 6$ eV.
Thus, we can write $H + H_\mathrm{SO}$ as following simple form: 
\begin{align}
\label{Matrix of HSO approx}
H + H_\mathrm{SO}
= \bordermatrix{
& \psi_{3, \uparrow}	&\psi_{4, \uparrow}	&\psi_{7, \downarrow}	&\psi_{5, \uparrow}	&\psi_{6, \uparrow}	&\psi_{8, \downarrow} \cr
& \epsilon_E^+	& -i\alpha_E^+	& e^{i \frac{3\pi}{4} } \gamma^+	&0	&0	&0	\cr
& i\alpha_E^+	& \epsilon_E^+	& e^{-i \frac{3\pi}{4} } \gamma^+	&0	&0	&0	\cr
& e^{-i \frac{3\pi}{4} } \gamma^+	& e^{i \frac{3\pi}{4} } \gamma^+	& \epsilon_{B_2}^+	&0	&0	&0	\cr
&0	&0	&0	& \epsilon_E^-	&-i\alpha_E^- 	& e^{i \frac{3\pi}{4} } \rho^-	\cr
&0	&0	&0	&  i \alpha_E^- 	& \epsilon_E^-	& e^{-i \frac{3\pi}{4} } \rho^-	\cr
&0	&0	&0	& e^{-i \frac{3\pi}{4} } \rho^-	& e^{i \frac{3\pi}{4} } \rho^-	& \epsilon_{B_2}^- 	\cr
}
\oplus (\mathrm{C.C.})
.
\end{align}
To diagonalize $H + H_\mathrm{SO}$, we obtain the eigenenergies and eigenstates described as follows:
\begin{align}
\label{Matrix of HSO diagonalized}
H + H_\mathrm{SO}
=\bordermatrix{
& \psi_\mathrm{SO}^{1+}	& \psi_\mathrm{SO}^{1-}	& \psi_\mathrm{SO}^{2+}	& \psi_\mathrm{SO}^{2-}	& \psi_\mathrm{SO}^{3+}	& \psi_\mathrm{SO}^{3-}	& \psi_\mathrm{SO}^{4+}	& \psi_\mathrm{SO}^{4-}	& \psi_\mathrm{SO}^{5+}	& \psi_\mathrm{SO}^{5-}	& \psi_\mathrm{SO}^{6+}	& \psi_\mathrm{SO}^{6-} \cr
& \epsilon_\mathrm{SO}^1	&	&	&	&	&	&	&	&	&	&	&	\cr
&	& \epsilon_\mathrm{SO}^1	&	&	&	&	&	&	&	&	&	&	\cr
&	&	& \epsilon_\mathrm{SO}^2	&	&	&	&	&	&	&	&	&	\cr
&	&	&	& \epsilon_\mathrm{SO}^2	&	&	&	&	&	&	&	&	\cr
&	&	&	&	& \epsilon_\mathrm{SO}^3	&	&	&	&	&	&	&	\cr
&	&	&	&	&	& \epsilon_\mathrm{SO}^3	&	&	&	&	&	&	\cr
&	&	&	&	&	&	& \epsilon_\mathrm{SO}^4	&	&	&	&	&	\cr
&	&	&	&	&	&	&	& \epsilon_\mathrm{SO}^4	&	&	&	&	\cr
&	&	&	&	&	&	&	&	& \epsilon_\mathrm{SO}^5	&	&	&	\cr
&	&	&	&	&	&	&	&	&	& \epsilon_\mathrm{SO}^5	&	&	\cr
&	&	&	&	&	&	&	&	&	&	& \epsilon_\mathrm{SO}^6	&	\cr
&	&	&	&	&	&	&	&	&	&	&	& \epsilon_\mathrm{SO}^6	\cr
}
\end{align}
\begin{align}
\label{SO_1+}
\psi_\mathrm{SO}^{1+} 
&= \frac{1}{ \sqrt{2} } \left( -\psi_{3, \uparrow} + i \psi_{4, \uparrow} \right)
, \\
\psi_\mathrm{SO}^{2+}
&= \frac{1}{ C_{\epsilon_\mathrm{SO}^2} }
\left[
\gamma^+ \left( i\psi_{3, \uparrow} - \psi_{4, \uparrow} \right)
	+ e^{i\frac{3\pi}{4}} \left( \frac{ \epsilon_E^+ - \epsilon_{B_2}^+ +  \alpha_E^+ }{2} - \delta\epsilon_\mathrm{SO}^+ \right) \psi_{7, \downarrow} 
\right]
, \\
\psi_\mathrm{SO}^{3+}
&= \frac{1}{ C_{\epsilon_\mathrm{SO}^3} }
\left[
\gamma^+ \left( i\psi_{3, \uparrow} - \psi_{4, \uparrow} \right)
	+ e^{i\frac{3\pi}{4}} \left( \frac{ \epsilon_E^+ - \epsilon_{B_2}^+ +  \alpha_E^+ }{2} + \delta\epsilon_\mathrm{SO}^+ \right) \psi_{7, \downarrow} 
\right]
, \\
\psi_\mathrm{SO}^{4+} 
&= \frac{1}{ \sqrt{2} } \left( \psi_{5, \uparrow} - i \psi_{6, \uparrow} \right)
, \\
\psi_\mathrm{SO}^{5+}
&= \frac{1}{ C_{\epsilon_\mathrm{SO}^5} }
\left[
\rho^- \left( i\psi_{5, \uparrow} - \psi_{6, \uparrow} \right)
	+ e^{i\frac{3\pi}{4}} \left( \frac{ \epsilon_E^- - \epsilon_{B_2}^- +  \alpha_E^- }{2} - \delta\epsilon_\mathrm{SO}^- \right) \psi_{8, \downarrow} 
\right]
, \\
\label{6+}
\psi_\mathrm{SO}^{6+}
&= \frac{1}{ C_{\epsilon_\mathrm{SO}^6} }
\left[
\rho^- \left( i\psi_{5, \uparrow} - \psi_{6, \uparrow} \right)
	+ e^{i\frac{3\pi}{4}} \left( \frac{ \epsilon_E^- - \epsilon_{B_2}^- +  \alpha_E^- }{2} + \delta\epsilon_\mathrm{SO}^- \right) \psi_{8, \downarrow} 
\right]
, \\
\psi_\mathrm{SO}^{j-} 
&= T\psi_\mathrm{SO}^{j+}
\end{align}
\begin{align}
\epsilon_\mathrm{SO}^1 
&= \epsilon_E^+ - \alpha_E^+
, \\
\epsilon_\mathrm{SO}^2
&=  \frac{ \epsilon_E^+ + \epsilon_{B_2}^+ +  \alpha_E^+ }{2} + \delta\epsilon_\mathrm{SO}^+
, \\
\epsilon_\mathrm{SO}^3
&=  \frac{ \epsilon_E^+ + \epsilon_{B_2}^+ +  \alpha_E^+ }{2} - \delta\epsilon_\mathrm{SO}^+
, \\
\epsilon_\mathrm{SO}^4 
&= \epsilon_E^- - \alpha_E^-
, \\
\epsilon_\mathrm{SO}^5
&=  \frac{ \epsilon_E^- + \epsilon_{B_2}^- +  \alpha_E^- }{2} + \delta\epsilon_\mathrm{SO}^-
, \\
\epsilon_\mathrm{SO}^6
&=  \frac{ \epsilon_E^- + \epsilon_{B_2}^- +  \alpha_E^- }{2} - \delta\epsilon_\mathrm{SO}^-
, \\
\delta\epsilon_\mathrm{SO}^+
&= \sqrt{
\left( \frac{ \epsilon_E^+ - \epsilon_{B_2}^+ +  \alpha_E^+ }{2}  \right)^2 + 2\left(\gamma^+ \right)^2
}
, \\
\delta\epsilon_\mathrm{SO}^-
&= \sqrt{
\left( \frac{ \epsilon_E^- - \epsilon_{B_2}^- +  \alpha_E^- }{2}  \right)^2 + 2\left(\rho^- \right)^2
}
,\\
C_{\epsilon_\mathrm{SO}^2} 
&= \sqrt{
2\left( \gamma^+ \right)^2 + \left( \frac{ \epsilon_E^+ - \epsilon_{B_2}^+ +  \alpha_E^+ }{2} - \delta\epsilon_\mathrm{SO}^+ \right)^2
}
,\\
C_{\epsilon_\mathrm{SO}^3} 
&= \sqrt{
2\left( \gamma^+ \right)^2 + \left( \frac{ \epsilon_E^+ - \epsilon_{B_2}^+ +  \alpha_E^+ }{2} + \delta\epsilon_\mathrm{SO}^+ \right)^2
}
,\\
C_{\epsilon_\mathrm{SO}^5} 
&= \sqrt{
2\left( \rho^- \right)^2 + \left( \frac{ \epsilon_E^- - \epsilon_{B_2}^- +  \alpha_E^- }{2} - \delta\epsilon_\mathrm{SO}^- \right)^2
}
,\\
C_{\epsilon_\mathrm{SO}^6} 
&= \sqrt{
2\left( \rho^- \right)^2 + \left( \frac{ \epsilon_E^- - \epsilon_{B_2}^- +  \alpha_E^- }{2} + \delta\epsilon_\mathrm{SO}^- \right)^2
}
.
\end{align}
Here, $T$ indicates the time reversal operator.
We omitted zero elements in this matrix for convenience.
These wave functions are also written by the $yz$, $zx$, and $xy$ orbitals and the molecular orbitals.
Taking into account the construction of the wave functions $\psi_3$, $\psi_4$, and $\psi_7$ in Table \ref{Coefficients of Wave Functions}, we show the norm of the coefficients of such wave functions (see Table \ref{Coefficients of Wave Functions from SO to yz zx at 0 T}).
\begin{table*}[htbp]
\caption{
Norm of the coefficients of $\psi_{yz, \uparrow}$, $\psi_{zx \uparrow}$, $\psi_{E(1+) \uparrow}$, $\psi_{E(1-) \uparrow}$, $\psi_{E(2+) \uparrow}$, $\psi_{E(2-) \uparrow}$, $\psi_{xy \uparrow}$, and $\psi_{B_2 \uparrow}$ that constitute the wave function $\psi_\mathrm{SO}^{i\pm}$ ($i = 1$ - $3$) at 0 T for $A$ site.
}
\begin{ruledtabular}
\label{Coefficients of Wave Functions from SO to yz zx at 0 T}
\begin{tabular}{c|cccccccc}

	&$\psi_{yz, \uparrow}$ 
		& $\psi_{zx, \uparrow}$ 
			& $\psi_{E(1+), \uparrow}$ 
				& $\psi_{E(1-), \uparrow}$
					& $\psi_{E(2+), \uparrow}$ 
						& $\psi_{E(2-), \uparrow}$ 
							&$\psi_{xy, \uparrow}$
								&$\psi_{B_2, \uparrow}$
\\
\hline
$\psi_\mathrm{SO}^{1+}$
	&0.66
		&0.66
			& 0.22
				& 0.22
					& 0.12
						& 0.12
							& 0
								&0
\\
$\psi_\mathrm{SO}^{1-}$
	&0
		&0
			& 0
				& 0
					& 0
						& 0
							& 0
								&0
\\
$\psi_\mathrm{SO}^{2+}$
	&0.66
		&0.66
			& 0.22
				& 0.22
					& 0.12
						& 0.12
							& 0
								&0
\\
$\psi_\mathrm{SO}^{2-}$
	&0
		&0
			& 0
				& 0
					& 0
						& 0
							& 0.10
								&0.03
\\
$\psi_\mathrm{SO}^{3+}$
	&0.07
		&0.07
			&0.02
				&0.02
					&0.01
						&0.01
							&0
								&0
\\
$\psi_\mathrm{SO}^{3-}$
	&0
		&0
			& 0
				& 0
					& 0
						& 0
							& 0.95
								&0.29
\end{tabular}
\end{ruledtabular}
\end{table*}

For these states, we can calculate the matrix elements of $\boldsymbol{O}_{x^2-y^2}$, $\boldsymbol{O}_{xy}$, and $\boldsymbol{P}_{z}$ as below:
\begin{align}
\label{Ov for SO calc}
\boldsymbol{O}_{x^2-y^2} 
&= \bordermatrix{
& \psi_\mathrm{SO}^{1+}	& \psi_\mathrm{SO}^{1-}	& \psi_\mathrm{SO}^{2+}	& \psi_\mathrm{SO}^{2-}	& \psi_\mathrm{SO}^{3+}	& \psi_\mathrm{SO}^{3-}	\cr
&0	&0	&-0.15 + 0.18i	&0	&-0.02 + 0.02i	&0	\cr
&0	&0	&0	&-0.15 - 0.18i	&0	&0.02 + 0.02i		\cr
&-0.15 - 0.18i	&0	&0	&0	&0	&0	\cr
&0	&-0.15 - 0.18i	&0	&0	&0	&0	\cr
&-0.02 - 0.02i	&0	&0	&0	&0	&0	\cr
&0	&0.02 - 0.02i	&0	&0	&0	&0	\cr
}
,	\\
\label{Oxy for SO calc}
\boldsymbol{O}_{xy} 
&= \left(
\begin{array}{cccccc}
0	&0	&0.20 + 0.16i	&0	&0.02 + 0.02i	&0	\cr
0	&0	&0	&0.20 - 0.16i	&0	&-0.02 + 0.02i	\cr
0.20 - 0.16i	&0	&0	&0	&0	&0	\cr
0	&0.20 + 0.16i	&0	&0	&0	&0	\cr
0.02 - 0.02i	&0	&0	&0	&0	&	\cr
0	&-0.02 - 0.02i	&0	&0	&0	&0	\cr
\end{array}
\right)
,	\\
\label{Pz for SO calc}
\boldsymbol{P}_z
&= \left(
\begin{array}{cccccc}
0	&0	&-0.31 - 0.25i	&0	&-0.03 - 0.03i	&0	\cr
0	&0	&0	&-0.31 + 0.25i	&0	&0.03 - 0.03i	\cr
-0.31 + 0.25i	&0	&0	&0	&0	&0	\cr
0	&-0.31 - 0.25i	&0	&0	&0	&0	\cr
-0.03 + 0.03i	&0	&0	&0	&0	&0	\cr
0	&0.03 + 0.03i	&0	&0	&0	&0	\cr
\end{array}
\right)
.
\end{align}
To simplify the calculations, we computed the matrix elements with $r_d^2 = 1$ and $r_p^2 = 1$.
These real values are considered to be incorporated into the experimentally determined $p$ and $g_\mathrm{XY}$.
These three states carry the off-diagonal elements, indicating that the mixing of the wave functions by the Zeeman effect provides the multipole degrees of freedom.

\section{
\label{Appendix_D}
Quantum states in magnetic fields
}

Here, we consider the Zeeman effects for the spin-dependent quantum states.
The Zeeman effect 
$H_\mathrm{Zeeman} = - \mu_\mathrm{B} \left( \boldsymbol{l} + 2 \boldsymbol{s} \right) \cdot \boldsymbol{B}$
for the in-plane magnetic field directions 
$\boldsymbol{B} =  \left(B_0 \cos \phi, B_0 \sin \phi, 0 \right)$
is written in the following form:
\begin{align}
H_\mathrm{Zeeman}
 = - \mu_\mathrm{B} B_0 \left( l_x \cos \phi + l_y \sin \phi \right)
-2\mu_\mathrm{B} B_0 \left( e^{-i \phi} s_+ + e^{i \phi} s_- \right)
\end{align}
Here, $s_+ = s_x + is_y$ and $s_- = s_x - is_y$ are ladder operators.
If the strong magnetic field is applied along the [110] direction, we deduce that the point group symmetry lowers from $D_{2d}$ to $D_2$.
In the $D_2$, both electric dipole $P_z$ and electric quadrupole $O_{xy}$ belong to the irrep $B_1$.
Even if the point group symmetry shows indeed lowering, we can still expect the FIEP and elastic softening.
For the $yz$, $zx$, and $xy$ orbitals of $3d$ electrons, the matrix of $H_\mathrm{Zeeman}$ is written as
\begin{align}
H_\mathrm{Zeeman}^d
= -\mu_\mathrm{B}B_0 \bordermatrix{
& \psi_{yz, \uparrow}	& \psi_{zx, \uparrow}	& \psi_{xy, \downarrow}	& \psi_{yz, \downarrow}	& \psi_{zx, \downarrow}	& \psi_{xy, \uparrow} \cr
&0	&0	&-i \sin\phi	&2e^{-i\phi}	&0	&0 \cr
&0	&0	&i \cos\phi	&0	&2e^{-i\phi}	&0 \cr
&i \sin\phi	&-i \cos\phi	&0	&0	&0	&2e^{-i\phi} \cr
&2e^{i\phi}	&0	&0	&0	&0	&-i \sin\phi \cr
&0	&2e^{i\phi}	&0	&0	&0	&i \cos\phi \cr
&0	&0	&2e^{i\phi}	&i \sin\phi	&-i \cos\phi	&
}.
\end{align}
The matrix of $H_\mathrm{Zeeman}$ for the $x$ and $y$ orbitals of $2p$ electrons is also written as
\begin{align}
H_\mathrm{Zeeman}^p
= -\mu_\mathrm{B}B_0 \bordermatrix{
& \psi_{x, \uparrow}	& \psi_{y, \uparrow}	& \psi_{x, \downarrow}	& \psi_{y, \downarrow} \cr
&0	&0	&2e^{-i\phi}	&0 \cr
&0	&0	&0	&2e^{-i\phi} \cr
&2e^{i\phi}	&0	&0	&0 \cr
&0	&2e^{i\phi}	&0	&0 \cr
}.
\end{align}

Thus, we obtain the matrix of $H + H_\mathrm{SO} + H_\mathrm{Zeeman}$ as follow:
\begin{align}
\label{Matrix of HSO + HZeeman}
&H + H_\mathrm{SO} + H_\mathrm{Zeeman}
\nonumber	\\
&= \bordermatrix{
& \psi_{3, \uparrow}	&\psi_{4, \uparrow}	&\psi_{7, \downarrow}	&\psi_{3, \downarrow}	&\psi_{4, \downarrow}	&\psi_{7, \uparrow} \cr
& \epsilon_E^+	& -i\alpha_E^+	& e^{i \frac{3\pi}{4} } \gamma^+	& -2 \mu_\mathrm{B} B_0 e^{-i \phi}	&0	& i \mu_\mathrm{B} B_0 g\phi_-	\cr
& i\alpha_E^+	& \epsilon_E^+	& e^{-i \frac{3\pi}{4} } \gamma^+	&0	& -2 \mu_\mathrm{B} B_0 e^{-i \phi}	&-i \mu_\mathrm{B} B_0 g\phi_+	\cr
& e^{-i \frac{3\pi}{4} } \gamma^+	& e^{i \frac{3\pi}{4} } \gamma^+	& \epsilon_{B_2}^+	&-i \mu_\mathrm{B} B_0 g\phi_-	& i \mu_\mathrm{B} B_0 g\phi_+	& -2 \mu_\mathrm{B} B_0 e^{-i \phi}	\cr
& -2 \mu_\mathrm{B} B_0 e^{i \phi} 	&0	&i \mu_\mathrm{B} B_0 g\phi_-	& \epsilon_E^+	&i\alpha_E^+ 	& -e^{-i \frac{3\pi}{4} } \gamma^+	\cr
&0	& -2 \mu_\mathrm{B} B_0 e^{i \phi}	&-i \mu_\mathrm{B} B_0 g\phi_+	&  -i \alpha_E^+ 	& \epsilon_E^+	& -e^{i \frac{3\pi}{4} } \gamma^+	\cr
&-i \mu_\mathrm{B} B_0 g\phi_-	&i  \mu_\mathrm{B} B_0 g\phi_+	& -2 \mu_\mathrm{B} B_0 e^{i \phi}	& -e^{i \frac{3\pi}{4} } \gamma^+	& -e^{-i \frac{3\pi}{4} } \gamma^+	& \epsilon_{B_2}^+
}.
\end{align}
Here, we set several matrix elements for the convenience as
\begin{align}
g &=  \frac{ 1 }{2 C_{\epsilon_E^+}C_{\epsilon_{B_2}} } 
	\left( \frac{ \Delta\epsilon_E }{2} \mp \delta\epsilon_E \right) \left( \frac{ \Delta\epsilon_{B_2} }{2} - \delta\epsilon_{B_2} \right)
,\\
\phi_\pm &= \cos\phi \pm \sin\phi
.
\end{align}
$H + H_\mathrm{SO} + H_\mathrm{Zeeman}$ is hard to diagonalize analytically.
So, we obtain the eigenstates and eigenenergies to diagonalize this Hamiltonian numerically by Julia language.
The field dependence of eigenenergies for $B//[110]$ is shown in Figs, \ref{EnergyScheme}(b) - \ref{EnergyScheme}(d).
Based on these calculations, for example, we can describe the diagonal elements of these multipole matrices in a magnetic field of 50 T at  $A$ site (see Table \ref{Diagonal elements at 50T}).
\begin{table*}[t]
\caption{
Diagonal elements of the electric multipoles $O_{x^2-y^2}$, $O_{xy}$, and $P_z$ for the quantum states $\psi_\mathrm{SO}^{1+'}$, $\psi_\mathrm{SO}^{1-'}$, $\psi_\mathrm{SO}^{2+'}$, $\psi_\mathrm{SO}^{2-'}$, $\psi_\mathrm{SO}^{3+'}$, and $\psi_\mathrm{SO}^{3-'}$ at 50 T for $A$ site.
}
\begin{ruledtabular}
\label{Diagonal elements at 50T}
\begin{tabular}{c|cccccc}

	&$\psi_\mathrm{SO}^{1+'}$
		& $\psi_\mathrm{SO}^{1-'}$ 
			& $\psi_\mathrm{SO}^{2+'}$
				& $\psi_\mathrm{SO}^{2-'}$
					& $\psi_\mathrm{SO}^{3+'}$ 
						& $\psi_\mathrm{SO}^{3-'}$ 
\\
\hline
\\
$O_{x^2-y^2}$
	&$-3.20 \times 10^{-2}$
		& $3.17 \times 10^{-2}$
			&$3.32 \times 10^{-2}$
				&$-3.28 \times 10^{-2}$
					&$-5.10 \times 10^{-5}$
						&$5.00 \times 10^{-5}$
\\
$O_{xy}$
	&$9.31 \times 10^{-3}$
		& $-9.25 \times 10^{-3}$
			&$-8.12 \times 10^{-4}$
				&$7.45 \times 10^{-4}$
					&$1.05 \times 10^{-4}$
						&$-0.96 \times 10^{-4}$
\\
$P_z$
	&$-1.43 \times 10^{-2}$
		& $1.42 \times 10^{-2}$
			&$1.25 \times 10^{-3}$
				&$-1.14 \times 10^{-3}$
					&$-1.61 \times 10^{-4}$
						&$1.47 \times 10^{-4}$
\end{tabular}
\end{ruledtabular}
\end{table*}

\section{
\label{Appendix_E}
Local and global coordinates of strains and quadrupoles
}

The strains for local coordinates are written by that for global coordinates as below:
\begin{align}
\label{Trasform_Strain}
&\left(
\begin{array}{ccc}
\varepsilon_{XX}^{A(B)}	&\varepsilon_{XY}^{A(B)}	&\varepsilon_{XZ}^{A(B)}	\cr
\varepsilon_{YX}^{A(B)}	&\varepsilon_{YY}^{A(B)}	&\varepsilon_{YZ}^{A(B)}	\cr
\varepsilon_{ZX}^{A(B)}	&\varepsilon_{ZY}^{A(B)}	&\varepsilon_{ZZ}^{A(B)}	\cr
\end{array}
\right)
\nonumber \\
&= \left(
\begin{array}{ccc}
\varepsilon_{xx} \cos^2 \kappa \mp \varepsilon_{xy} \sin 2\kappa + \varepsilon_{yy} \sin^2 \kappa
	&\varepsilon_{xy} \cos 2\kappa \pm \frac{1}{2} (\varepsilon_{xx} - \varepsilon_{yy}) \sin 2\kappa
		&\varepsilon_{zx} \cos \kappa \mp \varepsilon_{yz} \sin \kappa	\cr
\varepsilon_{xy} \cos 2\kappa \pm \frac{1}{2} (\varepsilon_{xx} - \varepsilon_{yy}) \sin 2\kappa
	&\varepsilon_{xx} \cos^2 \kappa \pm \varepsilon_{xy} \sin 2\kappa + \varepsilon_{yy} \sin^2 \kappa
		&\pm \varepsilon_{zx} \sin \kappa + \varepsilon_{yz} \cos \kappa	\cr
\varepsilon_{zx} \cos \kappa \mp \varepsilon_{yz} \sin \kappa
	&\pm \varepsilon_{zx} \sin \kappa + \varepsilon_{yz} \cos \kappa
		&\varepsilon_{zz}	\cr
\end{array}
\right)
\end{align}
Here, we set sign $= +(-)$ for $i = A(B)$.
These transformations lead to Eqs. (\ref{CoordinateTransformation_1}) and (\ref{CoordinateTransformation_2}).
Because of the transformation for the strains in Eq (\ref{Trasform_Strain}) and quadrupoles in Eqs (\ref{Define Quadrupole1}) - (\ref{Define Quadrupole4}), the quadrupole-strain coupling for the local coordinates in Eq. (\ref{HQS_v_local}) is described by the global coordinates as below:
\begin{align}
\label{H_QS_L_v}
H_\mathrm{QS}^\mathrm{L}
&= -g_{X^2-Y^2}O_{X^2-Y^2}^i \varepsilon_{X^2-Y^2}^i	\nonumber	\\
&= -g_{X^2-Y^2}
 \left( O_{x^2-y^2}\varepsilon_{x^2-y^2} \cos^2 2\kappa \mp O_{x^2-y^2}\varepsilon_{xy} \sin4 \kappa
 \mp O_{xy}\varepsilon_{x^-y^2} \sin4 \kappa + O_{xy}\varepsilon_{xy} \sin^2 2\kappa
 \right), \\
\label{H_QS_L_xy}
H_\mathrm{QS}^\mathrm{L}
&= -g_{XY}O_{XY}^B \varepsilon_{X^2-Y^2}^B	\nonumber	\\
&= -g_{X^2-Y^2}
\left( O_{x^2-y^2}\varepsilon_{x^2-y^2} \sin^2 2\kappa \pm O_{x^2-y^2}\varepsilon_{xy} \sin4\kappa
\pm O_{xy}\varepsilon_{x^-y^2} \sin4\kappa + O_{xy}\varepsilon_{xy} \cos^2 2\kappa
\right).
\end{align}
In addition, the derivative with respect to the strains and the electric field for global coordinates are written by those for local coordinates as below:
\begin{align}
\label{DerivativeTransformation_1}
\left(
\begin{array}{c}
\partial_{\varepsilon_{x^2 - y^2}}	\cr
\partial_{\varepsilon_{xy}}	\cr
\partial_{E_z}	\cr
\end{array}
\right)
= \boldsymbol{R}\left(2 \kappa \right) 
\left(
\begin{array}{c}
\partial_{\varepsilon_{X^2 - Y^2}^A}	\cr
\partial_{\varepsilon_{XY}^A}	\cr
\partial_{E_z}	\cr
\end{array}
\right)
, \\
\label{DerivativeTransformation_2}
\left(
\begin{array}{c}
\partial_{\varepsilon_{x^2 - y^2}}	\cr
\partial_{\varepsilon_{xy}}	\cr
\partial_{E_z}	\cr
\end{array}
\right)
= \boldsymbol{R}\left(-2 \kappa \right) 
\left(
\begin{array}{c}
\partial_{\varepsilon_{X^2 - Y^2}^B}	\cr
\partial_{\varepsilon_{XY}^B}	\cr
\partial_{E_z}	\cr
\end{array}
\right)
.
\end{align}
These derivatives lead to the elastic constants of Eqs. (\ref{Elastic_xy})  and (\ref{Elastic_v}) and the quadrupole susceptibility of Eq. (\ref{chi_Gamma}).

\end{widetext}


\end{document}